\documentclass{jfm}
\usepackage{graphicx}
\usepackage{epstopdf, epsfig}
\usepackage[dvipsnames]{xcolor}                                 
\usepackage{hyperref}                                           
\usepackage{subcaption}                                         
\usepackage{lipsum}                                             
\usepackage{xargs}                                              
\usepackage{color}                                              
\usepackage{comment}                                            
\usepackage{amsmath}                                            
\usepackage{xspace}                                             
\usepackage[normalem]{ulem}
\usepackage[english]{babel}
\usepackage{blindtext}
\usepackage[textwidth=0.7in,
  textsize=tiny,
  linecolor=OliveGreen!25,
  backgroundcolor=OliveGreen!25,
  bordercolor=OliveGreen!25]{todonotes}                         
\reversemarginpar

\newcommand{\karman}{von K\'{a}rm\'{a}n\xspace}                 
\newcommand{\intd}{\,\textrm{d}}                                
\newcommand{\pdiff}[2]{\frac{\partial #1}{\partial #2}}

\newcommand{\uu}{$\langle u'^2 \rangle$\xspace}
\newcommand{\vv}{$\langle v'^2 \rangle$\xspace}
\newcommand{\ww}{$\langle w'^2 \rangle$\xspace}
\newcommand{\uv}{$\langle u'v' \rangle$\xspace}

\newcommand{\uup}{$\langle u'^2 \rangle^+$\xspace}

\newcommand{\uvp}{$\langle u'v' \rangle^+$\xspace}

\definecolor{green}{RGB}{0,120,0} 
 
\def\A180{{                                                     
    \setbox0\hbox{------}                                       
    \rlap{\hbox to \wd0{\hss\footnotesize$\bigcirc$\hss}}\box0
}}

\def\B550{{                                                     
    \setbox0\hbox{------}                                       
    \rlap{\hbox to \wd0{\hss$\square$\hss}}\box0
}}

\def\C1000{{                                                    
    \setbox0\hbox{------}                                       
    \rlap{\hbox to \wd0{\hss$\triangledown$\hss}}\box0
}}

\def\D2000{{                                                    
    \setbox0\hbox{------}                                       
    \rlap{\hbox to \wd0{\hss$\triangle$\hss}}\box0
}}

\def\E5200{{                                                    
    \setbox0\hbox{------}                                       
    \rlap{\hbox to \wd0{\hss$\bigcirc$\hss}}\box0
}}

\shorttitle{Spectral analysis of the budget equation in high-$Re$  channel flows}
\title{Spectral analysis of the budget equation in turbulent channel flows at high \Rey}
\author{Myoungkyu Lee\aff{1}\aunote{Present address: Combustion Research Facility, Sandia National Laboratories, Livermore, CA 94550, USA} and Robert D. Moser\aff{1,2}\corresp{\email{rmoser@ices.utexas.edu}}}
\shortauthor{M. Lee and R. D. Moser}
\affiliation{\aff{1}Center for Predictive Engineering and Computational Sciences, Insititute for Computational Engineering and Sciences, The University of Texas at Austin, TX 78712, USA 
\aff{2}Department of Mechnanical Engineering, The University of Texas at Austin, TX 78712, USA
}

\begin{document}

\maketitle

\begin{abstract}
The transport equations for the variances of the velocity components
are investigated using data from direct numerical simulations of
incompressible channel flows at friction Reynolds number ($Re_\tau$)
up to $Re_\tau = 5200$.  Each term in the transport equation has been
spectrally decomposed to expose the contribution of turbulence at
different length scales to the processes governing the flow of energy
in the wall-normal direction, in scale and among components.

The outer-layer turbulence is dominated by very large-scale streamwise
elongated modes, which are consistent with the very large-scale
motions (VLSM) that have been observed by many others. The presence of
these VLSMs drive many of the characteristics of the turbulent energy
flows. Away from the wall, production occurs primarily in these
large-scale streamwise-elongated modes in the streamwise velocity, but
dissipation occurs nearly isotropically in both velocity components
and scale. For this to happen, the energy is transferred from the
streamwise elongated modes to modes with a range of orientations
through non-linear interactions, and then transferred to other
velocity components. This allows energy to be transferred more-or-less
isotropically from these large scales to the small scales at which
dissipation occurs. The VLSMs also transfer energy to the wall-region,
resulting in a modulation of the autonomous near-wall dynamics and
the observed Reynolds number dependence of the near-wall velocity
variances.

The near-wall energy flows are more complex, but are consistent with
the well-known autonomous near-wall dynamics that give rise to streaks
and streamwise vortices. Through the overlap region between outer and
inner layer turbulence, there is a self-similar structure to the
energy flows. The VLSM production occurs at spanwise scales that grow
with $y$. There is transport of energy away from the wall over a range
of scales that grows with $y$. And, there is transfer of energy to
small dissipative scales which grow like $y^{1/4}$, as expected from
Kolmogorov scaling. Finally, the small-scale near-wall processes
characterised by wavelengths less that 1000 wall units are largely
Reynolds number independent, while the larger-scale outer layer
process are strongly Reynolds number dependent. The interaction
between them appears to be relatively simple.
\end{abstract}


\section{Introduction}
\label{sec:intro}
High Reynolds number ($Re$) wall-bounded turbulence is an important
physical phenomenon, and has been challenging to study experimentally,
computationally and theoretically. The presence of the wall causes
strong anisotropy and inhomogeneity which invalidates the simple
Kolmogorov description of turbulence. Also, the length and time scales
of the turbulence near the wall get smaller, relative to outer scales,
as $Re$ increases, which increases demands on instrumentation or
computational resources in experimental or computations studies.
However, the study of wall-bounded turbulence at high $Re$ has greatly
advanced with the remarkable improvements in experimental techniques
and computing power over the last few decades.

Perhaps the most important characteristic of wall-bounded turbulence
at high $Re$ is the scale-separation between the near-wall and outer
turbulence. \citet{Millikan:1938tv} argued that the mean velocity
profile in the overlap region has a logarithmic variation with
wall-normal distance at infinitely high $Re$. Even though his
prediction is generally accepted, the universality of the \karman
constant, $\kappa$, has remained controversial
\citep{Nagib:2008bx,Marusic:2010bn,Monkewitz:2017el,Luchini:2018eh}.
Nonetheless, the existence of a logarithmic mean velocity is an
indication of the separation of scales, and has been observed in
experimental and computational studies. The variance of the velocity
components, or turbulent kinetic energy (TKE), also exhibit
near-wall/outer scale separation.  For example,
\citet{Hutchins:2007kd} show that small-scales dominate near the wall,
while large scales dominate away from the wall. Further, the
contribution of the large-scale outer turbulence is shown to grow with
increasing $Re$, causing the streamwise velocity variance to have both
an inner-layer peak and an outer-layer peak
\citep{Fernholz:1996tm,Morrison:2004ir}.  Moreover, based on his
``attached'' eddy hypothesis, \citet{Townsend:1976uj} predicted that,
at high $Re$ and in the overlap region between inner and outer layers,
the streamwise and spanwise velocity variances decrease
logarithmically with wall-normal distance, and that the wall-normal
variance and streamwise/wall-normal velocity covariance are
constants. However, there is only limited experimental evidence
\citep{Marusic:2010bb,Hultmark:2012ce} for the logarithmic variation
of the streamwise variance at very high $Re$ (say $Re_\tau \gtrsim
20000$). Here, $Re_\tau$ is the friction Reynolds number based on the
friction velocity ($u_\tau\equiv\sqrt{\tau_w/\rho}$), the channel
half-width ($\delta$) and the kinematic viscosity ($\nu$), where
$\tau_w$ and $\rho$ are the mean wall shear stress and fluid density,
respectively.  Agreement with Townsend's predictions for the spanwise
velocity variance, has been observed in direct numerical simulation
(DNS) \citep{Lee:2015er}.

The dynamics of near-wall turbulence is relatively well-understood,
since it can be studied with flows at relatively low $Re$. Streamwise
vortices create streaks, which become unstable and wavy. In turn, the
wavy streaks stretch the streamwise vorticity creating new streamwise
vortices, and ultimately new streaks. This ``autonomous'' or
``self-sustaining'' cycle occurs at the small scales characteristic of
the near-wall, and is maintained independent of the outer-layer
turbulence
\citep{Hamilton:1995vu,Jeong:1997uj,Jimenez:1999wf,Schoppa:2002dq}.
At small scales, energy is transferred away from the wall through a
self-similar ejection mechanism
\citep{delAlamo:2004bd,delAlamo:2006db,Jimenez:2007kz}.  Also, the
large-scales in the outer layer transfer energy to the near-wall
region, leaving ``footprints'' that are effectively large-scale
modulations of the small-scale near-wall turbulence
\citep{Ganapathisubramani:2012dh}.  This modulation by the outer layer
increases with $Re$ , so that the contributions of large scales in the
near-wall region also increases with $Re$
\citep{DeGraaff:2000wm,Hutchins:2007ty,Marusic:2010bb,Marusic:2010hy}.
Here, to better understand this interplay between the inner- and
out-layer turbulence at high Reynolds number, we seek to ``map'' the
exchange of energy in the wall-normal direction, in scale and among
components.

Of the many methods of studying wall-bounded turbulence, we focus here
on the Reynolds stress transport equations (RSTE here after). The RSTE
provide detailed information regarding the life cycle of the
turbulence, which includes production, transport, and
dissipation. However, there have only been a limited number of studies
of the terms in the RSTE because their computation requires
high-fidelity three-dimensional velocity fields. For example, in the
case of incompressible channel flows, \citet{Mansour:1988vz} first
performed an analysis of terms in the RSTE using the data from the
$Re_\tau=180$ direct numerical simulation (DNS) of
\citet{Kim:1987ub}. Two decades later, \citet{Hoyas:2008jl} analysed
the $Re$ dependencies of the RSTE terms using DNS data from channel
flow at up to $Re_\tau =2003$. Both these studies focused on the
profiles of each RSTE term and their scaling.

More than fifty years ago, \citet{Lumley:1964fp} suggested the
analysis of spectral budget equations in wall-bounded turbulence.  One
of the objectives of such a study would be to analyze the nonlinear
transfer of energy between scales, as characterised by the exchange of
energy among Fourier modes. In the usual Kolmogorov picture of
turbulence, the net transfer of energy is from large scales (low
wavenumbers) to small scales (high wavenumbers). However,
\citet{Lumley:1964fp} conjectured that the energy will be transferred
from small scales to large, as distance from the wall increases, a
so-called ``inverse energy transfer''. Thirty years later,
\citet{Domaradzki:1994ky} performed a spectral analysis of the RSTE
using DNS data of channel flow at $Re_\tau = 210$ and observed inverse
energy transfers from small to large scales. Later,
\citet{Bolotnov:2010hxa} performed spectral analysis of RSTE at
$Re_\tau = 180$ including viscous transport and dissipation
terms. However \citet{Bolotnov:2010hxa} did not observe an inverse
energy transfers. As will be seen in
\S\ref{subsec:inter-scale_transfer}, identifying inverse transfers
from one-dimensional spectra can be miss-leading.  Recently,
\citet{Mizuno:2015wz,Mizuno:2016kx} performed RSTE spectral analysis
using channel flow DNS data at up to $Re_\tau=1000$, and observed
length-scale dependencies in the transport of TKE in the wall-normal
direction. In parallel, \citet{Lee:2015wz} observed the same phenomena
in DNS data up to $Re_\tau = 5200$.

In noteworthy related studies: \citet{Richter:2015ipa} performed
spectral analysis of the TKE budget equations using DNS of
particle-laden turbulent plane Couette flows at up to $Re_\tau = 900$
to investigate the $Re$ and $St$ (Stokes number) dependencies of the
production and dissipation of TKE. \citet{Aulery:2016cg} performed
spectral analysis of the TKE budget in anisothermal turbulent channel
flow up to $Re_\tau = 395$, and also observed inverse energy
transfers. Finally,
\citet{Cimarelli:2013ke,Cimarelli:2015iy,Cimarelli:2016bt} studied the
evolution equation of the second order structure function as a
function $({\bf r},y)$ where ${\bf r}$ is the separation vector in the
structure function and $y$ is the distance from the wall. They
observed two distinct production mechanism at different wall-normal
distances and inverse energy transfers.

Up to now, most analysis of RSTE has been done in terms of TKE or the
streamwise velocity variance, in relatively low $Re$ turbulence. The
goal of this paper is to study the detailed flow of energy in $y$,
scale, and among velocity components in high-Reynolds-number
turbulence.  To pursue this, we use the DNS data of \citep{Lee:2015er}
at $Re_\tau = 5200$, which exhibits many features of high $Re$
wall-bounded turbulent flows; such as, logarithmic variation of the
mean-velocity profile and scale separations in the energy spectra. The
remainder of this paper is organised as follows: the details of the
analysis method, including the derivation of the spectral budget
equation and visual representation of two-dimensional spectral
densities in log-polar coordinates are provided in
\S\ref{sec:method}. The spectral density of TKE components and their
production, scale-transfer, wall-normal transport and dissipation are
explored in \S\ref{sec:results} as functions of the wall-normal
distance and scale. Finally, a summary and conclusions are provided in
\S\ref{sec:conclusion}.

\section{Analysis details}
\label{sec:method}
In the following discussion, the velocity components in the streamwise
($x$), wall-normal ($y$) and spanwise ($z$) directions are denoted as
$u$, $v$ and $w$, respectively, and when using index notation, these
directions are labeled 1, 2 and 3, respectively. The expected value is
 denoted with angle brackets (as in $\langle \cdot \rangle$), and
upper case $U$ and $P$ indicate the mean velocity and pressure, so that
$\langle u_i \rangle = U_i$. The velocity and pressure fluctuations
are indicated with primes, e.g. $u_i = U_i + u'_i$, and as usual, a
superscript ``+'' signifies non-dimensionalisation with the kinematic
viscosity $\nu$ and the friction velocity $u_\tau$.

\subsection{Simulation data}
Direct numerical simulations of pressure-driven incompressible channel
flows at different Reynolds numbers were performed \citep{Lee:2015er},
and the data from these simulations are used in the analysis reported
here. The simulation parameters for each case are summarised in
table~\ref{table:simulation_parameters}. Note that in addition to the
simulations reported in \citet{Lee:2015er} a new simulation at
$Re_\tau \approx 2000$ was performed for completeness and consistency,
and is also included here. In the direct numerical simulations, the
wall-normal velocity-vorticity formulation of \citet{Kim:1987ub} was
used to solve the Navier-Stokes equations with Fourier-Galerkin
discretisations in the $x$ and $z$ directions and a seventh order
b-splines representation in the $y$ direction. See
\citet{Lee:2013kv,Lee:2014ta,Lee:2015er} for more details on the
simulations and the numerical methods.
\begin{table}
  \begin{center}
  \def~{\hphantom{0}}
    \begin{tabular}{c c c c c c c c c c}
     Case name & $Re_\tau$ & $N_x$ & $N_y$ & $N_z$ & $\Delta x^+$ & $\Delta z^+$ & $\Delta y^+_{w}$ & $\Delta y^+_{c}$ & $T u_\tau / \delta$ \\
     R550      & ~544      & ~1536 & ~384  & 1024  & ~8.9         & 5.0          & 0.019            & ~4.5             & 13.6~ \\
     R1000     & 1000      & ~2304 & ~512  & 2048  & 10.9         & 4.6          & 0.019            & ~6.2             & 12.5~ \\
     R2000     & 1995      & ~4096 & ~768  & 3072  & 12.2         & 6.1          & 0.017            & ~8.2             & 11.5~ \\ 
     R5200     & 5186      & 10240 & 1536  & 7680  & 12.7         & 6.4          & 0.498            & 10.3             & ~7.80 \\
    \end{tabular}
  \caption{Summary of simulation parameters; $\Delta x$ and $\Delta z$
  are in terms of Fourier modes for spectral methods.  $\Delta y_w$
  and $\Delta y_c$ are b-spline knot spacing at wall and center of
  channel, respectively. $\delta$ - Channel half width, $Re_\tau =
  u_\tau \delta/\nu$, $Tu_\tau/\delta$ - Total simulation time without
  transition. In all cases, the domain size in the $x$ and $z$
  directions is $8\pi\delta$ and $3\pi\delta$ respectively.}
  \label{table:simulation_parameters}
  \end{center}
\end{table}

\subsection{Mathematical formulations}

In this section, evolution equations for velocity spectral densities
are derived. Equations for the velocity fluctuation are  obtained by
introducing the Reynolds decomposition in to the Navier-Stokes
equations and are given by:
\begin{equation}
  \pdiff{u_i'}{t}  = 
  - U_k\pdiff{u_i'}{x_k} 
  - u_k' \pdiff{U_i}{x_k}
  - \pdiff{u_k' u_i'}{x_k} 
  + \pdiff{\overline {u_k' u_i' }}{x_k} 
  - \frac{1}{\rho}\pdiff{p'}{x_i} 
  + \nu \frac{\partial^2 u_i'}{\partial x_k \partial x_k}
\label{eq:vel_fluctuation}
\end{equation}
Consider two points separated from each other in only the horizontal
directions, with coordinates ${\bf x}$ and $\tilde{{\bf x}}$, where
$y=\tilde y$. Also, let $r_x=\tilde x- x$ and $r_z=\tilde z- z$ be the
separation between the points in the streamwise and spanwise
directions. Further, let the velocities at the points ${\bf x}$ and
$\tilde{{\bf x}}$ be ${\bf u}$ and $\tilde{{\bf u}}$,
respectively. The evolution equation for the two-point correlation
tensor $R_{ij} = \langle u_i \tilde{u}_j \rangle$ can then be obtained
from (\ref{eq:vel_fluctuation}) as
\begin{equation}
\begin{split}
\pdiff{R_{ij}}{t}(r_x,y,r_z) =
& \left\langle u'_i \pdiff{\tilde{u}'_j}{t} 
+ \tilde{u}'_j \pdiff{u'_i}{t}  \right\rangle \\
=& \underbrace{- \langle \tilde{u}'_j u'_k \rangle \pdiff{U_i}{x_k}-\langle   u'_i \tilde{u}'_k \rangle \pdiff{U_j}{\tilde{x}_k}  }_{\displaystyle R^P_{ij}} \;\;
\underbrace{-\left\langle \tilde{u}'_j\pdiff{u_k' u_i'}{x_k}  +
u'_i\pdiff{{\tilde u_k' \tilde u_j'}}{\tilde{x}_k} \right\rangle }_{\displaystyle R^T_{ij}}\\
&\underbrace{-\frac{1}{\rho}\left\langle \tilde{u}'_j \pdiff{p'}{x_i} + u'_i \pdiff{\tilde{p}'}{\tilde{x}_j} \right\rangle}_{\displaystyle R^{\Pi}_{ij}}
\;\;
+\underbrace{\nu \left\langle \tilde{u}'_j \frac{\partial^2 u'_i }{\partial x_k \partial x_k}  +  u'_i\frac{\partial^2 \tilde{u}'_j}{\partial \tilde{x}_k \partial \tilde{x}_k}\right\rangle}_{\displaystyle R^{\nu}_{ij}} 
\end{split}
\label{eq:2pt_corr_evol}
\end{equation} 
where the terms labeled $R^P_{ij}$, $R^T_{ij}$, $R^\Pi_{ij}$ and
$R^\nu_{ij}$ are interpreted as production, turbulent convection,
pressure  and viscous terms, respectively. Terms in
(\ref{eq:2pt_corr_evol}) are functions of only $r_x$, $y$ and $r_z$
due to the homogeneity in horizontal ($x$ and $z$) directions. Note
that the Reynolds stress transport equation, also known as the budget
equation for $\langle u_i' u_j' \rangle$, is a special case
of (\ref{eq:2pt_corr_evol}) where $r_x = r_z = 0$. Equation
(\ref{eq:2pt_corr_evol}) can be simplified by taking advantage of the
following relationships, where $\alpha$ signifies an index for one of
the wall-parallel direction (i.e. 1 or 3).
\begin{subequations}	
\begin{equation}
\frac{\partial u_i}{\partial \tilde{x}_\alpha} = \frac{\partial \tilde{u}_j}{\partial x_\alpha} =0
\end{equation}
\begin{equation}
\frac{\partial U}{\partial x_\alpha} =\frac{\partial U}{\partial \tilde{x}_\alpha} =0
\end{equation}
\begin{equation}
  V=W=0
\end{equation}
\begin{equation}
  y=\tilde{y}
\end{equation}
\begin{equation}
\left. \frac{\partial}{\partial \tilde{x}_\alpha} \right\vert_{x_\alpha}
= - \left. \frac{\partial}{\partial x_\alpha} \right\vert_{\tilde{x}_\alpha} 
= \frac{\partial}{\partial r_\alpha}
\end{equation}
\label{eq:2pt_corr_rule}
\end{subequations}

The $R^P$ term is interpreted as production because it arises from
interaction between the fluctuations and the mean velocity gradient
and results in a net transfer of energy from the mean to the turbulent
fluctuations. Since only the streamwise component of the mean velocity
is non-zero, and because it varies only in the wall-normal direction,
$R^P$ can be written
\begin{equation}
R^P_{ij} = - \left(\langle \tilde{u}'_j v' \rangle\delta_{1i} + \langle
u'_i \tilde{v}' \rangle\delta_{1j}\right) \pdiff{U}{y}
\end{equation}
Further for zero separation, symmetry requires that
$R^P_{13}=R^P_{31}=0$, leaving non-zero production of only the
streamwise velocity variance and the Reynolds shear stress, as is well
known.

To facilitate the interpretation of the DNS data, it will also be
useful to decompose the $R^T_{ij}$, $R^\Pi_{ij}$ and $R^\nu_{ij}$
terms, so that different effects can be isolated. In general, such
decompositions are not unique \citep{Lumley:1975hu}, but we select the
following decompositions of $R^T_{ij}$, $R^\Pi_{ij}$ and $R^\nu_{ij}$,
because they are consistent with the definitions used by
\citet{Mansour:1988vz} and many others for the Reynolds stress
transport equations when $r_x$ and $r_z$ approach zero.

First, the turbulent term $R^T_{ij}$ can be decomposed into two terms
as follows:
\begin{equation}
R^T_{ij} = R_{ij}^{T^{\bot}} + R_{ij}^{T^\|} 
\label{eq:turb_decomposition}
\end{equation}
where $R_{ij}^{T^\bot}$ and $R_{ij}^{T^\|}$ satisfy:
\begin{equation}
  \int_0^\delta R_{ij}^{T^{\bot}} \intd y =  0, \quad \forall \; (r_x, r_z)
  \label{eq:turb_vertical_condition}
\end{equation}
\begin{equation}
  \lim_{\substack{r_x \rightarrow 0 \\ r_z \rightarrow 0}} \displaystyle R^{T^{\|}}_{ij} = 0, \quad \forall \; y
  \label{eq:turb_horizontal_condition}
\end{equation}
These conditions allow $R_{ij}^{T^\bot}$ to be interpreted as the
transport of the two-point correlation in the wall-normal direction
and $R_{ij}^{T^\|}$ to be interpreted as transfer across scales.  The
decomposition that satisfies (\ref{eq:turb_vertical_condition}) and
(\ref{eq:turb_horizontal_condition}), and is consistent with the
definitions of \citet{Mansour:1988vz} and \citet{Mizuno:2016kx} is:
\begin{equation}
R_{ij}^{T^{\bot}} = -\frac{1}{2}\left( \pdiff{\langle \tilde{u}'_j
u'_i v'\rangle}{y} + \pdiff{\langle{u}'_i \tilde
u'_j \tilde v'\rangle}{y} \right)
\label{eq:turb_vertical}
\end{equation}
and 
\begin{equation}
\begin{split}
R_{ij}^{T^\|} =& \pdiff{\langle \tilde{u}'_j u'_i u'\rangle}{r_x}
- \pdiff{\langle{u}'_i \tilde u'_j \tilde u'\rangle}{r_x}
+ \pdiff{\langle \tilde{u}'_j u'_i w'\rangle}{r_z}
- \pdiff{\langle{u}'_i \tilde u'_j \tilde w'\rangle}{r_z} \\
& -\frac{1}{2}\left( \pdiff{\langle \tilde{u}'_j u'_i
v'\rangle}{y} + \pdiff{\langle{u}'_i \tilde u'_j \tilde
v'\rangle}{y} \right) +\left\langle u'_i
v'\pdiff{\tilde{u}'_j}{y} \right\rangle + \left\langle \tilde
u'_j \tilde v'\pdiff{{u}'_i }{y}\right\rangle
\end{split}
\label{eq:turb_horizontal}
\end{equation}

The pressure term $R^\Pi_{ij}$ is decomposed into two terms
\begin{equation}
R^\Pi_{ij} = -\frac{1}{\rho}\left\langle \tilde{u}'_j \pdiff{p'}{x_i} + u'_i \pdiff{\tilde{p}'}{\tilde{x}_j} \right\rangle  = R^{\Pi^d}_{ij} + R^{\Pi^s}_{ij}.
\label{eq:press_decomposition}
\end{equation}
where $R^{\Pi^d}_{ij}$ and $R^{\Pi^s}_{ij}$ satisfy:
\begin{equation}
  \int_0^\delta R_{ij}^{\Pi^d} \intd y =  0, \quad \forall \; (r_x, r_z)
\label{eq:press_diffusion_condition}
\end{equation}
\begin{equation}
  R_{ii}^{\Pi^s} =  0, \quad \forall \; (r_x, y, r_z)
\label{eq:press_strain_condition}
\end{equation}
These conditions allow $R_{ij}^{\Pi^d}$ to be interpreted as pressure
transport of the two-point correlation in the $y$ direction, and
$R_{ij}^{\Pi^s}$ as the energy exchange between velocity components.
The decomposition is chosen to be:
\begin{equation}
R^{\Pi^d}_{ij}=\frac{1}{\rho}\left( \pdiff{\langle \tilde u_jp\rangle}{y}\delta_{i2}+\pdiff{\langle
u_i\tilde p\rangle}{\tilde y}\delta_{j2} \right)
\label{eq:press_diffusion}
\end{equation}
\begin{equation}
R^{\Pi^s}_{ij}=-\frac{1}{\rho}\left\langle p\pdiff{\tilde u_j}{\tilde
x_i} + \tilde p\pdiff{u_i}{x_j}\right\rangle,
\label{eq:press_strain}
\end{equation}
where $R_{ij}^{\Pi^s}$ becomes the pressure-strain correlation, as
defined by \citet{Lumley:1975hu}, when $r_x\rightarrow 0$ and
$r_z\rightarrow 0$. In (\ref{eq:press_strain}), terms involving
derivatives in the horizontal directions can be rewritten in terms of
derivatives with respect to the separations using
(\ref{eq:2pt_corr_rule}e), so in particular $\langle p (\partial
{\tilde u_j}/\partial {\tilde x_i}) \rangle= \partial \langle p\tilde
u_j\rangle/\partial r_i $ for $i=1$ or 3, and $\langle\tilde p
(\partial {u_i}/ \partial {x_j})\rangle =-\partial {\langle\tilde
  pu_i\rangle}/ \partial {r_j}$ for $j=1$ or 3. Again, the pressure
term decomposition is not unique. For example, the following
alternative decomposition satisfies
(\ref{eq:press_diffusion_condition}) and
(\ref{eq:press_strain_condition}) \citep{Lumley:1975hu}.
\begin{equation}
R_{ij,\text{alt}}^{\Pi^s} = R_{ij}^\Pi - \frac{1}{3}\mathbf{tr}(R_{ij}^\Pi) \delta_{ij} 
\end{equation}
\begin{equation}
R_{ij,\text{alt}}^{\Pi^d} = R_{ij}^\Pi - R_{ij,\text{alt}}^{\Pi^s}
\end{equation}
Here, the differences between decompositions are not important because
$R_{ii}^\Pi$ (=$R_{22}^{\Pi^d}$) is considerably smaller than the
other terms.

Note that the pressure term in the Navier-Stokes equations arises as
the impact of the continuity constraint on the (non-linear) convection
terms. One can thus interpret the sum of the convection and pressure
terms in the Navier-Stokes equations as the complete non-linear term,
which is the divergence-free projection of the convection term.
Pushing this interpretation through to the two-point correlation
equation considered here leads to combining the terms based on the
rapid pressure \citep{Lumley:1979fz} with the production term, and
those arising from the slow pressure with the turbulent term. The
energy exchange between components represented by $R_{ij}^{\Pi^s}$ has
a completely different character than the production and turbulent
terms, so for our purposes here, it is most useful to analyze it
separately. On the other hand, the wall-normal transport arising from
$R_{ij}^{\Pi^d}$ has the same character as $R_{ij}^{T^\bot}$, so it is
appropriate to combine these two terms to obtain:
\begin{equation}
R_{ij}^{\mathcal{N}}=R_{ij}^{T^\bot}+R_{ij}^{\Pi^d}
\end{equation}
which when $r_x\rightarrow0$ and $r_z\rightarrow0$ becomes the
non-linear transport term $\mathcal{N}_{ij}=T_{ij}+\Pi_{ij}^d$. In
channel flow $\Pi_{11}^d=\Pi_{33}^d=0$, so of the diagonal terms, it
is only $\mathcal{N}_{22}$ that differs from $T_{22}^\bot$. Further,
$\Pi_{22}^d$ includes contributions from the rapid pressure, in
addition to the slow pressure that is appropriate to combine with
$T^\bot_{22}$, but in the channel flow studied here, the rapid
contribution to $\Pi_{22}^d$ is small.

Finally, the viscous term, $R^\nu_{ij}$, can be decomposed as
\begin{equation}
  R^{\nu}_{ij}  = \nu \left\langle \tilde{u}'_j \frac{\partial^2 u'_i }{\partial x_k \partial x_k}  +  u'_i\frac{\partial^2 \tilde{u}'_j}{\partial \tilde{x}_k \partial \tilde{x}_k}\right\rangle = R^D_{ij} - R^\epsilon_{ij}
  \label{eq:viscous_decomposition}
\end{equation}
where $R^D_{ij}$ and $R^\epsilon_{ij}$ satisfy:
\begin{equation}
  \int_0^\delta R_{ij}^D \intd y =  0, \quad \forall \; (r_x, r_z)
  \label{eq:viscous_diffusion_condition}
\end{equation}
\begin{equation}
  \lim_{\substack{r_x \rightarrow 0 \\ r_z \rightarrow
  0}} \displaystyle R^\epsilon_{\alpha \alpha} \geq 0, \quad \forall \;
  y
  \label{eq:viscous_dissipation_condition}
\end{equation}
These conditions allow $R^D$ to be interpreted as viscous transport of
the two-point correlation in the $y$ direction, and $R^\epsilon$ to be
interpreted as the viscous dissipation. Consistent with
\cite{Mansour:1988vz}, we define $R^D$ and $R^\epsilon$ as:
\begin{equation}
  R^D_{ij} =  \nu \frac{\partial^2 \langle u'_i \tilde{u}'_j \rangle}{\partial y^2}
  \label{eq:viscous_diffusion}
\end{equation}
\begin{equation}
R^\epsilon_{ij} = 2\nu \left( \frac{\partial^2 \langle u'_i \tilde{u}'_j \rangle}{\partial r_x^2}  +  \left \langle \frac{\partial u'_i}{\partial y }\frac{\partial \tilde{u}'_j}{\partial y }\right\rangle +   \frac{\partial^2 \langle u'_i \tilde{u}'_j \rangle}{\partial r_z^2}\right )  
\label{eq:viscous_dissipation}
\end{equation}

In summary, then, the evolution equation for the two-point correlation
tensor with separation in directions parallel to the wall can be
written
\begin{equation}
\pdiff{R_{ij}}{t} = R_{ij}^P + R_{ij}^D + R_{ij}^{\Pi^s} +  R_{ij}^{\mathcal{N}} + R_{ij}^{T^\|} - R_{ij}^\epsilon
\label{eq:2pt_corr_eq_final}
\end{equation}

The evolution equation for the spectrum tensor $E_{ij}(k_x,y,k_z)$
($k_x$ and $k_z$ are the wave numbers) can be obtained by Fourier
transforming (\ref{eq:2pt_corr_eq_final}) in $r_x$ and $r_z$ to give
\begin{equation}
\pdiff{E_{ij}}{t}(k_x,y,k_z) = E_{ij}^P + E_{ij}^D + E_{ij}^{\Pi^s} +  E_{ij}^{\mathcal{N}} + E_{ij}^{T^{\|}} - E_{ij}^\epsilon
\label{eq:E_evol_eq_final}
\end{equation}
where each term on the right hand side of (\ref{eq:E_evol_eq_final})
is the Fourier transform of the corresponding term in
(\ref{eq:2pt_corr_eq_final}). When (\ref{eq:2pt_corr_eq_final}) is
evaluated in the limit $r_x\rightarrow 0$ and $r_z\rightarrow 0$, or
equivalently when (\ref{eq:E_evol_eq_final}) is integrated over the
wavenumbers $k_x$ and $k_z$, the result is the Reynolds stress
transport equation,
\begin{equation}
\begin{split}
\lim_{\substack{r_x \rightarrow 0 \\ r_z \rightarrow 0}} \pdiff{R_{ij}}{t}  &= \iint \pdiff{E_{ij}}{t} \intd k_x \intd k_z \\
&= \pdiff{\langle u_i' u_j' \rangle}{t} (y) = P_{ij} + D_{ij} + \Pi^s_{ij} + \mathcal{N}_{ij} - \epsilon_{ij}
\end{split}
\label{eq:RSTE}
\end{equation}
where each of the terms in (\ref{eq:2pt_corr_eq_final}) or
(\ref{eq:E_evol_eq_final}) gives rise to a term in the Reynolds stress
equation as commonly written and interpreted \citep{Mansour:1988vz} as
mentioned above. The terms on the right hand side of
(\ref{eq:E_evol_eq_final}), except for $E_{ij}^{T^\|}$, can therefore
be interpreted as the spectral decomposition of the terms in the
Reynolds stress transport equation. In this paper, these terms are
studied to discover the scale dependence of the phenomena responsible
for the Reynolds stress dynamics in high $Re$ channel flows.

\subsection{Visual representation of data}
\label{subsec:explain_spectra}
In the following subsections, the spectral structure of the terms in
the transport equations for the diagonal elements of the Reynolds
stress tensor (the velocity variances) are examined. Generally,
logarithmic scales will be used in $y$ and the wavenumbers ($k_x$ and
$k_z$). Profiles in $y$ of terms in the transport equations will
generally be ``premultiplied'' by $y$ (e.g. $y^+P^+_{11}$ in
figure~\ref{fig:1d_E_P_uu}), since the integral $\intd\log y$ of the
premultiplied quantity is the integral $\intd y$ of the original
quantity. In this way, the plot shows the log-density of the quantity
(e.g. production); that is, the distribution of the quantity in the
logarithmic coordinates.

\begin{figure}
  \begin{center} \includegraphics[width=.9\textwidth]{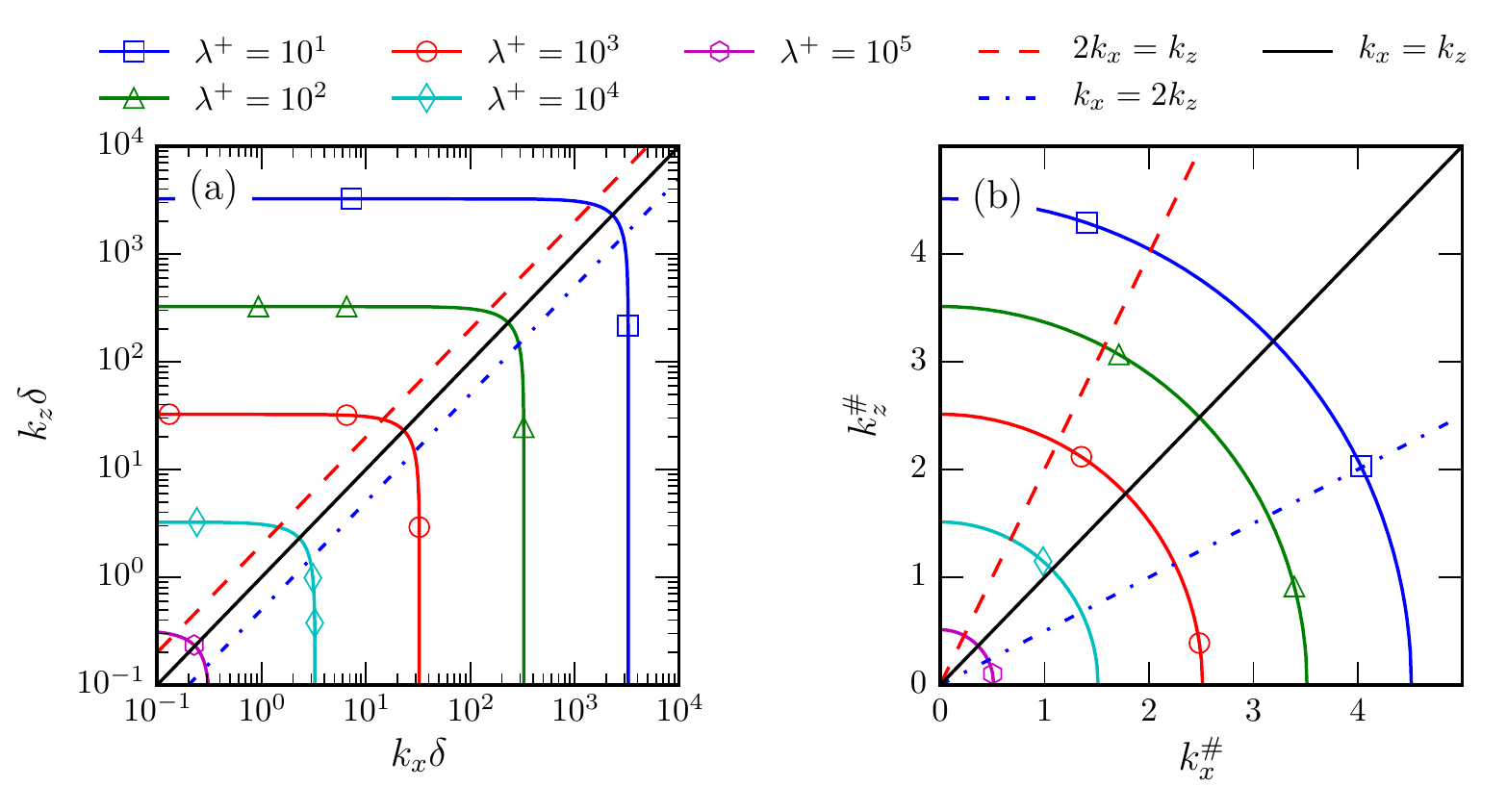} \end{center} \caption{Comparison
    of (a) log-Cartesian and (b) log-polar coordinates for
    visualisation of two-dimensional spectra. In (a), the $\lambda^+$
    contours are shown for for the R5200 case. In (b),
    $k_x^\#=(k_x/k)\log_{10}(k/k_{\textrm{ref}})$ and
    $k_z^\#=(k_z/k)\log_{10}(k/k_{\textrm{ref}})$, where
    $k=(k_x^2+k_z^2)^{1/2}$ and
    $k_{\textrm{ref}}\delta=Re_\tau/50000$ ($k_{\textrm{ref}}^+=1/50000$).
    }  \label{fig:polar_coord_explain}
\end{figure}

As with profiles in $y$, one-dimensional spectra are plotted on
logarithmic scales, and so the spectral density are premultiplied by
the wavenumber ($k_x$ or $k_z$) to yield the spectral
log-density. Where again, the integral of the spectral log-density in
the visualised logarithmic space, yields the integral of the spectral
density in wavenumber. In the visualisation of a two-dimensional
spectrum in a two-dimensional wavenumber space ($k_x,k_z$), however,
there are several ways that a logarithmic scaling can be
introduced. An obvious generalisation of the one-dimensional spectrum
approach is to simply plot the log-spectral density in a
two-dimensional ``logarithmic Cartesian'' space, where the spectral
log-density is just $k_xk_z$ times the normal spectral density. This
has been done by several previous authors
\citep[\eg][]{Jimenez:2008ed,Mizuno:2011bx,Chandran:2017ct,Kraheberger:2018ga}. However,
this has some shortcomings; two of which are: 1) the orientation of
the two-dimensional wavevector $(k_x,k_z)$ is distorted, making
interpretation of the spectrum in terms of orientation and alignment
of the Fourier modes difficult. Indeed, all lines of constant
$k_z/k_x$ are mapped to lines with slope of 1, offset from the
$k_z=k_x$ line by $\log k_z/k_x$ (see
figure~\ref{fig:polar_coord_explain}a), obscuring how well aligned
modes are with the $x$ and $z$ directions; and, 2) vector wavenumbers
with the same magnitude $k=\sqrt{k^2_x+k_z^2}$ do not map onto a
circle (see figure~\ref{fig:polar_coord_explain}a), making it
difficult to assess scale isotropy in the spectrum.

Both of these shortcomings can be overcome by representing the
magnitude of the wavenumber $k$ on a logarithmic scale, rather than
the individual wavenumber components, while preserving orientation in
the logarithmic coordinate. To accomplish this, consider the spectral
density $E$ expressed in ``polar'' coordinates in wavenumber, that is
$k_x=k\cos\theta$ and $k_z=k\sin\theta$. Then the integral of
$E(k_x,k_z)=E(k\cos\theta,k\sin\theta)$ in $k_x$ and $k_z$ can be
rewritten
\begin{equation}
  \int_{k_x>0}\int_{k_z>0} E\intd k_x\intd k_z=\int_{\theta=0}^{\pi/2}\int_{k>0}
  E\,k\intd k\intd\theta
  \label{eq:kint}
\end{equation}
Now, we want to plot in a space that is logarithmic in $k$, so define
$\xi=\log(k/k_{\rm ref})$, where $k_{\rm ref}$ is an arbitrary
reference wavenumber, which must be smaller than the smallest non-zero
wavenumber included in the spectrum. The ``logarithmic polar''
coordinates would be $\xi$ and $\theta$, while the associated
Cartesian coordinates would be $k^\#_x=\xi\cos\theta=\xi k_x/k$ and
$k^\#_z=\xi\sin\theta=\xi k_z/k$. Re-expressing the integral
(\ref{eq:kint})in terms of $\xi$ yields
\begin{equation}
  \int_{k_x>0}\int_{k_z>0} E\intd k_x\intd k_z=\int_{\theta=0}^{\pi/2}\int_{\xi>0}
  \frac{k^2}{\xi}
  E\,\xi\intd\xi\,d\theta=\int_{k^\#_x>0}\int_{k^\#_z>0}
  \frac{k^2}{\xi} E \intd k^\#_x\intd k^\#_z
  \label{eq:rhoint}
\end{equation}
So, the desired two-dimensional spectral density in a space that is
logarithmic in $k$ is plotted as a function of $k^\#_x$ and $k^\#_z$,
and to preserve the integral in this transformed space, the log-polar
spectral density should be $k^2/\xi$ times the normal spectral
density. As shown in the figure~\ref{fig:polar_coord_explain}b, in
this space, lines of constant $k_z/k_x$ have slopes of $k_z/k_x$, and
contours of constant $k$ are circles. The contours of constant $k$ are
labeled according to their wavelength $\lambda=2\pi/k$, which will be
done throughout to facilitate connection with well-known length
scales.

In the remainder of the paper, all quantities that are essentially
densities, such as Reynolds stress components, terms in the RSTE and
their spectra, will be plotted in logarithmic coordinates in $y$
and/or wavenumber. They will be plotted as densities in the
logarithmic coordinates used, so that the visualisation depicts the
distribution of the relevant quantity in those coordinates. Further,
in the case of two-dimensional spectra, log-polar coordinates as
defined above will be used, with
$k_{\textrm{ref}}\delta=Re_\tau/50000$ ($k^+_{\textrm{ref}}=1/50000$),
as shown in figure~\ref{fig:polar_coord_explain}b, and log-polar
spectral densities will be depicted. Using a $k^+_{\textrm{ref}}$ that
is independent of Reynolds number will allow spectral features that
scale in wall units to be directly compared for different
wavenumbers. The DNS calculations naturally yield two-dimensional
spectra represented on a Cartesian grid in $k_x$ and $k_z$. To produce
the log-polar spectra, these are interpolated onto a Cartesian grid in
$k_x^\#$ and $k_z^\#$ with grid spacings of 0.01, and the results are
smoothed using a Gaussian kernel with $\sigma=0.06$.  In what follows,
the terms ``spectrum'' and ``spectra'' will be used to mean the
spectral density associated with the logarithmic space in which it is
plotted. Finally, unless otherwise indicated, all plotted and
visualised quantities are normalised in wall units.

\section{Results}
\label{sec:results}
In this section, the results of a detailed spectral analysis of the
transport equations for the velocity variances in channel flow are
presented. We begin with the velocity variance spectra themselves,
followed by analysis of the terms in the transport equations, as
described in \S\ref{sec:method}. Finally, we discuss the Reynolds
number independence of the small-scale contributions to the transport
equations.

As described in \S\ref{subsec:explain_spectra}, the quantities
visualised in this paper are densities, in the logarithmic space in
which they are plotted. For the velocity variances, these are energy
densities, representing the distribution of energy in $y$ and the
wavenumber space. However, for the terms on the right hand side of the
transport equations (\ref{eq:E_evol_eq_final}), these are densities of
the sources (when positive) and sinks (when negative) of the velocity
variances. Of course, the only true source of turbulent energy is
production, and the only true sink is dissipation. The other terms
transport energy in $y$ and transfer it between scales and between
components. For these terms, we will commonly refer to regions of
positive density as ``recipient'' regions and regions of negative
density as ``donor'' regions, to reflect the fact that energy is being
redistributed \emph{from} donor (negative) regions \emph{to} recipient
(positive) regions. In the following, detailed interpretation of where
energy is being transferred from and to will depend on the
characteristics of the individual terms, and indeed they have been
formulated to allow such interpretation to the extent possible.

\subsection{Energy spectra}
\label{subsec:energy_spectra}
\begin{figure}
  \begin{center}
    \includegraphics[width=\textwidth]{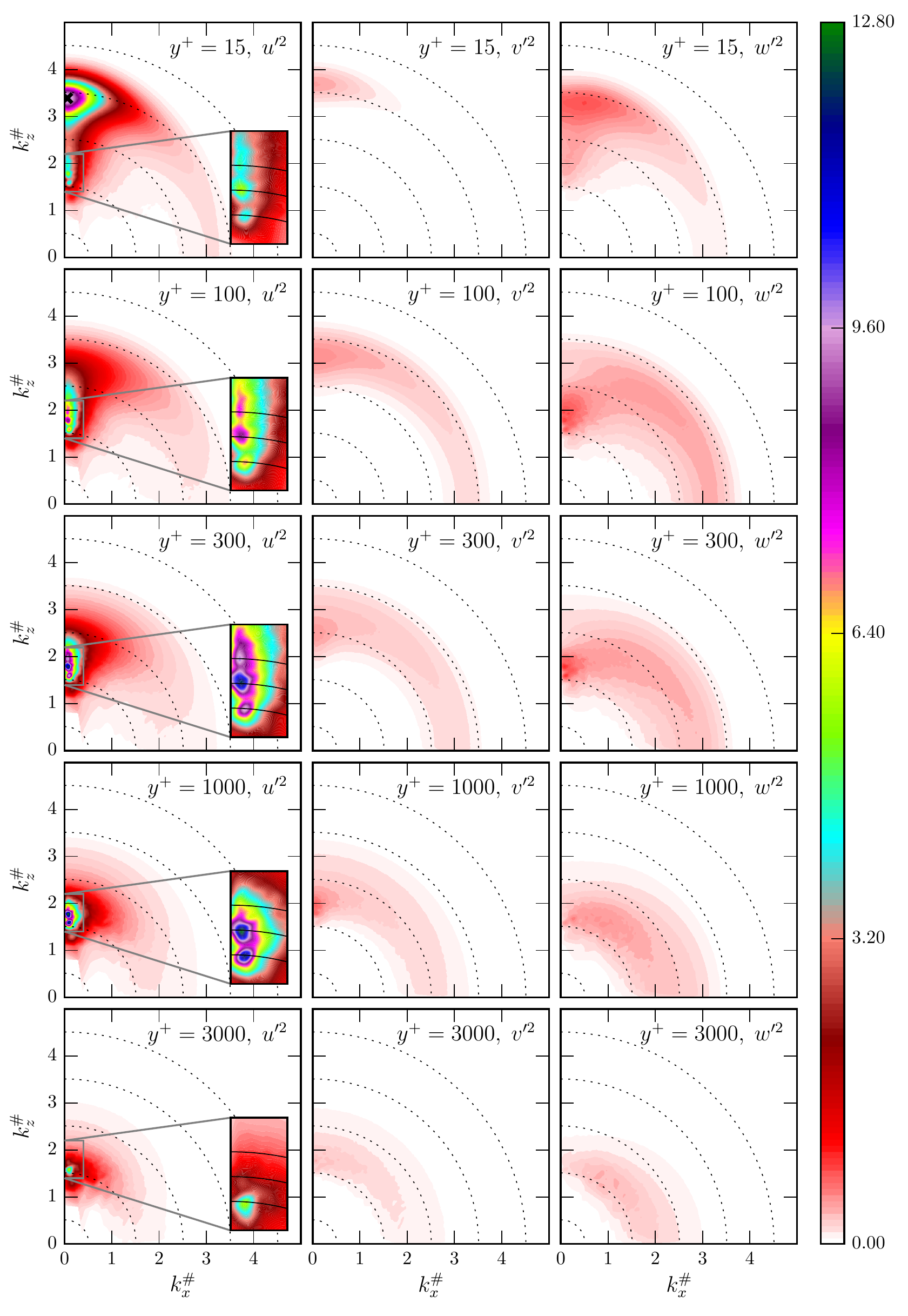}
  \end{center}
  \caption{Two-dimensional spectra of \uu, \vv and \ww in log-polar
    coordinates, as defined in figure~\ref{fig:polar_coord_explain},
    from R5200. $\lambda^+=10$ on the outer-most dotted circle and
    increases by a factor of 10 for each dotted circle moving
    inward. In the magnified inserts, $\lambda/\delta =
      0.67$ on the top-most solid line and increases by a factor of 1.5 for each solid line moving downward.}
  \label{fig:2d_uu}
\end{figure}
 
\begin{figure}
  \begin{center}
    \includegraphics[width=0.7\textwidth]{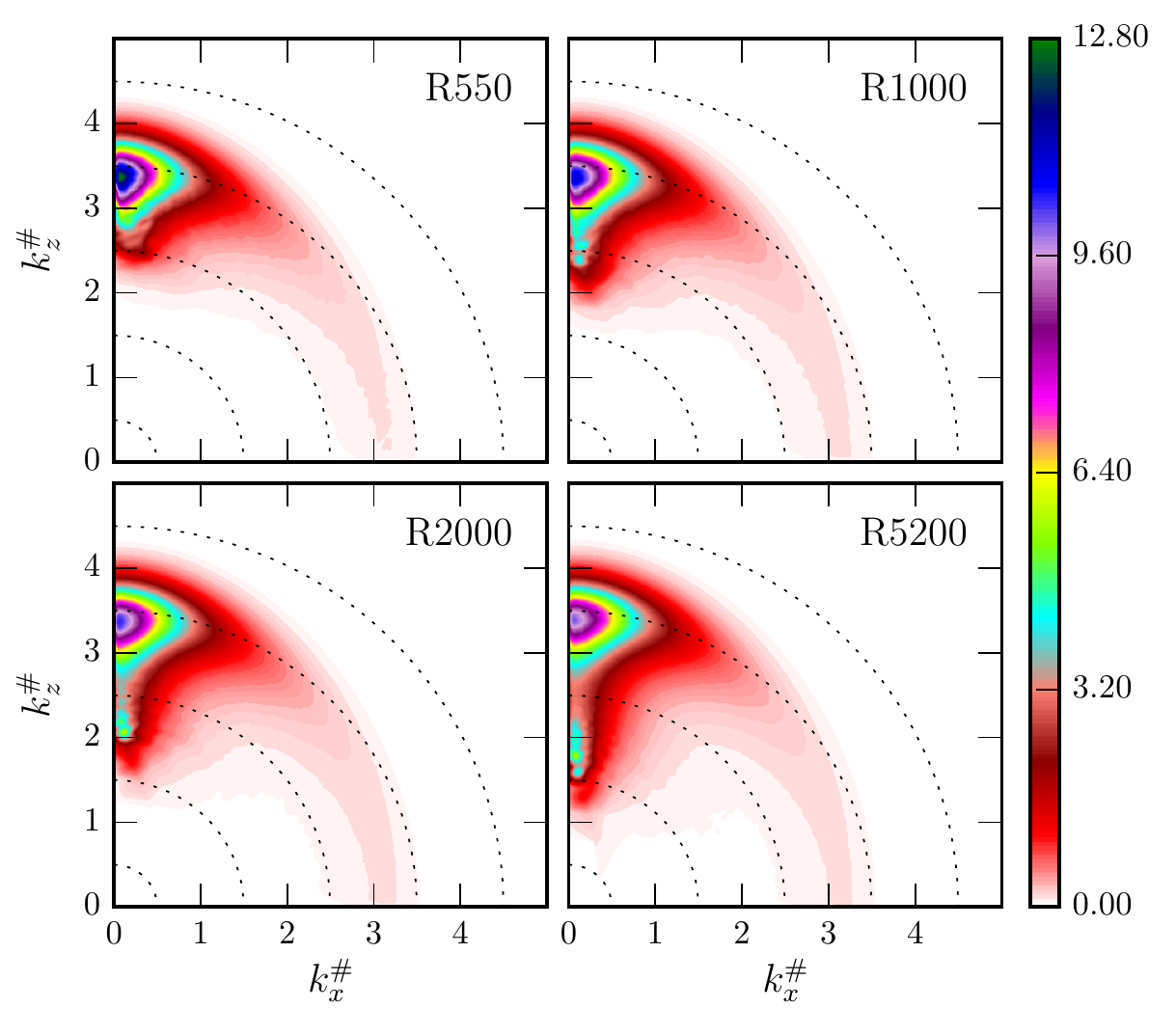}
  \end{center}
  \caption{Two-dimensional spectra of \uu in log-polar
    coordinates, as defined in figure~\ref{fig:polar_coord_explain},
    at $y^+=15$. $\lambda^+=10$ on the outer-most dotted circle and
    increases by a factor of 10 for each dotted circle moving
    inward.}
  \label{fig:2d_uu_y_plus_15}
\end{figure}

The primary objective of the current study is to characterize the flow
of turbulent energy underlying the distribution of energy in high
Reynolds number wall-bounded turbulence. It is useful to begin this
exploration by examining this energy distribution. To this end, the
two dimensional \uu, \vv and \ww energy spectra are plotted at several
$y$-locations in the log-polar space described above
(figure~\ref{fig:2d_uu}). The most obvious feature of these spectra is
the extreme concentration of energy in the spectrum of \uu into
regions along the $k_z^\#$ axis, that is Fourier modes with
$k_z/k_x\gtrsim10$. These are associated with structures that are
strongly elongated in the $x$ direction (inclination angle less that
$6^\circ$). Near the wall ($y^+=15$), this is expected as a
consequence of the well-known low- and high-speed near-wall
streaks. At $y^+=15$, the high energy region is distributed around
$\lambda^+\approx 100$, consistent with expectations for the spacing
of the streaks. Note however, that the energy density peaks along the
axis at $\lambda^+\approx 140$ and the energy distribution extends to
significantly larger spanwise scales. The triangular region of high
energy density that extends to approximately $\lambda^+=1000$ (marked
with a $\times$ in figure~\ref{fig:2d_uu}) is a consequence of the
autonomous near-wall dynamics described by
\citet{Hamilton:1995vu,Jimenez:1999wf}, and is similar in the
two-dimensional \uu spectra at lower Reynolds numbers (see
figure~\ref{fig:2d_uu_y_plus_15}). Note, however, that the magnitude
of the \uu spectral peak is slowly decreasing with increasing Reynolds
number, so that the peak at $Re_\tau=5200$ is approximately 30\% lower
than at $Re_\tau=550$.  This will be discussed in more detail below.

Farther from the wall ($y^+>100$), there are again strong peaks, but
at much longer wavelengths ($1000<\lambda^+<10000$). Notice that at
$y^+=300$, there are actually three distinct peaks in the spectrum,
all located along the $k_z^\#$ axis, with (spanwise) wavelengths
$\lambda^+\approx 3400$, 5200 and 7800, or
$\lambda/\delta\approx0.67$, 1.0 and 1.5, which are separated by about
a factor of 1.5. At $y^+=100$, all three peaks are present, though the
long wavelength one (lower on the figure) is much weaker. Farther from
the wall, at $y^+=1000$, the short wavelength peak is absent, and the
other two are about the same strength, and at $y^+=3000$, only the
longest wavelength peak is present. This suggests that there is a
discrete hierarchy of spanwise streak scales, with each dominant in a
different range of $y$ (longer $\lambda$'s at larger $y$'s).  Note
that similar discrete peaks in spanwise scales are also observed in
DNS of pipe flows at $Re_\tau = 3000$. \citep[See figure 10
  in][]{Ahn:2015gp}. Previously published one-dimensional spanwise \uu
spectra have indicated the presence of discrete spectral peaks in
channels, such as figures~2b in \citet{delAlamoJimenez2003} and and 8b
in \citet{Lee:2015er}, though low Reynolds number ($Re_\tau=550$)
effects are likely in \citet{delAlamoJimenez2003}.
  
The mechanism by which such a discrete scale hierarchy might arise is
not clear. It is probably not a direct consequence of the finite
computational domain size, since the spanwise domain size is about six
times the largest dominant wavelength. In addition, the R550 case
exhibits discrete peaks in the spanwise one-dimensional spectra (not
shown) at approximately the same values of $k_z\delta$ as in
\citet{delAlamoJimenez2003}, despite the fact that the spanwise domain
sizes are significantly different ($3\pi\delta$ here and $4\pi\delta$
in \citet{delAlamoJimenez2003}). This provides support for the
hypothesis that these discrete spectra are not driven by domain size
effects. However, the low Reynolds number in these cases makes this
far from definitive.  On the other hand, the fact that the domain size
is \emph{only} six times the longest wavelength means that the
variation of the spectra near the long-wavelength peak is not well
resolved. For example, there are just three points in the discrete
spectrum between the two longest wavelength peaks. It is thus possible
that misleading plotting artifacts are present, though it seems that
the wave-space resolution should be sufficient to at least distinguish
these peaks.

One possible physical explanation, is that a combination or phase
locking of smaller structures could be creating larger aggregates, as
suggested by \citet{Tomkins:2003dr}.

In addition to the dominance of these spanwise scales far from the
wall, the outer scales appear to be imprinted on the spectra nearer to
the wall, including at $y^+=15$, where three weak peaks are present at
the same wavenumbers. This suggests that these features of the
near-wall spectral structure are imposed from the outer flow,
consistent with suggestions by \citet{Hutchins:2007kd,Marusic:2010hy},
and would therefore be different at different Reynolds numbers. As is
apparent in figure~\ref{fig:2d_uu_y_plus_15}, this is indeed the
case. At the lower Reynolds numbers, there are fewer of these weak
large-wavelength peaks. Since the shorter wavelength features of the
spectra are similar at different Reynolds numbers due to the
autonomous near-wall dynamics, these outer flow imprints on the
near-wall region appear to be responsible for the Reynolds number
dependence of the near-wall peak of \uu as suggested by
\cite{Hutchins:2007ty,Marusic:2010bb}. Indeed, applying a high-pass
filter that eliminates wavenumbers with $\lambda^+>1000$ also
eliminates the near-wall Reynolds number dependence of \uu (see
\S\ref{subsec:small_scale_universality}).

The large-scale features of the near-wall turbulence that are imposed
from the outer flow will also interact non-linearly with the
autonomous near-wall turbulence. Particularly, the large scale
fluctuations in the streamwise velocity implied by the spectrum will
result in large-scale variations in the near-wall shear. The
autonomous near-wall dynamics will therefore by modulated by these
variations as suggested by \citet{Hutchins:2007kd,Marusic:2010hy},
resulting particularly in local variations in the dominant
wavelength. This results in a reduction of magnitude of the
small-scale spectral peak with increasing Reynolds number, as
described above, due to increasing large-scale fluctuations with
Reynolds number. However, because the near-wall scaling for energy is
linear in shear, these large scale modulations in shear do not result
in a net change in the turbulent energy associated with the near-wall
autonomous dynamics. Thus, increasing Reynolds number will not affect
small-scale energy, consistent with the high-pass filtering
observations described above. The reduction with Reynolds number of
the magnitude of the small-scale spectral peak is difficult to see in
figure~\ref{fig:2d_uu_y_plus_15}, but was confirmed in plots of the
difference in spectra at different Reynolds number (not shown)

Near the wall, the \vv and\ww spectra are also strongest in the
wavenumber regimes in which \uu is so strongly peaked, that is modes
with large $k_z/k_x$ (elongated in the $x$ direction). However, the
spectra are not nearly so strongly peaked as in the \uu
spectrum. Farther from the wall, there is a similar structure with
weak concentrations of energy in modes for which \uu is strongly
peaked, though this appears to get weaker with increasing $y$, until
at $y^+=3000$ ($y/\delta\approx0.6$), it is not discernible. In all
the velocity components, some energy is also distributed more-or-less
isotropically over a range of $\lambda$ of a decade or more (an
approximately circular band in figure~\ref{fig:2d_uu}). However, the
range in $\lambda^+$ of this band depends on $y$, and for $y^+\ge100$,
the $\lambda^+$ in the center of the band is approximately
proportional to $y$, as would be expected in the scale similarity
range connected with the log-law. The spectra thus appear to be a
combination of classical turbulence that is nearly isotropic in scale
and component, combined with the strongly streaky streamwise velocity
discussed above along with corresponding streaky fluctuations in \vv
and \ww.

\subsection{Production}
\label{subsec:prod}
\begin{figure}
  \centering
  \includegraphics[width=0.5\textwidth]{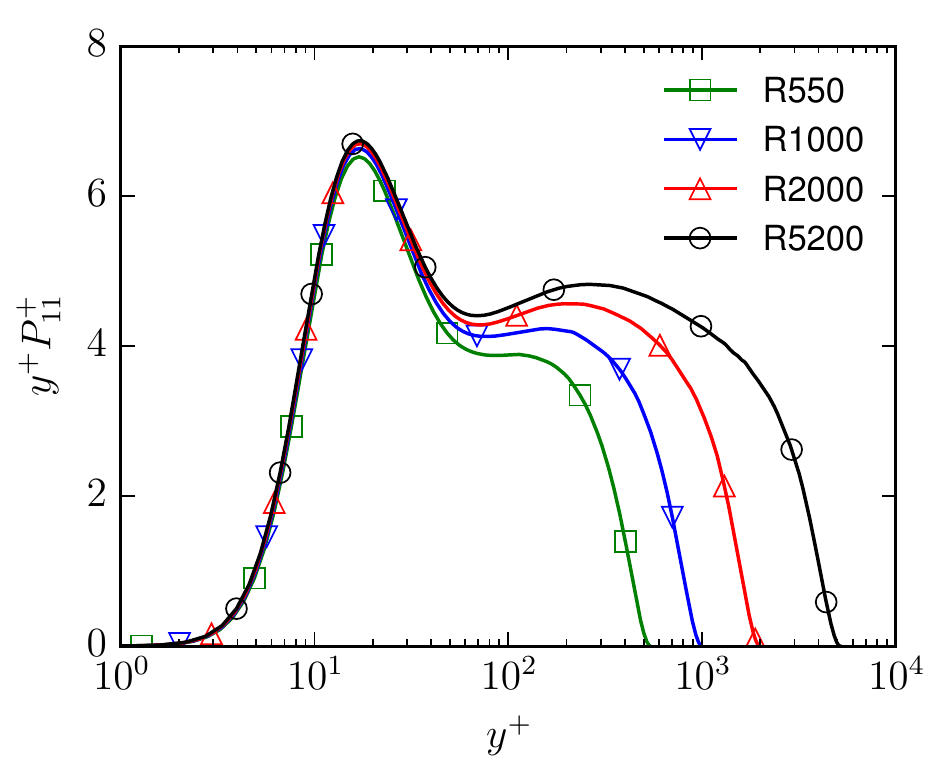}
  \caption{Profiles of \uu production as log-densities.}
  \label{fig:1d_prof_prod}
\end{figure}

\begin{figure}
  \begin{center}
    \includegraphics[width=\textwidth]{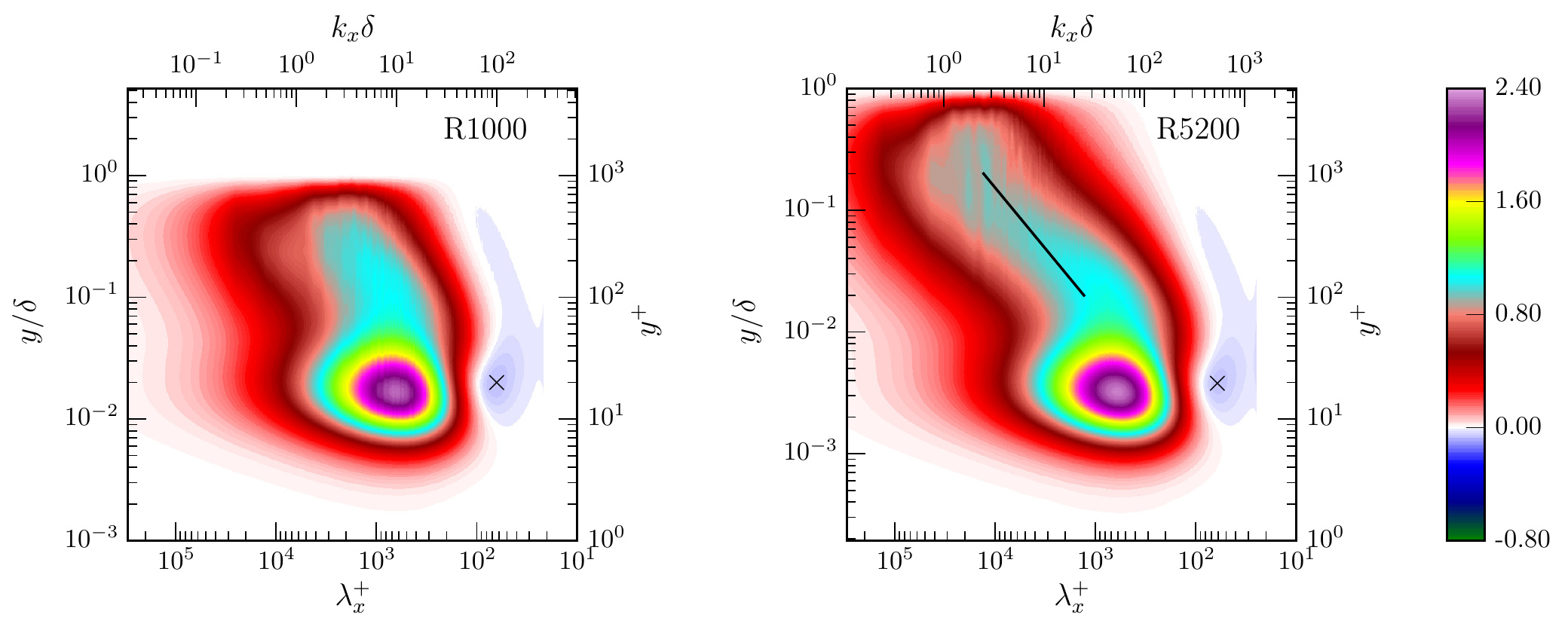}
    \includegraphics[width=\textwidth]{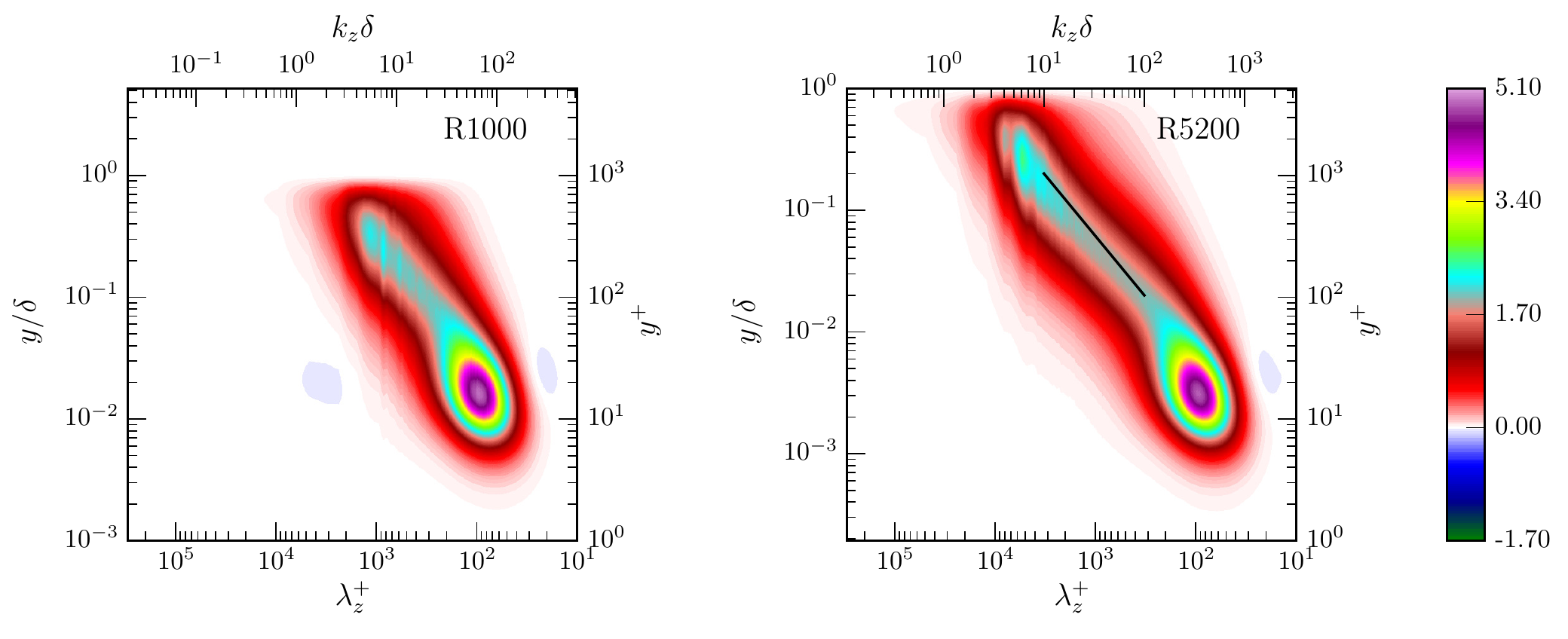}
  \end{center}
  \caption{One-dimensional streamwise and spanwise spectral densities
    of $P_{11}$. Solid lines are at $k_x y = 0.5$ and $k_z y = 2$. The
    symbol $\times$ marks features described in the text.}
  \label{fig:1d_E_P_uu}
\end{figure}

\begin{figure}
  \centering
  \includegraphics[width=\textwidth]{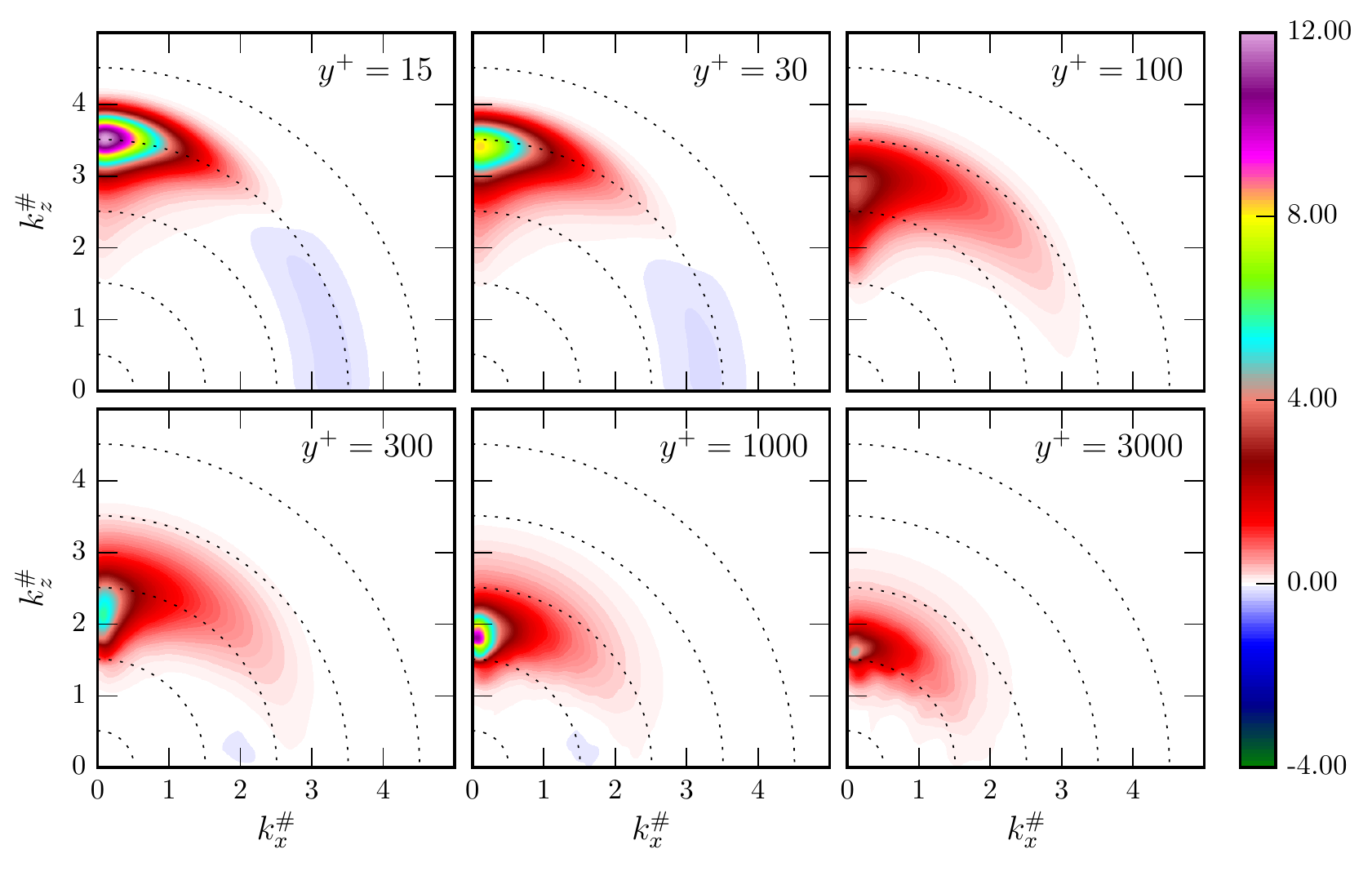}
  \caption{Two-dimensional spectral density of $P_{11}$ in log-polar
    coordinates, as defined in figure~\ref{fig:polar_coord_explain},
    from R5200. $\lambda^+=10$ on the outer-most dotted circle and
    increases by a factor of 10 for each dotted circle moving
    inward.}
  \label{fig:2d_P_uu_new}
\end{figure}

The one-dimensional profile of production of \uu is shown in
figure~\ref{fig:1d_prof_prod}, which because the production of \vv and
\ww are zero, is also twice the production of turbulent kinetic
energy.  There are only weak $Re$ dependencies in the production up to
$y^+\approx70$. On the other hand, the contribution to the production
from regions away from the wall (say $y^+>70$) increases continuously
with the $Re$, with approximately 50\% of \uu production occurring in
the region $y^+>70$ at $Re_\tau=2000$. This is not just because the
domain is getting larger in plus units, but also because in the outer
region the production is increasing with Reynolds number at constant
$y^+$. From the definition of $P^+_{11}$, in the log region,
\begin{equation}
y^+ P_{11}^+
= - 2 y^+ \langle u'v' \rangle^+ \frac{\partial U^+}{\partial y^+}\approx - \frac{2}{\kappa} \langle u'v' \rangle^+
\label{eq:yPk_definition}
\end{equation}
where $\kappa$ is the \karman constant, which is approximately 0.384
in the R5200 case \citep{Lee:2015er}. Presuming $\kappa$ does not
change as $Re$ increases, $y^+ P_{11}^+$ depends only on $\langle
u'v'\rangle^+$ in the log region. The theoretical minimum of \uvp at
infinite $Re$ is $-1$, so the maximum of $y^+P^+_{11}$ for $y^+ > 70$
approaches $2/\kappa\approx 5.21$ as $Re$ increases. The maximum value
of $y^+ P_{11}^+$ in R5200 is 4.83 which is 92.7$\%$ of the
theoretical maximum. Also, asymptotically for large $Re$, $-\langle
u'v' \rangle_{\text{max}} \approx1-2/\sqrt{\kappa Re_\tau}$
\citep{Lee:2015er,Afzal:1982uk}. Hence, a flow at $Re_\tau \approx
10^5$ is required for the outer peak of $y^+ P_{11}^+$ to be within
1\% of theoretical maximum, as for \uvp. This means that the
contribution of the outer flow to the production keeps increasing as
$Re$ increases and the production in the overlap region scales as
$y^{-1}$ for large Reynolds number, i.e.  $P_{11}^+=5.21/y^+$, as
noted by \citet{Hoyas:2008jl}.

The $y$- and $k$-premultiplied one-dimensional spectral density of \uu
production is shown in figure~\ref{fig:1d_E_P_uu}. The appearance of
the \uu production spectra are inherently similar to those of \uv
presented in \citet{Lee:2015er} because in the log region,
\begin{equation}
y^+ E^{P+}_{11} = - 2 y^+ E_{12}^+ \frac{\partial U^+}{\partial y^+} =
-\frac{2}{\kappa} E_{12}^+
\end{equation}
The production spectra of R1000 and R5200 are similar in several
respects up to $y^+\approx 70$.  First, the inner peaks are located at
$\lambda_x^+ \approx 600$, $\lambda_z^+ \approx 100$ and $y^+ \approx
15$. Second, similar to \uv, there is weak negative production of \uu
around $\lambda_x^+ \approx 60$ and $y^+\approx20$ in the streamwise
spectral density (marked with $\times$). Finally, the streamwise
spectral density contours are similar in R1000 and R5200 up to $y^+
\approx 70$, while $Re$ dependencies are stronger for $y^+ > 70$. An
outer peak in the spanwise spectrum occurs at $k_z\delta \approx 6$
and $y/\delta \approx 0.3$ in both cases, but R5200 has a much
stronger outer peak. The outer peak in the streamwise spectrum of
R5200 is diffuse, making a definitive determination of its location
difficult (figure~\ref{fig:1d_E_P_uu}). The distinction of inner and
outer peaks in the \uu production spectra is consistent with the work
of \citet{Cimarelli:2013ke}, and as they noted, the existence of outer
scale production suggests a self-sustaining mechanism of the large
scale motion. In the overlap region, the wavenumbers at which the
production spectra are maximum scale as $y^{-1}$ from $y^+ \approx
100$ to $y/\delta \approx 0.2$ in R5200. This is because the
premultiplied spectral density of \uv in the overlap region scales as
$y^{-1}$, consistent with the observation of
\citet{Jimenez:2008ed}. As noted above, the \uu production approaches
a $y^{-1}$ scaling in the overlap region, which would imply a plateau
in figure~\ref{fig:1d_prof_prod} at high $Re$.
 
More insight into the spectral structure of the production can be
gained from the two-dimensional spectra shown in
figure~\ref{fig:2d_P_uu_new}. Unsurprisingly, the production is
concentrated along the $k_z^\#$ axis, as in the \uu spectra
(figure~\ref{fig:2d_uu}), with spectral peaks that correspond to the
peaks in \uu spectra. However, there are several significant
differences. Very near the wall ($y^+=15$ or 30), there are no
discrete long-wave spectral peaks along the $k_z^\#$ axis as there are
in the \uu spectra. This is expected since these \uu spectral peaks
appeared to be imposed from the outer-region turbulence (see
discussion in \S\ref{subsec:energy_spectra}), and so the energy in
them would be transported in $y$ or transferred in scale, rather than
produced locally. In addition, farther from the wall, there is only
one dominant production spectral peak at each $y$-location shown,
indicating that the more complex multi-peak characteristics of the \uu
spectra at these locations are also likely due to wall-normal
transport. An examination of the two-dimensional production spectra as
a continuous function of $y$ shows that the production peak at
$\lambda/\delta\approx 1.5$ is dominant for $y/\delta>0.35$, and the
one at $\lambda /\delta\approx 1.0$ is dominant from
$0.1<y/\delta<0.35$.  The lower bound of the region where there is a
$\lambda/\delta\approx 0.67$ peak is not distinct. In a region around
the change in dominant production scale, two distinct peaks are
present.  These discrete production peaks are consistent with the
discrete set of spectral peaks in the \uu spectra.

Other interesting features of the two-dimensional production spectra
are the regions of negative production, especially near the wall. This
is clearly the negative production that gives rise to the negative
regions in the one-dimensional $k_x$ production spectra. Near the wall
($y^+=15$), negative production occurs over a range of scale around
$\lambda^+=100$, for modes with wavevectors oriented more in the
streamwise direction than the spanwise direction. Farther from the
wall, negative production occurs with wavevectors oriented more
strongly in the streamwise direction (i.e. structure more elongated in
the spanwise direction), until at $y^+=300$ and 1000, there is only a
very small region of negative production with very strong spanwise
elongation. While this negative production is undeniably weak, it is
curious, and the underlying mechanisms or structures responsible for
it are not clear. Further, the mechanisms may be different in the
near-wall and outer regions.

\subsection{Dissipation}
\label{subsec:diss}
\begin{figure}
  \centering
  \includegraphics[width=\textwidth]{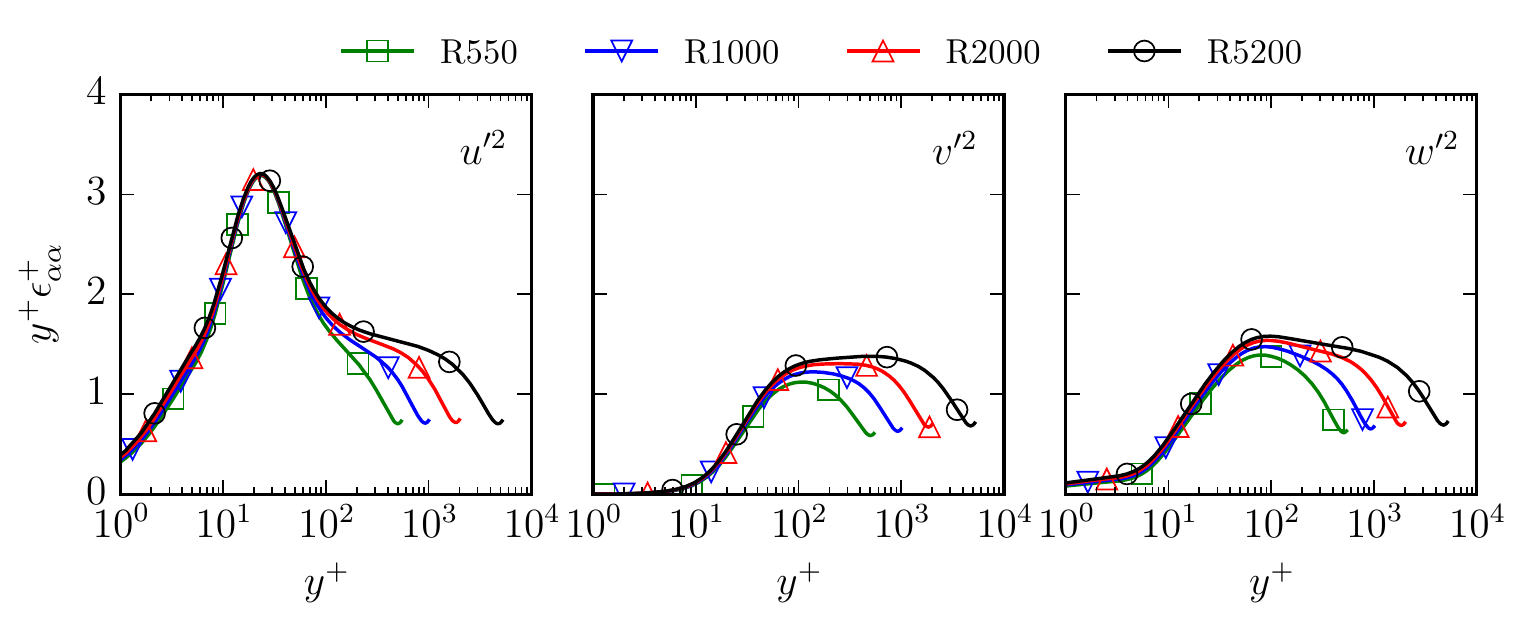} 
  \caption{Profiles of \uu, \vv, 
    and \ww dissipation as log-densities.}
  \label{fig:1d_diss}
\end{figure}

The one-dimensional profiles of the velocity variance dissipations are
shown in figure~\ref{fig:1d_diss}.  There are only weak $Re$
dependencies in these profiles for $y^+ < 70$. For $y^+>100$ or so and
$y/\delta<0.2$ (the overlap region), all three dissipation profiles
are approaching a plateau as $Re$ increases.  This is as expected,
under the assumptions of: 1) a constant stress region (\uvp$\approx
-1$) and logarithmic velocity profile in the overlap region, 2)
equality of production and dissipation of turbulent kinetic energy,
and 3) isotropy of the dissipation. With these assumptions, the
production and therefore dissipation of kinetic energy in the overlap
region is given simply by $y^+\epsilon^+_K=1/\kappa\approx 2.6$,
assuming $\kappa=0.384$ as reported in \citet{Lee:2015er}. None of the
above assumptions is exactly true. For example, a much higher Reynolds
number is required for the Reynolds stress to approach one in the
overlap region, and it appears that the equality of production and
dissipation is an approximation that does not improve with increasing
Reynolds number \citep{Lee:2015er}.  Within these limitations, the
profiles in figure~\ref{fig:1d_diss} are in reasonable agreement with
this prediction.  For $y/\delta > 0.2$ the premultiplied dissipation
profiles of all components decrease for large $y$, though at the
center of channel they increase slightly. This increase is a
consequence of multiplying the dissipation by $y$ since dissipation is
symmetric and non-zero at the center.

\begin{figure}
  \centering
  \includegraphics[width=\textwidth]{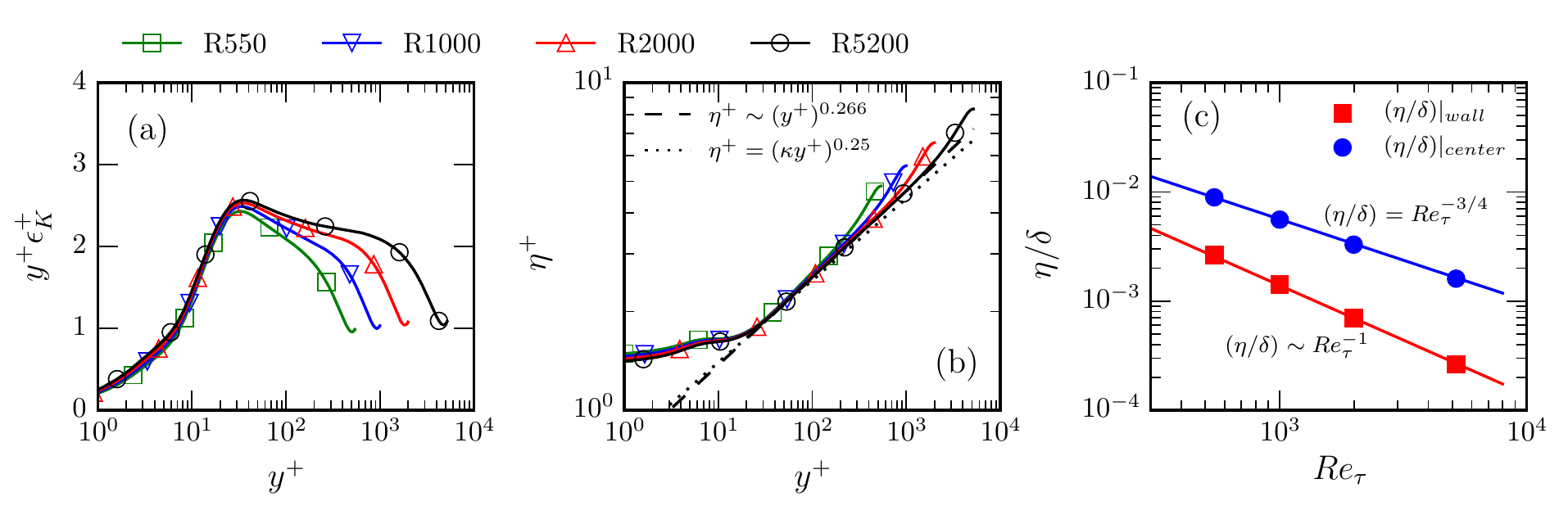} 
  \caption{Profiles of kinetic energy dissipation $\epsilon_K$ (a) as
    log-densities, and Kolmogorov length scale
    $\eta=(\nu^3/\epsilon_K)^{\frac{1}{4}}$  (b) as a function of
    $y^+$, and (c) at the wall and channel center as a function of
    Reynolds number.}
  \label{fig:kolmogorov_scale}
\end{figure}

The same assumptions also yield an estimate for the Kolmogorov length
scale $\eta = (\nu^3/\epsilon_k)^{1/4}$ \citep{Tennekes:1972vb}. That
is,
\begin{equation}
\eta^+ = \left(\epsilon_K^+\right)^{-1/4} \approx
\left(P_K^+\right)^{-1/4} \approx (\kappa y^+)^{1/4}
\label{eq:kolmogorov_scale}
\end{equation}
Profiles of the dissipation of TKE and the Kolmogorov scale at
different $Re$ are plotted in figure~\ref{fig:kolmogorov_scale}a,b. In
the overlap region, the $\eta$ profiles collapse reasonably well to a
power-law dependence on $y^+$, with an exponent of 0.266 which is only
slightly different from 1/4. Again, a disagreement that is consistent
with the limitations of the underlying assumptions.  From
figure~\ref{fig:kolmogorov_scale}c, it is clear that $\eta / \delta$
scales as $Re_\tau ^{-1}$ and $Re_\tau^{-\frac{3}{4}}$ at the wall and
the centerline respectively, which is consistent with expectations
from the law of the wall and Kolmogorov scaling.

\begin{figure}
  \begin{center}
    \includegraphics[width=\textwidth]{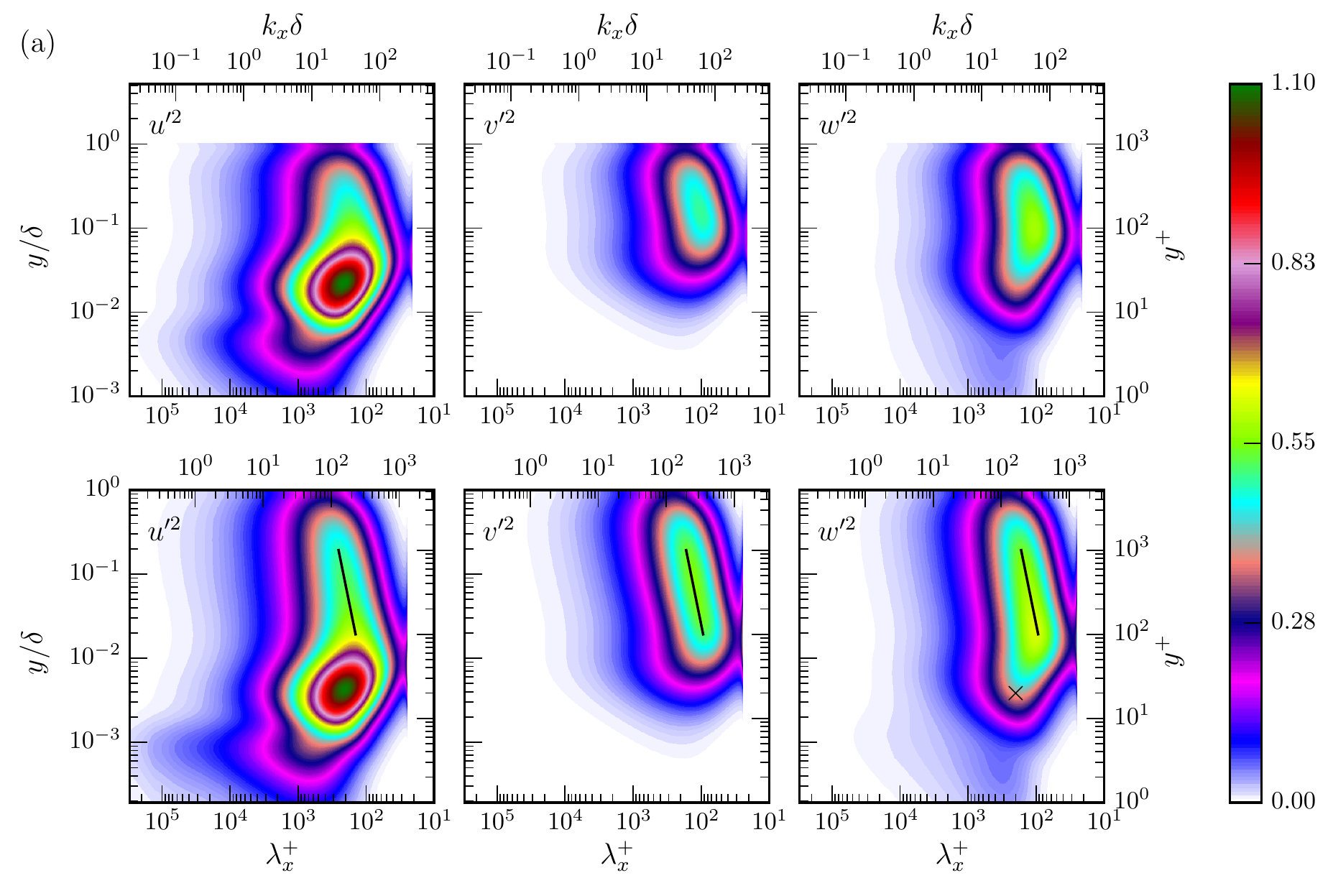}\\
    \includegraphics[width=\textwidth]{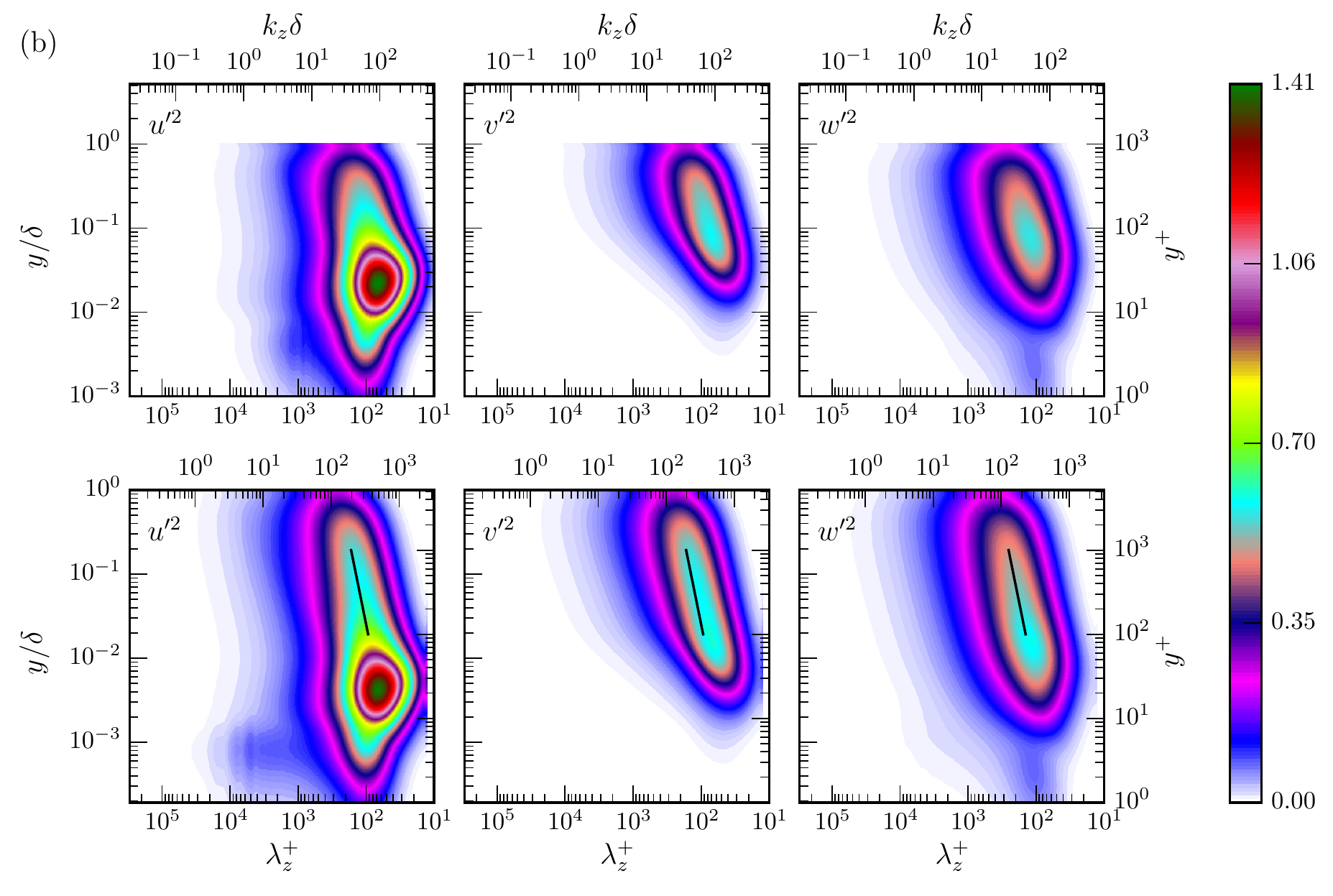}
  \end{center}
  \caption{One-dimensional spectral density of \uu, \vv, and \ww
    dissipation. Solid lines are at $k_xy^{1/4}\delta^{3/4}=85$, 130
    and 130 (streamwise spectra) and $k_zy^{1/4}\delta^{3/4}=130$, 130
    and 85 (spanwise spectra) for \uu, \vv and \ww, respectively. The
    symbol $\times$ marks a feature discussed in the text.}
  \label{fig:1d_E_E}
\end{figure}

The dissipation spectral densities from R1000 and R5200 are shown in
figure~\ref{fig:1d_E_E}. In both cases, the \uu dissipation spectra,
$E^\epsilon_{11}$, have peaks at $y^+ \approx 20$, $\lambda_x^+\approx
200$ and $\lambda_z^+ \approx 70$. Notice that very close to the wall
($y^+<5$), the streamwise \uu dissipation spectrum has a relatively
broad high wavenumber peak, while the spanwise spectral peak in this
region is narrower and centered around $\lambda^+_z\approx 100$. This
is consistent with dissipation of the near-wall streaks due to the
no-slip condition at the wall. Further, notice that near the wall, the
spectral range of \uu dissipation extends to higher $\lambda^+$
(larger scales) with increasing $Re$, up to $\lambda_x^+\sim 10^5$ and
$\lambda_z^+\sim 10^4$ at $Re_\tau=5200$. This is presumably due to
the impact of the increasingly large (in `+' units) streamwise
elongated outer scales on the wall region, consistent with
\citet{Jimenez:2012ef}.

As is apparent in figure~\ref{fig:1d_diss}, the \vv dissipation is
negligible near the wall because the wall-normal gradient of $v'$ and
thus the dissipation must be zero at the wall, and this is reflected
in the spectra, $E^\epsilon_{22}$. The profiles of \ww dissipation are
similar to those for $\langle v'^2 \rangle$, with the exception of
near the wall ($y^+<10$), since \ww dissipation is not zero at the
wall. The spectra ($E^\epsilon_{22}$ and $E^\epsilon_{33}$) are also
similar, with the most obvious difference being a weak region of \ww
dissipation around $y^+=20$ and $\lambda_x^+=200$ (marked with
$\times$) that roughly overlaps the region where $E^\epsilon_{11}$ has
its strong peak.  These near-wall dissipation spectra are consistent
with the known structure of near-wall turbulence, particularly the
presence of streamwise vortices with limited streamwise extent.

In the overlap region, the location of the dissipation spectral peak
of all three components varies as approximately $\lambda\sim y^{1/4}$
(see the solid lines in figure~\ref{fig:1d_E_E}). This is consistent
with the observation above (\ref{eq:kolmogorov_scale}) that the
Kolmogorov scale varies approximately like $y^{1/4}$.

\begin{figure}
  \centering
  \includegraphics[width=\textwidth]{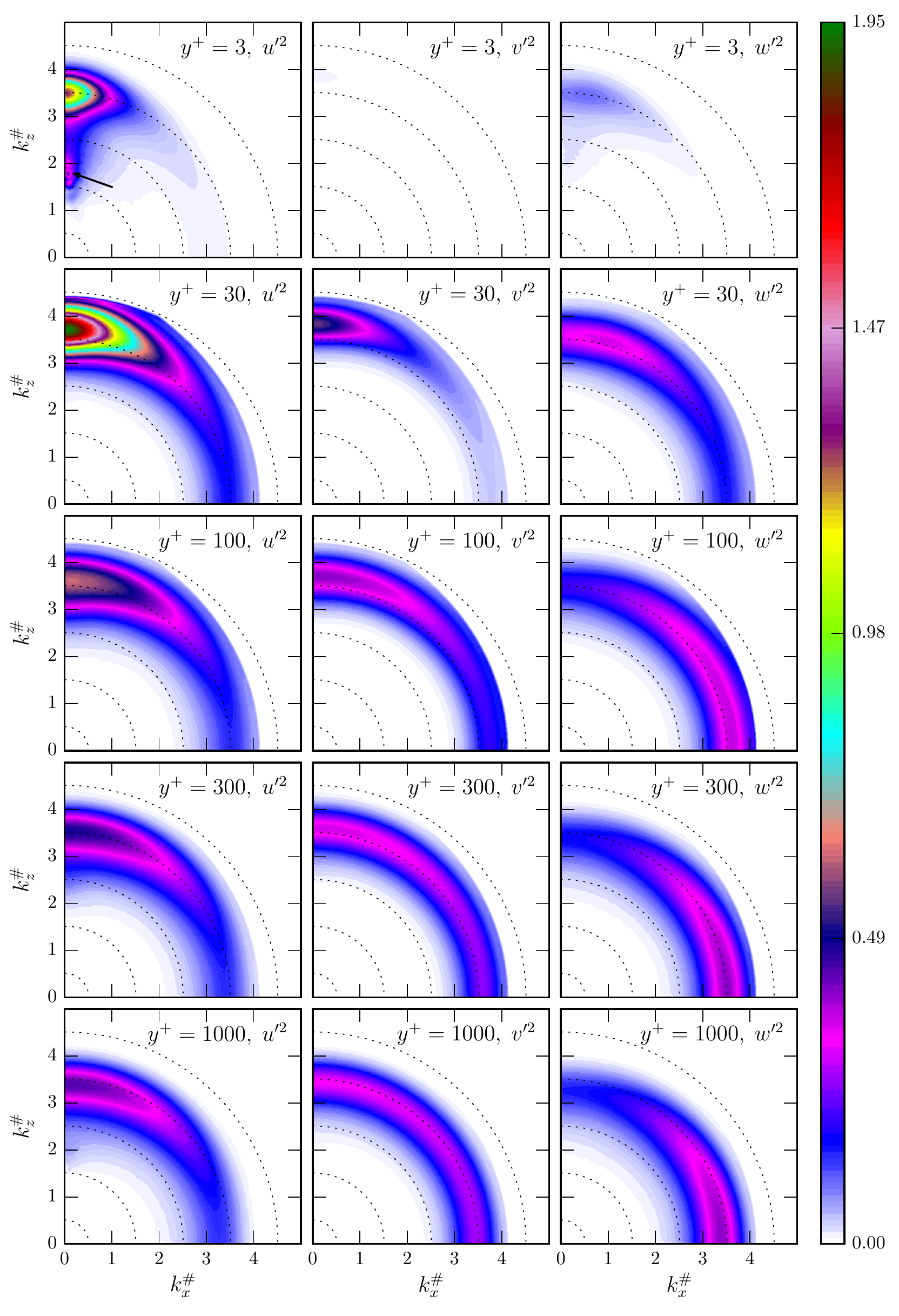}
  \caption{Two-dimensional spectral density of \uu, \vv, and \ww
    dissipation in log-polar coordinates, as defined in
    figure~\ref{fig:polar_coord_explain}, from R5200. $\lambda^+=10$
    on the outer-most dotted circle and increases by a factor of 10
    for each dotted circle moving inward. The arrow in the \uu
    spectrum at $y^+=3$ marks a feature discussed in the text.}
  \label{fig:2d_E_uu}
\end{figure}

The two-dimensional dissipation spectra in figure~\ref{fig:2d_E_uu}
provide more detailed information regarding the dissipation scales.
Very near the wall ($y^+=3$), the spectrum has the same structure as
the near-wall velocity spectra (see figure~\ref{fig:2d_uu}), which is
as expected because the dissipation this close to the wall is
dominated by the wall-normal derivatives \citep{Antonia:1991vo}, and
the velocity is zero at the wall. Note particularly, that even the
large-scale spectral peaks along the $k_z^\#$ axis as described for
the \uu spectra (see section~\ref{subsec:energy_spectra}) are present
in the \uu dissipation spectrum for this reason (marked with an
arrow).  With increasing distance from the wall, the dissipation
occurs more isotropically in scale, in a band of modes with
wavelengths that increase slowly with $y$ (like $y^{1/4}$), consistent
with the structure of the one-dimensional dissipation spectra. By
$y^+=1000$, the \vv dissipation spectrum is nearly isotropic in scale,
but the \uu and \ww spectra do not appear to be. However, this is
simply due to the difference in contributions between longitudinal and
transverse derivatives to the dissipation, because of
continuity. Particularly, in isotropic turbulence, the variance of
longitudinal derivatives is half that of transverse derivatives. The
stronger \uu (\ww) dissipation spectra along the $k_z^\#$ ($k_x^\#$)
axis is consistent with this characteristic of isotropic turbulence.

\begin{figure}
  \centering
  \includegraphics[width=\textwidth]{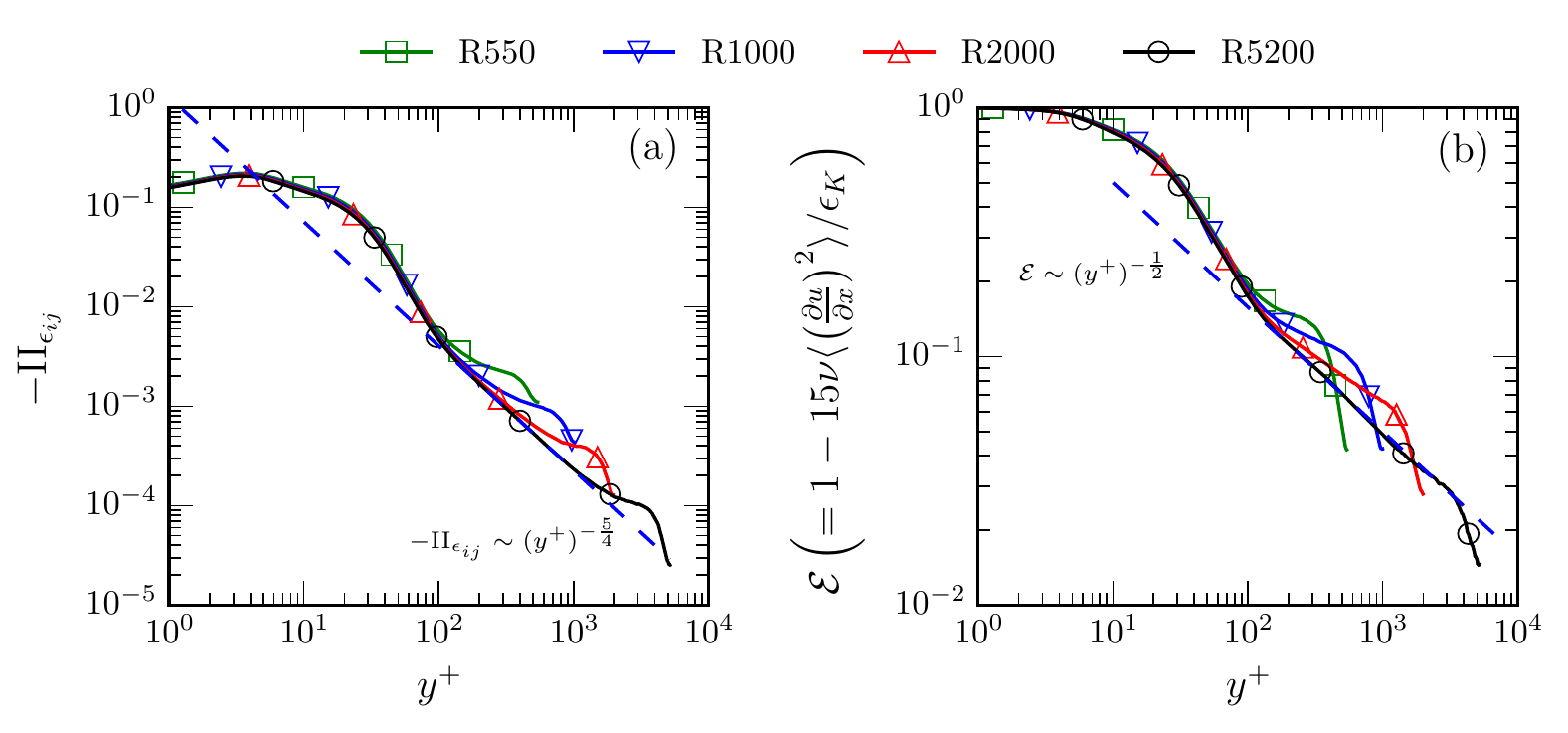}
  \caption{Isotropy of dissipation: (a)
    second invariant of the deviatoric part of the dissipation tensor; and,
    (b) Relative error of estimating the dissipation from the variance of
    $\partial u/\partial x$ using isotropy.}
  \label{fig:E_isotropy}
\end{figure}

It appears then that the dissipation processes are anisotropic near
the wall and become isotopic in scale as $y$ increases. Anisotropy is
expected near the wall because of the presence of the wall, and local
isotropy is widely assumed in modeling wall-bounded flow away from the
wall. To characterize the component anisotropy of the dissipation, the
second invariant of the deviatoric part of the dissipation tensor,
$\textrm{II}_{\epsilon_{ij}}$, is computed as suggested by
\citet{Antonia:1991vo,Antonia:1994ca}.
\begin{equation}
\textrm{II}_{\epsilon_{ij}} = -\frac{1}{2} d_{ij} d_{ji}\qquad\mbox{with}\qquad
d_{ij} = \frac{\epsilon_{ij}}{2\epsilon_K} - \frac{\delta_{ij}}{3},
\end{equation}
Also of interest is the error committed when using an assumption of
isotropy in both scale and components to estimate the dissipation of
kinetic energy from the variance of $\partial u/\partial x$, as might
be done with experimental data.  These characterisations of anisotropy
are shown as a function of $y^+$ in figure~\ref{fig:E_isotropy}a and
b. As expected, anisotropy of dissipation decreases with distance from
the wall. In particular, in the overlap region ($y^+\approx 100$ to
$y/\delta \approx 0.2$) $\textrm{II}_{\epsilon_{ij}}$ varies as
$-y^{-5/4}$ while the relative error in estimating dissipation from
the streamwise velocity gradient varies like $y^{-1/2}$.

\subsection{Inter-component energy transfer}
\label{subsec:pressure_strain}
\begin{figure}
  \centering
  \includegraphics[width=\textwidth]{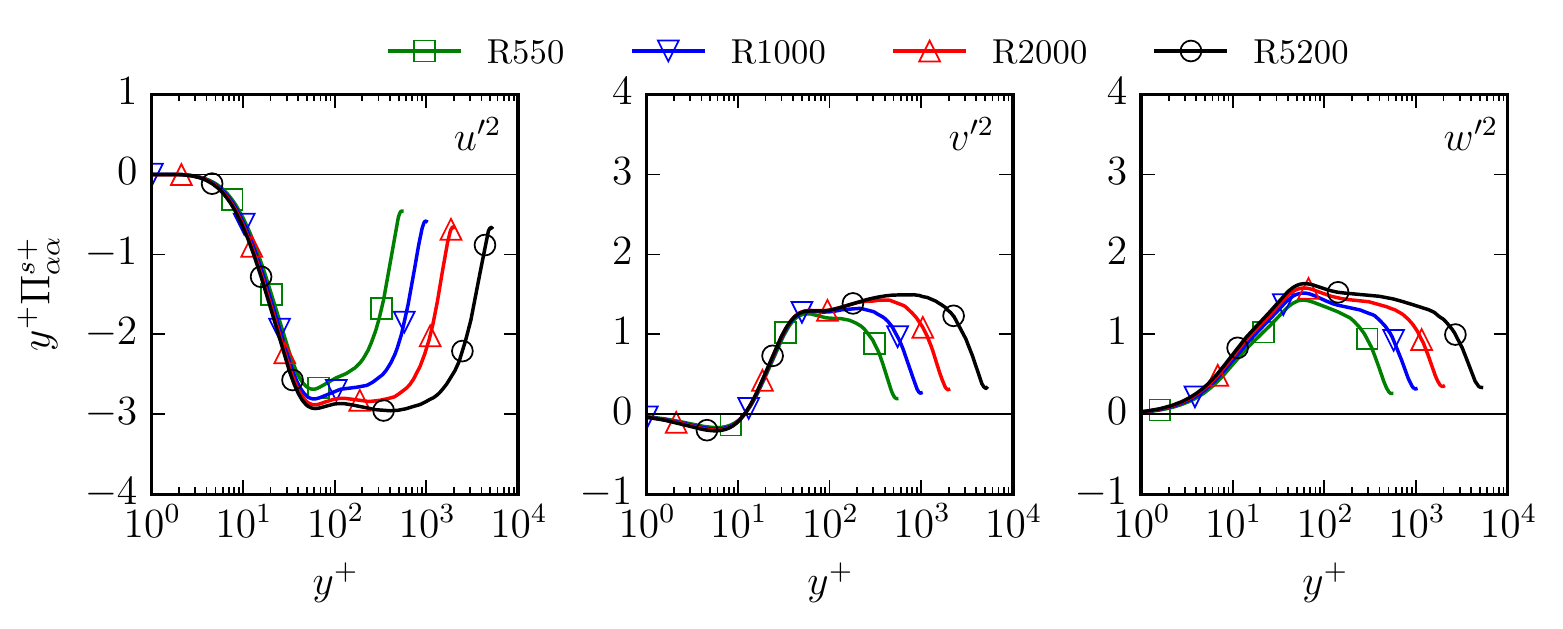} 
  \caption{Profiles of the pressure strain terms for \uu, \vv,
  and \ww, as log-densities.}
  \label{fig:1d_press_transport}
\end{figure}

The pressure-strain correlation $\Pi^s$ exchanges energy between the
velocity components (it has zero trace). As is well known, energy
production occurs only in \uu but the dissipation occurs in all three
components, and $\Pi^s$ enables this by redistributing energy from \uu
to the other components. This is shown in
figure~\ref{fig:1d_press_transport}, where the streamwise component
$\Pi^s_{11}$ is negative except for $y^+<1.5$, where it is only
slightly positive, while the $\Pi^s_{22}$ and $\Pi^s_{33}$ components
are positive almost everywhere. The one exception is that $\Pi^s_{22}$
is negative for $y^+<12$. This is due to the well-known ``splat''
effect \citep{Mansour:1988vz,Perot:1995vu}, in which fluid moving
toward the wall is blocked by the wall and must turn to flow parallel
to the wall, transferring energy from the vertical to horizontal
velocity fluctuations. Because vertical velocity fluctuations very
near the wall are primarily associated with streamwise vortices, this
energy transfer is primarily from \vv to \ww.

It is interesting to note that in the components parallel to the wall
($\Pi^s_{11}$ and $\Pi^s_{33}$), there is a weak Reynolds number
dependence in the profiles near the wall, while the profile of the
wall-normal component ($\Pi^s_{22}$) is Reynolds number independent
out to $y^+$ as large as 100. This is presumably because there are
processes in the near-wall exchange of energy between \uu and \ww that
are mediated by outer-region turbulent structures that do not involve
\vv.  Also interesting is that a broad outer peak in the profiles
appears to develop with increasing Reynolds number, in both the
streamwise and wall-normal components ($\Pi^s_{11 }$ and $\Pi^s_{22}$)
at $y/\delta\approx0.1$ and 0.2 respectively. These are in the outer
parts of the overlap region. However, no such outer peak seems to be
developing in the spanwise component. These differences suggest that
the mechanisms for exchange of energy to \vv and \ww are different.

\begin{figure}
  \begin{center}
    \includegraphics[width=\textwidth]{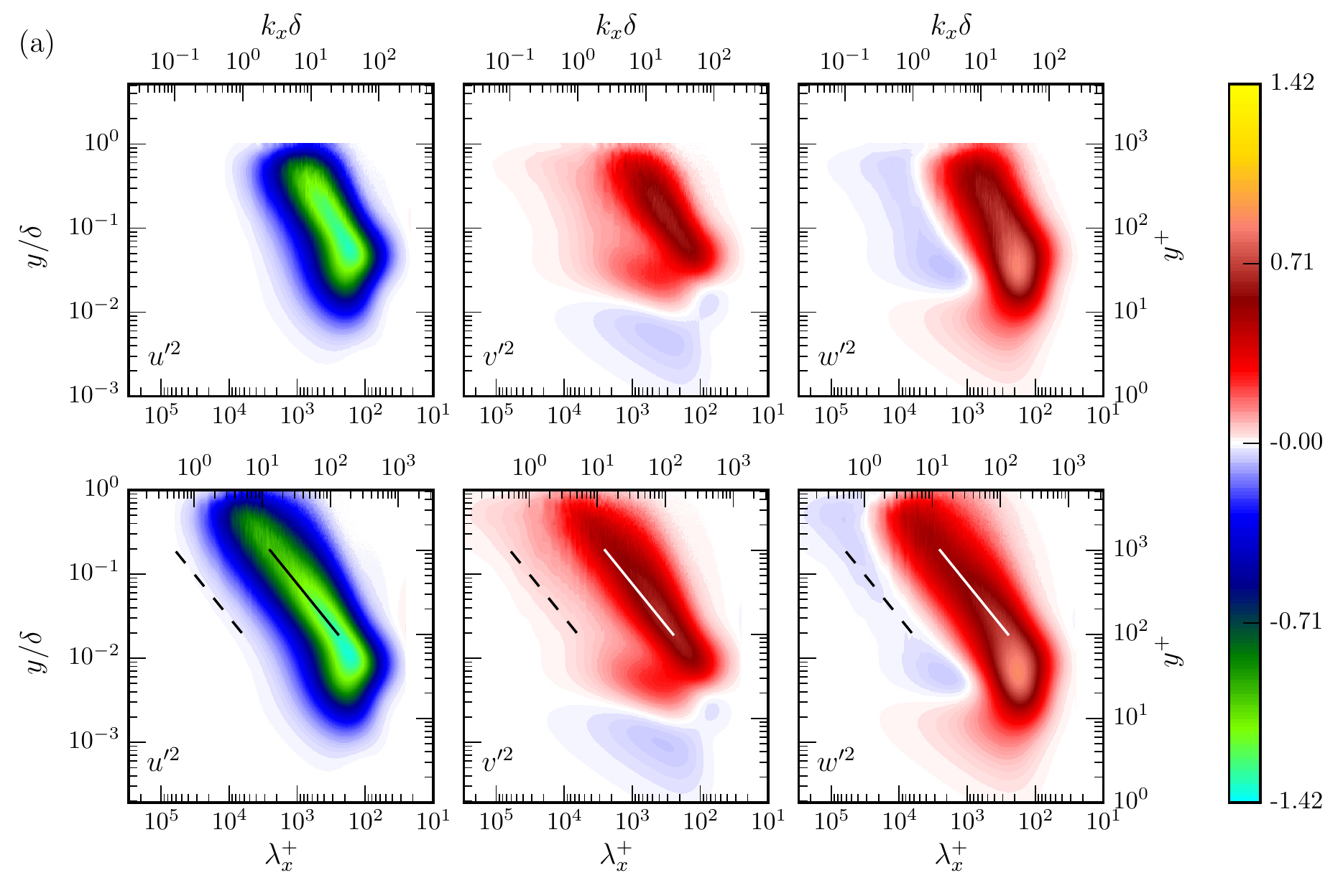}\\
    \includegraphics[width=\textwidth]{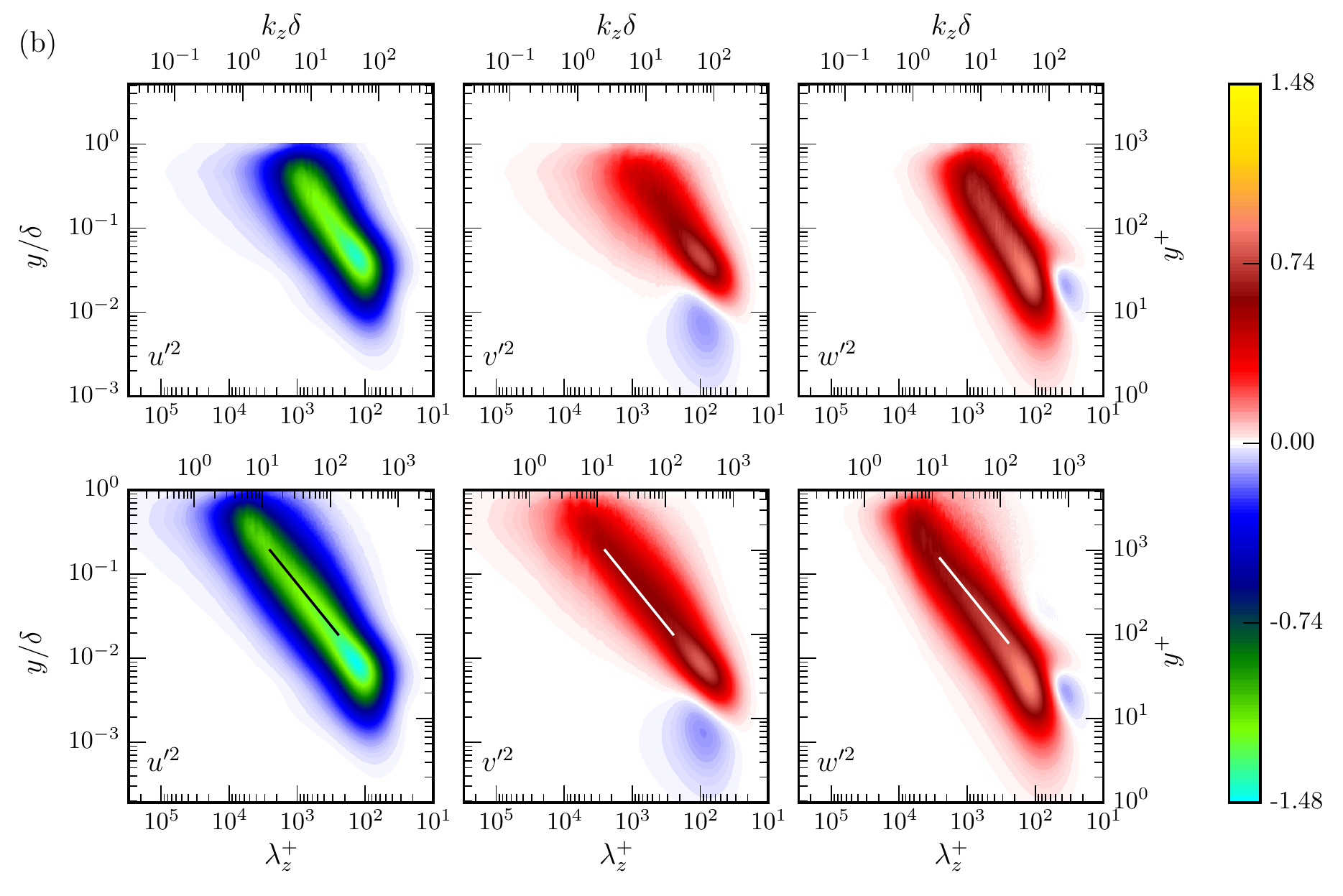}
  \end{center}
  \caption{One-dimensional spectral density of the \uu, \vv and \ww
    pressure-strain terms. Solid lines are at $ky=2.5$. Dashed lines in streamwise spectra are at $k_x y = 0.1$.
    }
  \label{fig:1d_E_PI_s}
\end{figure}

\begin{figure}
  \begin{center}
    \includegraphics[width=\textwidth]{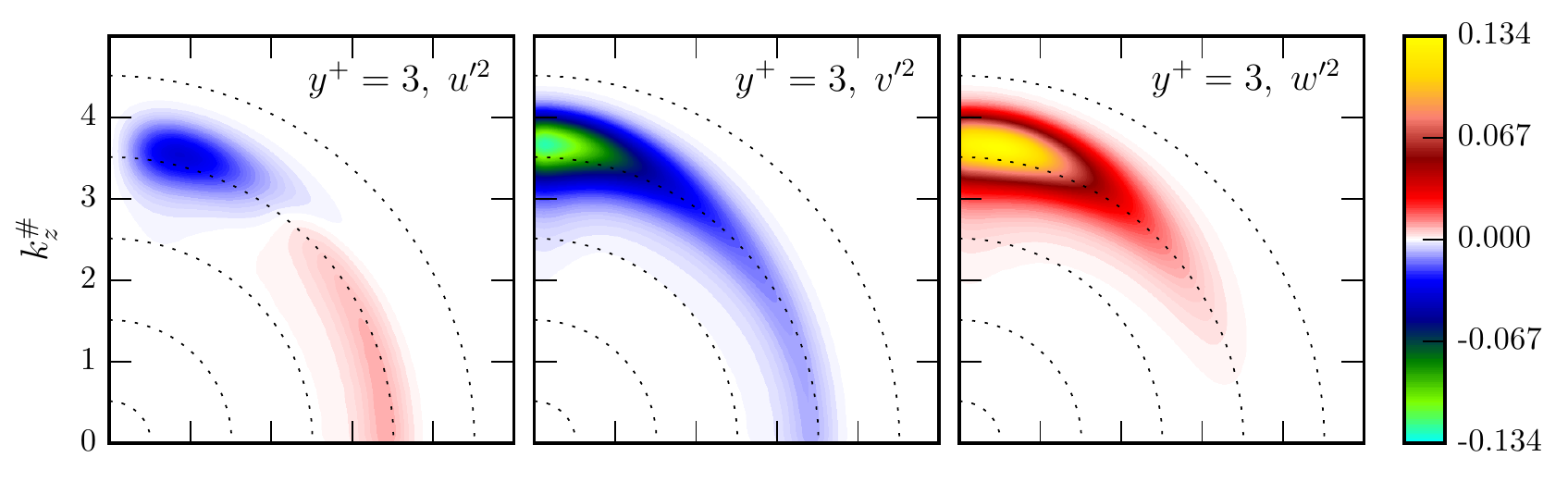} \\[-0.7em]
    \includegraphics[width=\textwidth]{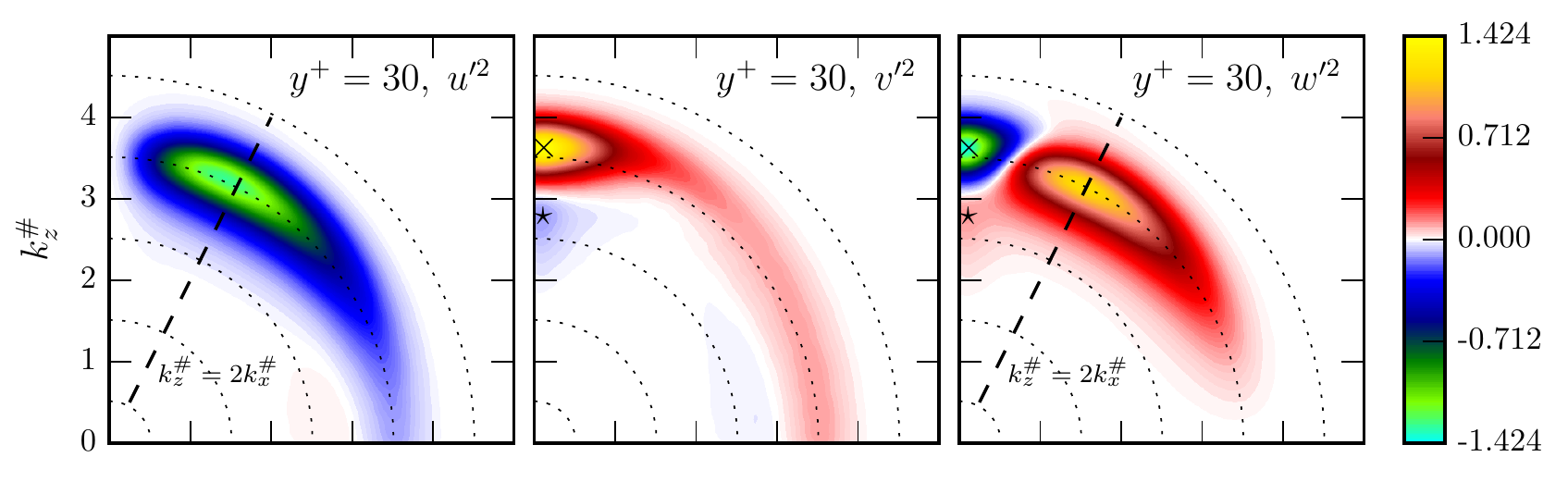} \\[-0.7em]
    \includegraphics[width=\textwidth]{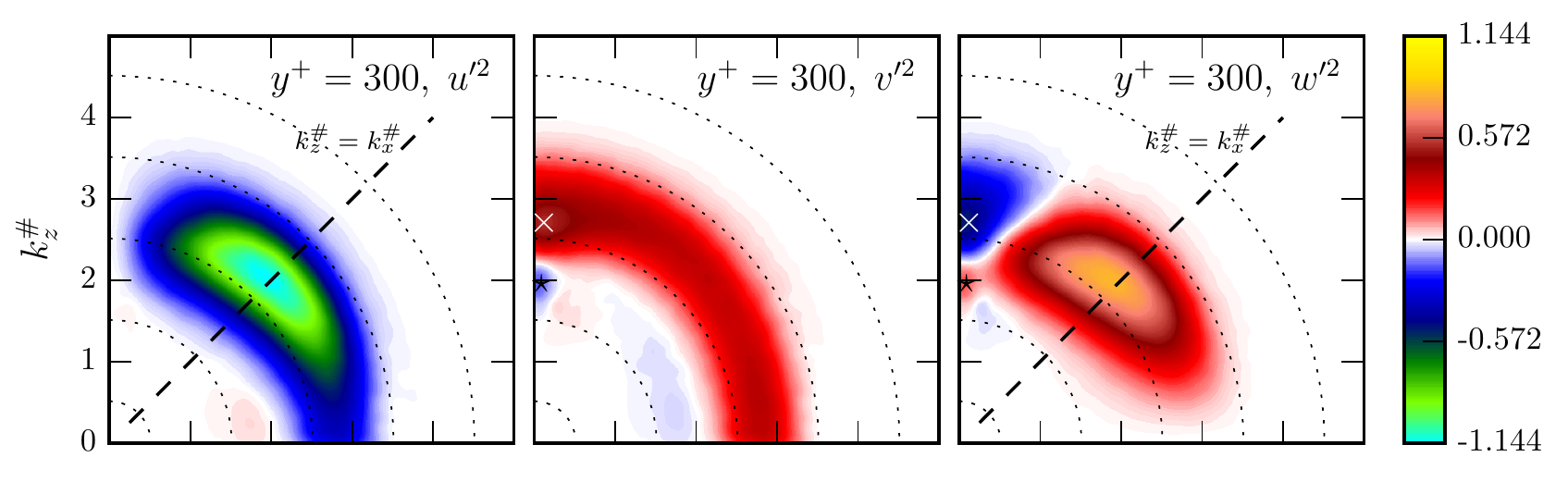} \\[-0.7em]
    \includegraphics[width=\textwidth]{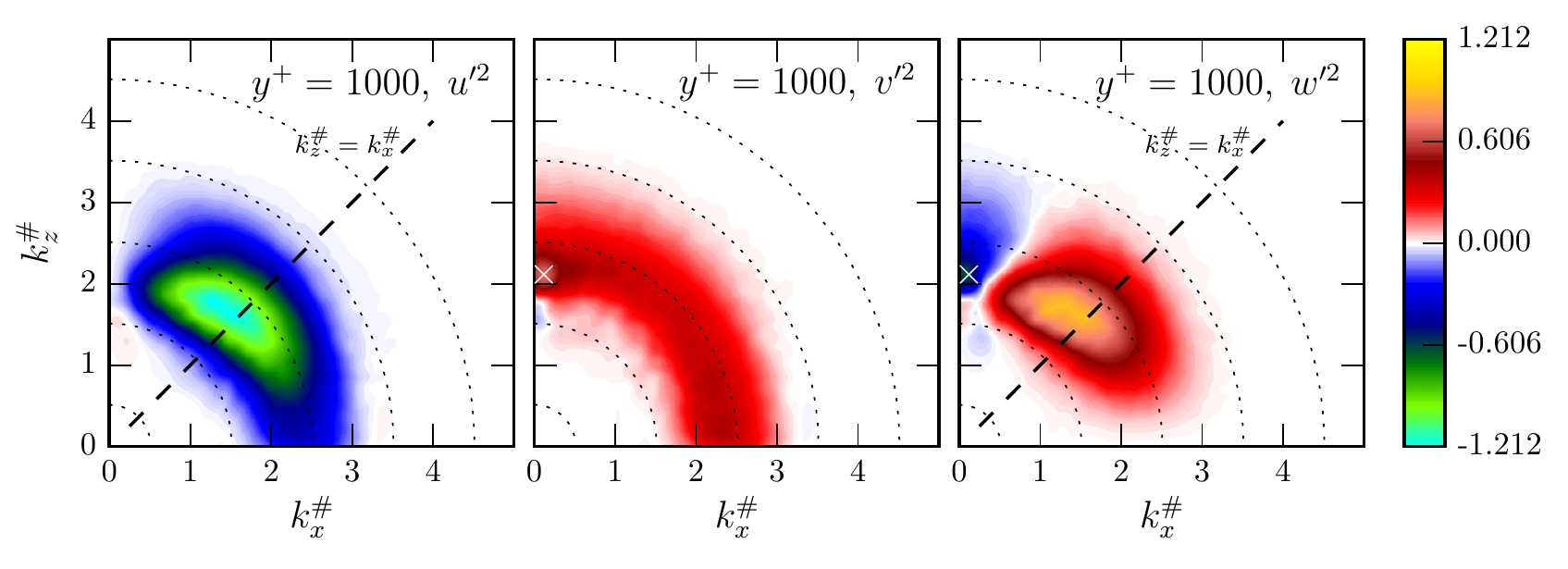}
  \end{center}
  \caption{Two-dimensional spectral density of the \uu, \vv and \ww
    pressure-strain terms in log-polar coordinates as defined in
    figure~\ref{fig:polar_coord_explain}, from R5200. $\lambda^+=10$
    on the outer-most dotted circle and increases by a factor of 10
    for each dotted circle moving inward. Symbols ($\times$ and $\star$)
    mark features discussed in the text.}
  \label{fig:2d_E_PI_s}
\end{figure}

Just as the pressure-strain term in the two-point correlation has zero
trace (\ref{eq:press_strain_condition}), so does the pressure-strain
spectrum:
\begin{equation}
  E_{ii}^{\Pi^s} = 0 \quad \forall \;(k_x, y, k_z)
  \label{eq:press_strain_energy_condition}
\end{equation}
Thus, the one- and two-dimensional pressure-strain spectra can be
interpreted as inter-component energy transfer at given scales
(wavenumbers) and wall-normal locations.  The one-dimensional spectra
of the pressure strain terms are shown in
figure~\ref{fig:1d_E_PI_s}. The dominant feature of these spectra, for
both Reynolds numbers and in both spectral directions ($k_x$ and
$k_z$), is a transfer of energy from \uu to \vv and \ww over a range
of wavenumbers that scale with $y^{-1}$.  This feature, which extends
from $y^+\approx 30$ to $y/\delta\approx 0.5$, is centered on the
high-wavenumber side of the region of \uu production (see
figure~\ref{fig:1d_E_P_uu}) in $k_x$, and is centered on the maximum
production region in $k_z$.  Indeed over this range in $y$, the
pressure-strain spectral peaks of \uu occur at $k_xy\approx
k_zy\approx 2.5$, while the production peaks occur at $k_xy\approx
0.5$ and $k_zy\approx2$, with significant production occurring at
wavenumbers as low as $k_xy\approx 0.06$. There is thus a range of
large scales (low wavenumbers) in $x$ at which there is significant
production of \uu and insignificant transfer of energy to the other
components. The same is true near the wall, where \uu production at
scales that are long in $x$ results in no transfer to other
components. This low $x$-wavenumber production occurs with much higher
$z$-wavenumbers, as is clear from figure~\ref{fig:2d_P_uu_new}. The
production of streamwise velocity fluctuations that are elongated in
$x$ in this way will not significantly impact continuity, and will
therefore not generate a significant pressure response or
inter-component transfer. This suggests that pressure redistribution
should be dominated by wavenumbers with $k_x$ of the same order as
$k_z$, which is consistent with the one-dimensional spectra in
figure~\ref{fig:1d_E_PI_s}, though the two-dimensional spectra
indicate that things are a bit more complex, especially near the wall.

Another noteworthy feature of the one-dimensional spectra is the
weakly negative region in the streamwise spectrum of $\Pi^s_{33}$ that
extends from the center of the channel to $y^+\approx30$ (marked with
a dashed line in figure~\ref{fig:1d_E_PI_s}). This energy is
transferred to \vv, as indicated by the fact that there is a
corresponding positive region in the $\Pi^s_{22}$ spectrum (also
marked with a dashed line), but not $\Pi^s_{11}$.  It appears that the
source of the \ww energy at these wavenumbers is inter-scale transfer
from the dominant positive region in figure~\ref{fig:1d_E_PI_s} to
larger scales (see discussion in
section~\ref{subsec:inter-scale_transfer}).

Two-dimensional pressure-strain spectra at several $y$-locations are
shown in figure~\ref{fig:2d_E_PI_s}. One common feature of the spectra
at all $y$-locations is that the dominant inter-component transfer
occurs in a circular band of wavenumbers centered around a value of
$k$ that increases with $y$. Beyond $y^+\approx300$, the dominant
wavelength scales with $y$ with the center of the dominant zone at
$ky\approx 2.5$, consistent with the one-dimensional spectra. The
width of the dominant zone also increases with $y$, presumably because
of the increase in local Reynolds number.  Another common feature is
that at the $k_x^\#=0$ and $k_z^\#=0$ axes, the $\Pi^s_{11}$ and
$\Pi^s_{33}$ spectra, respectively, go to zero. This is because for
modes with no variation in $x$ ($z$) the fluctuations in $u$ ($w$)
have no impact on continuity, and so do not produce a pressure
response.

Very near the wall ($y^+=3$ in figure~\ref{fig:2d_E_PI_s}), the
spectra are dominated by the splat effect discussed above. Here, the
energy is primarily transferred from \vv to \ww for fluctuations that
are strongly aligned in the streamwise direction, with the strongest
transfers occurring at $k_x=0$. Curiously there is also a weak
transfer from \uu to \ww in these streamwise elongated wavenumbers.
Wavenumbers associated with spanwise elongation also produce
transfers, but in this case from \vv to \uu. This is consistent with
expectations for elongated regions of motion toward the wall, that the
fluid turns in the direction perpendicular to the elongation, like a
two-dimensional stagnation point flow.

Except for very near the wall ($y^+=3$), the dominant transfer is from
\uu to \vv and \ww over a broad range of wavenumber orientations
centered around $k_x\approx k_z$ farther from the wall ($y^+=300$ and
1000), and centered around a more streamwise elongated orientation
($k_z\approx 2k_x$) at $y^+=30$ (see dashed guide lines in
figure~\ref{fig:2d_E_PI_s}). This transfer is strongest to \ww, except
for the spanwise elongated modes ($k_z$ near zero). In addition,
except very near the wall, there is a strong transfer from \ww to \vv
where $k_z\gg k_x$ (marked with $\times$) as well as transfer from \vv
to \ww at larger scales (smaller $k_z^\#$, marked with $\star$).
Transfer between \vv and \ww like this would be expected from
streamwise oriented vortices. It is this structure that is responsible
for the negative regions in figure~\ref{fig:1d_E_PI_s}.  It is also is
interesting that there is weak transfer from \vv to \uu and \ww even
away from the wall ($y^+=30$ and 300) for $\lambda^+\gg y^+$.  As with
the splat effect the transfer is toward \uu (\ww) for modes that are
elongated in $z$ ($x$). Indeed, for such large wavelength modes, even
these $y$-locations are essentially ``close'' to the wall, so it may
indeed be a manifestation of the splat effect.

The scale-dependent inter-component transfer described here is
complicated, with only the near-wall process described in terms of the
structural features of the turbulence. We should aspire to conceptual
models of wall-bounded turbulence in the buffer and overlap regions
that are consistent with these observations.

\subsection{Wall-normal energy transport}
As is well-known, there are two wall-normal energy transport
mechanisms; a linear mechanism due to viscous diffusion, and a
nonlinear mechanism due to turbulent fluctuations. These linear and
non-linear transport mechanisms are investigated in the following
subsections.
\subsubsection{Linear mechanism: Viscous diffusion}
\label{subsubsec:wall-normal_linear}
\begin{figure}
  \centering
  \includegraphics[width=\textwidth]{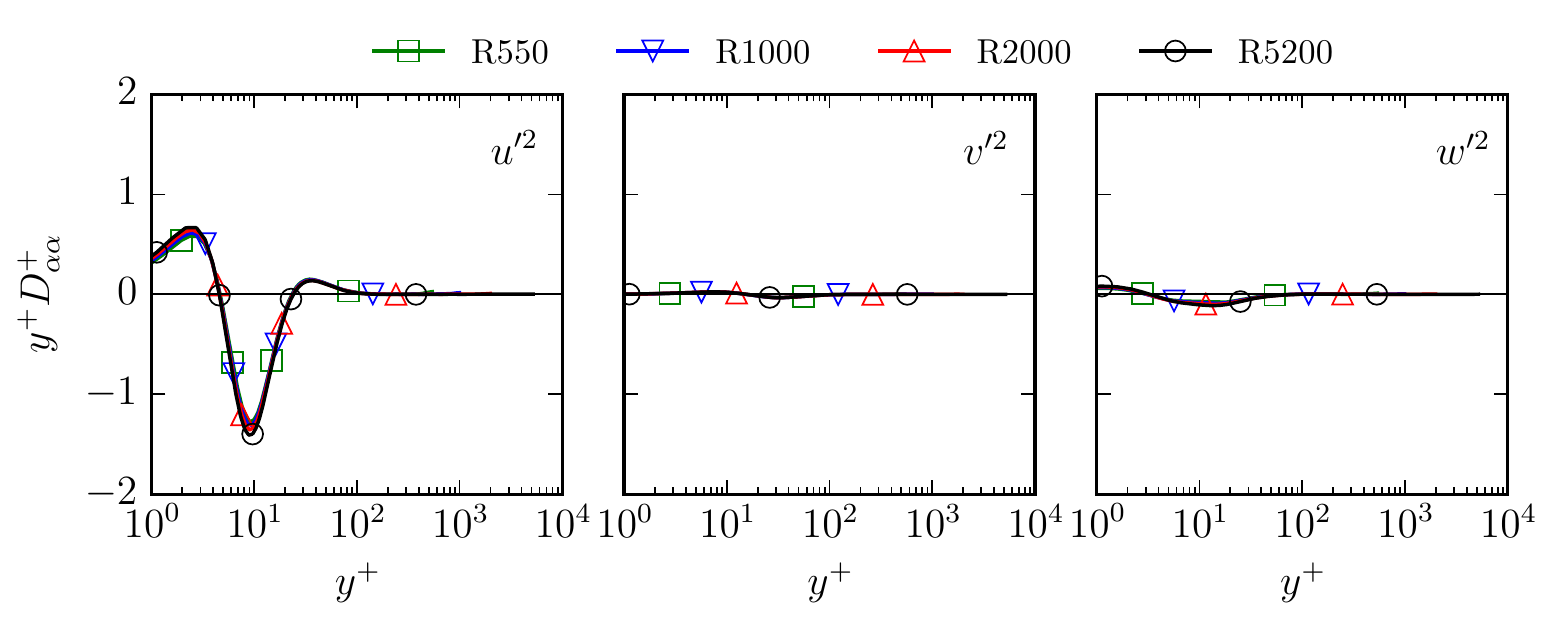} 
  \caption{Profiles of \uu, \vv, and \ww viscous transport as
    log-densities.}
  \label{fig:1d_visc_transport}
\end{figure}

\begin{figure}
  \begin{center}
    \includegraphics[width=\textwidth]{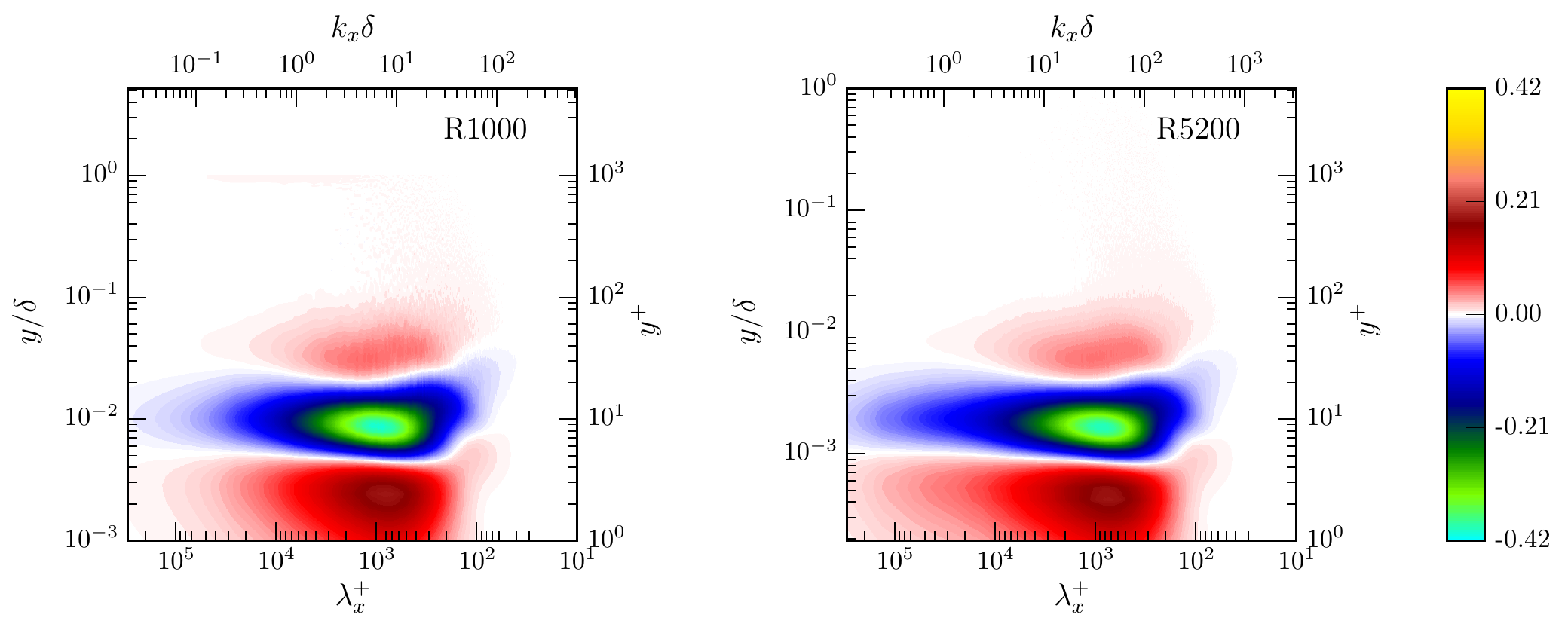}
    \includegraphics[width=\textwidth]{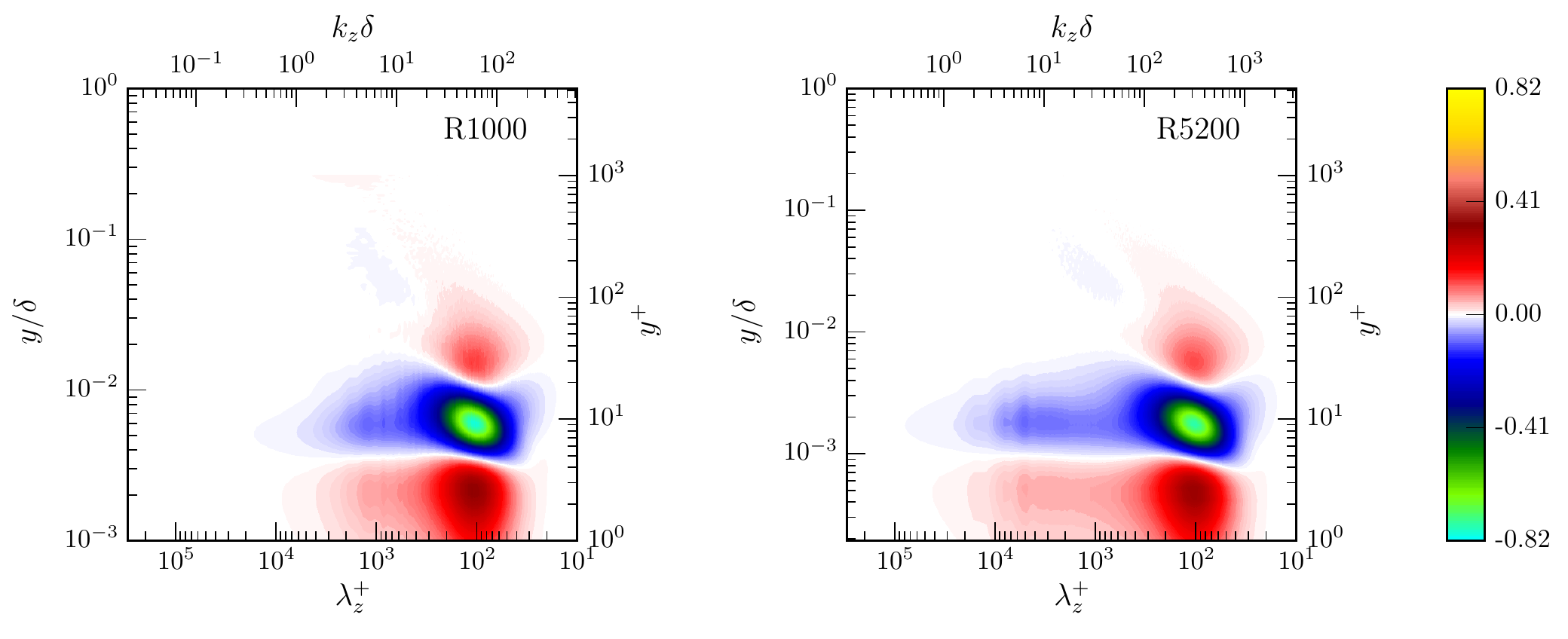}
  \end{center}
  \caption{One-dimensional streamwise and spanwise spectral densities
    of $D_{11}$.}
  \label{fig:1d_E_D}
\end{figure}

\begin{figure}
  \begin{center}
    \includegraphics[width=0.7\textwidth]{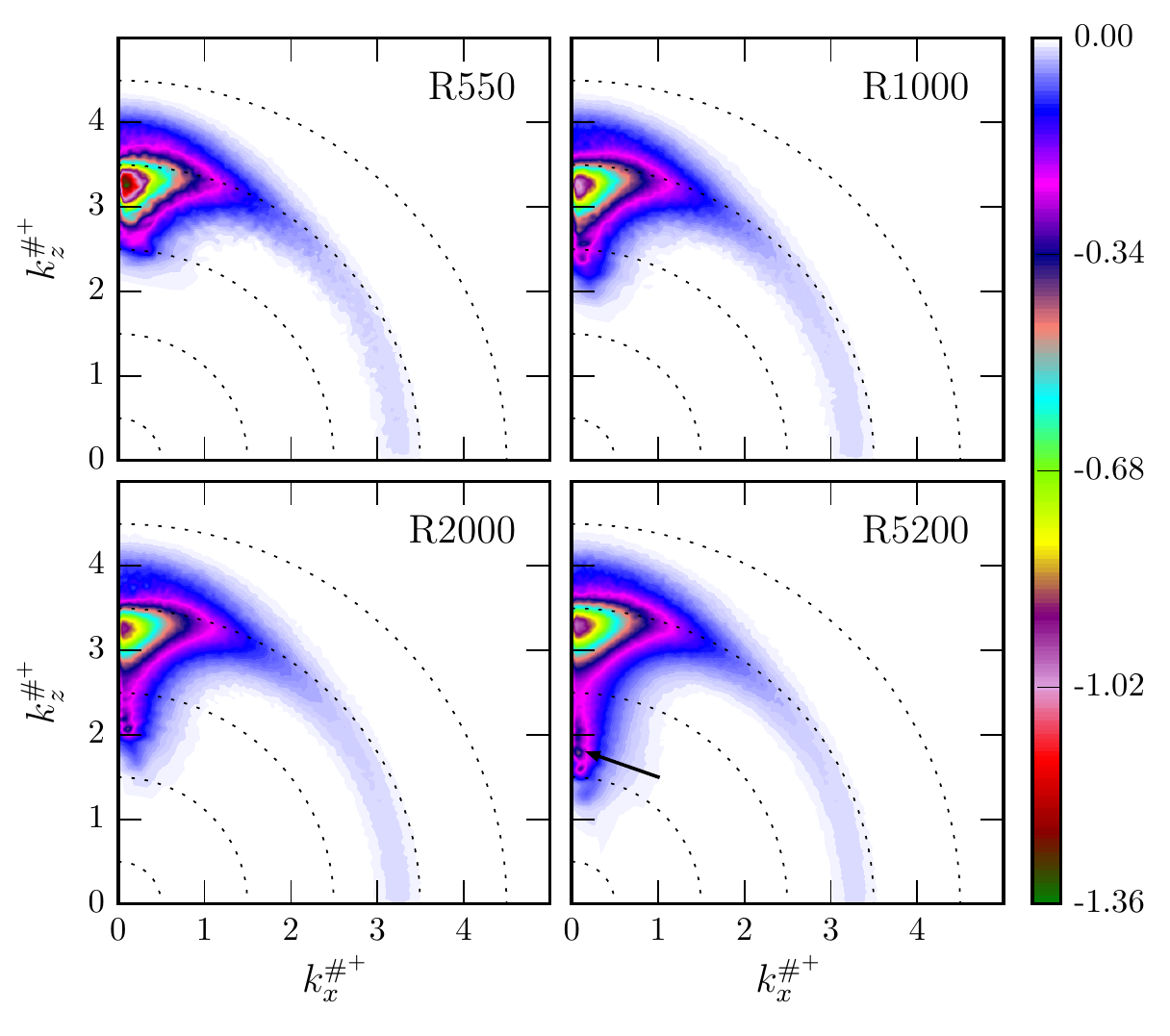}
  \end{center}
  \caption{Two-dimensional spectral density of $D_{11}$ in log-polar
    coordinates, as defined in figure~\ref{fig:polar_coord_explain},
    at $y^+=15$. $\lambda^+=10$ on the outer-most dotted circle and
    increases by a factor of 10 for each dotted circle moving
    inward. The arrow marks a feature discussed in the text.}
  \label{fig:2d_E_D_uu}
\end{figure}

The one dimensional profile of viscous transport in the $y$ direction
is shown in figure~\ref{fig:1d_visc_transport}. The transport of \vv
and \ww is negligible compared to the viscous transport of $\langle
u^{\prime2}\rangle$. There is a weak $Re$ dependence in $D^+_{11}$
near the wall ($y^+<100$), which is absent further from the wall. This
weak $Re$ dependence can be understood in the context of the
one-dimensional diffusion spectra, which are shown in
figure~\ref{fig:1d_E_D}.  The dominant small-scale structure of these
spectra appears the same at the two Reynolds numbers shown. For both
Reynolds numbers, the spectral peaks occur at approximately
$\lambda_x^+\approx 800$ and $\lambda_z^+\approx 100$. However, the
contribution from the large scales increases with $Re$. There even
appear to be weak large-scale local extrema at $k_z\delta\approx7$ in
the spanwise spectra at $y^+\approx 3 $ and $y^+\approx10$ though it
is not obvious in figure~\ref{fig:1d_E_D}.

The reason for the Reynolds number dependence is clearer in the
two-dimensional $D_{11}$ spectra at $y^+=15$ shown in
figure~\ref{fig:2d_E_D_uu} for flows at different Reynolds numbers.
These spectra have the same structure as the \uu spectra at the same
$y$-location and Reynolds numbers in figure~\ref{fig:2d_uu_y_plus_15},
except that here the spectrum is negative because this $y$-location is
near the negative peak of the \uu transport in
figure~\ref{fig:1d_visc_transport}. The structural similarity includes
the large wavelength region of large magnitude along the $k_z^\#$
axis, which is the primary feature that depends on Reynolds number
(marked with an arrow). As discussed in
section~\ref{subsec:energy_spectra}, this spectral region arises due
to large-scale outer-flow structures that are imposed on the near-wall
turbulence, which also explains the Reynolds number dependence. A
quantitative analysis of the effects of these large-scale features on
Reynolds number dependence is provided in
\S~\ref{subsec:small_scale_universality}.

\subsubsection{Nonlinear mechanism: Turbulent \& Pressure-driven wall-normal transport}
\label{subsubsec:wall-normal_nonlinear}
\begin{figure}
  \centering
  \includegraphics[width=\textwidth]{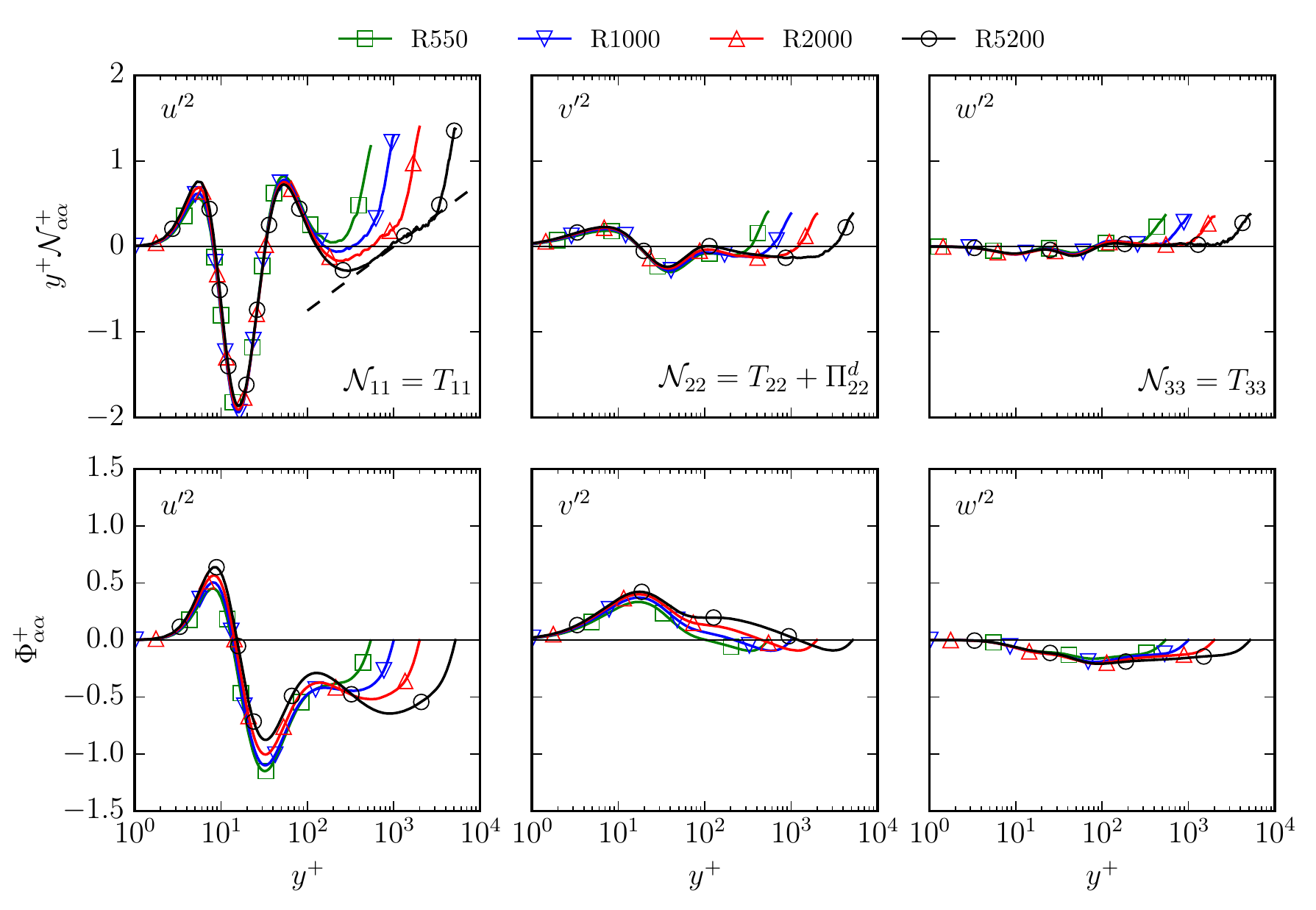} 
  \caption{Profile of \uu, \vv and \ww non-linear transport, as
    log-densities and the nonlinear flux $\Phi_{\alpha\alpha}$ as
    defined in (\ref{eq:turb_transport_flux}).}
  \label{fig:1d_turb_transport}
\end{figure}

\begin{figure}
  \begin{center}
    \includegraphics[width=\textwidth]{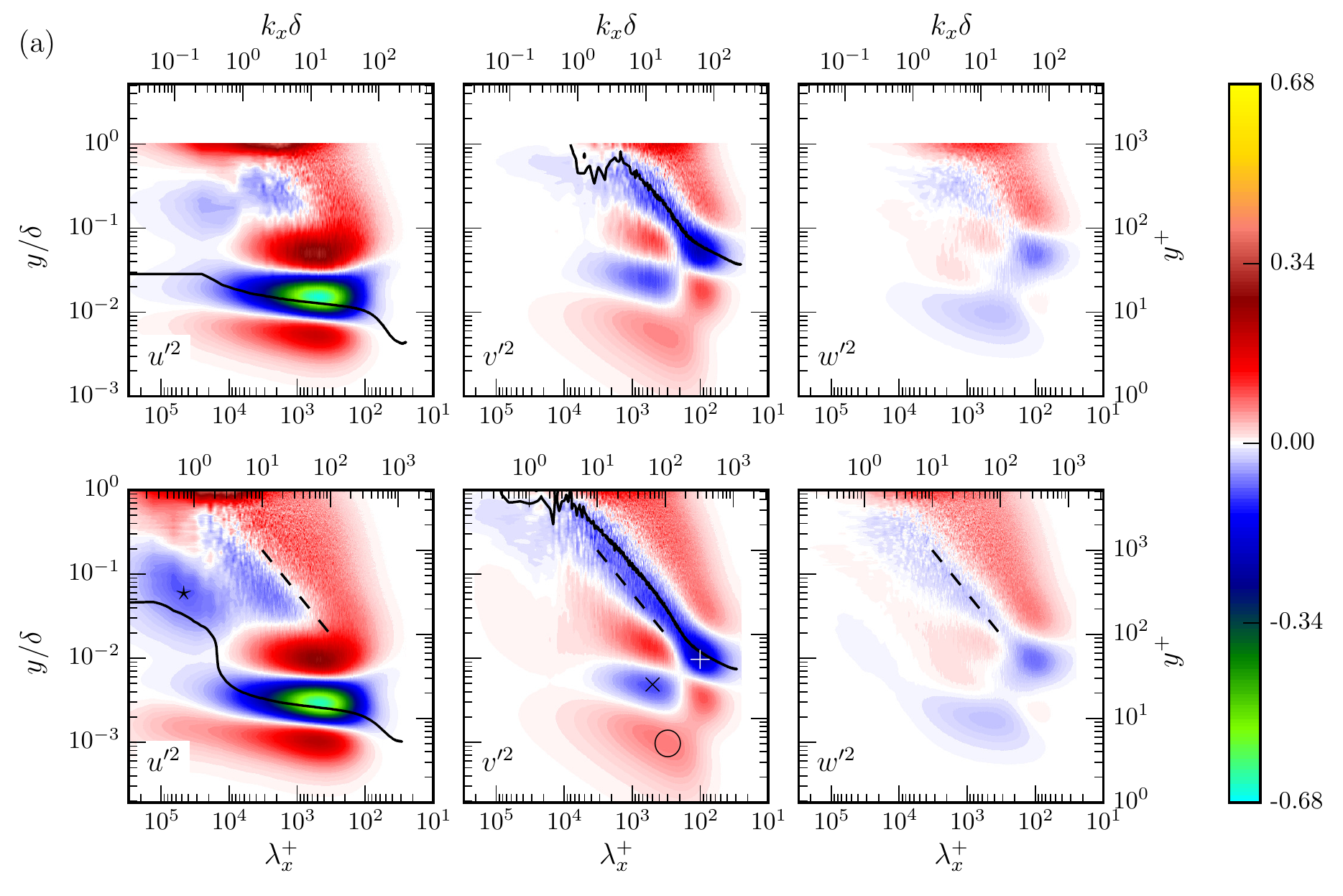}\\
    \includegraphics[width=\textwidth]{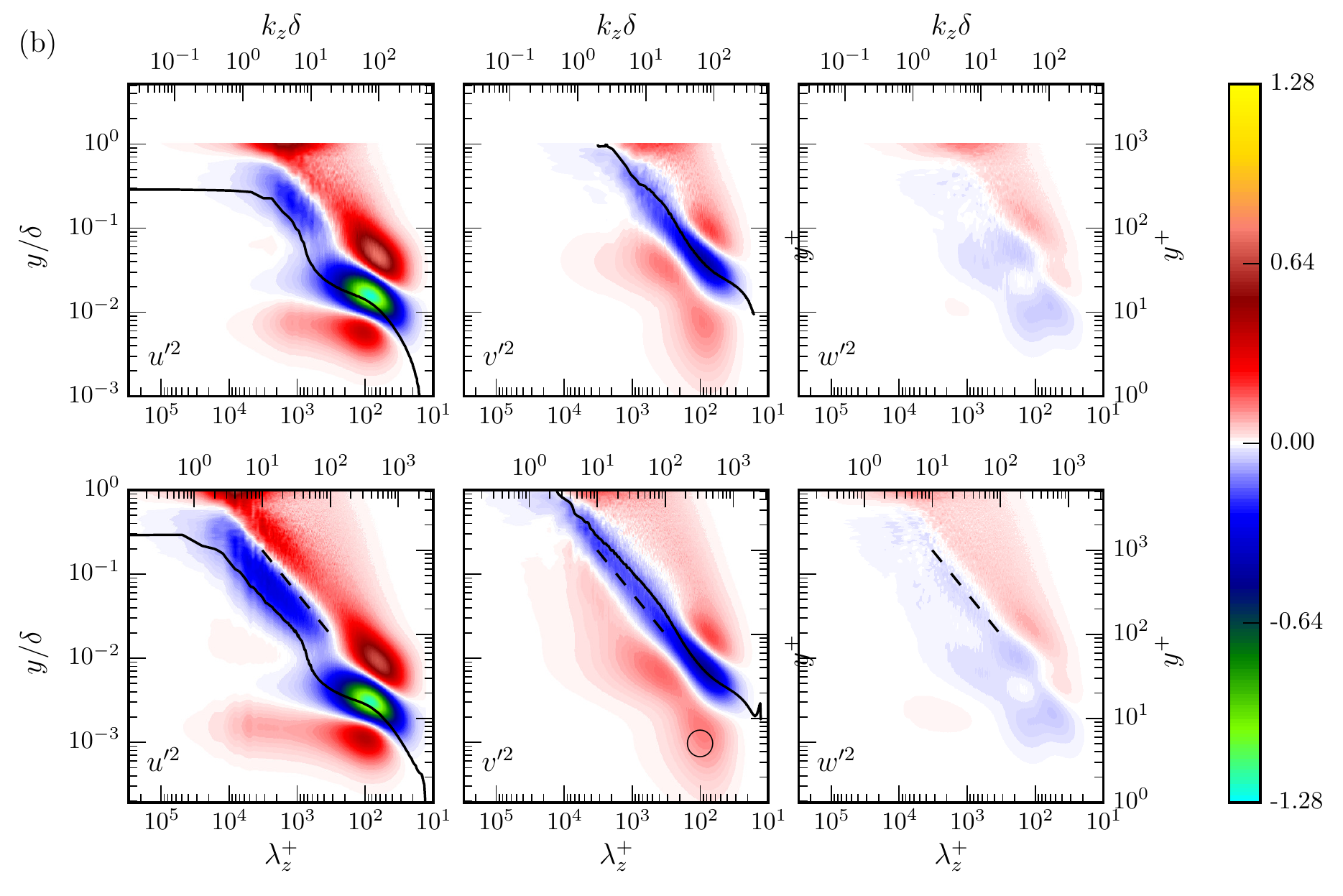}
  \end{center}
  \caption{One-dimensional spectral density of
    $\mathcal{N}_{\alpha\alpha}$. Solid curves are where
    $E_{\alpha\alpha,x}^{\Phi_{\bot}}=0$ or
    $E_{\alpha\alpha,z}^{\Phi_{\bot}}=0$, and dashed lines indicate
    $k_x\sim y^{-1}$ or $k_z\sim y^{-1}$. Symbols ($\times$ and
    $\circ$) mark features discussed in the text.}
  \label{fig:1d_E_N_y}
\end{figure}

\begin{figure}
  \includegraphics[width=\textwidth]{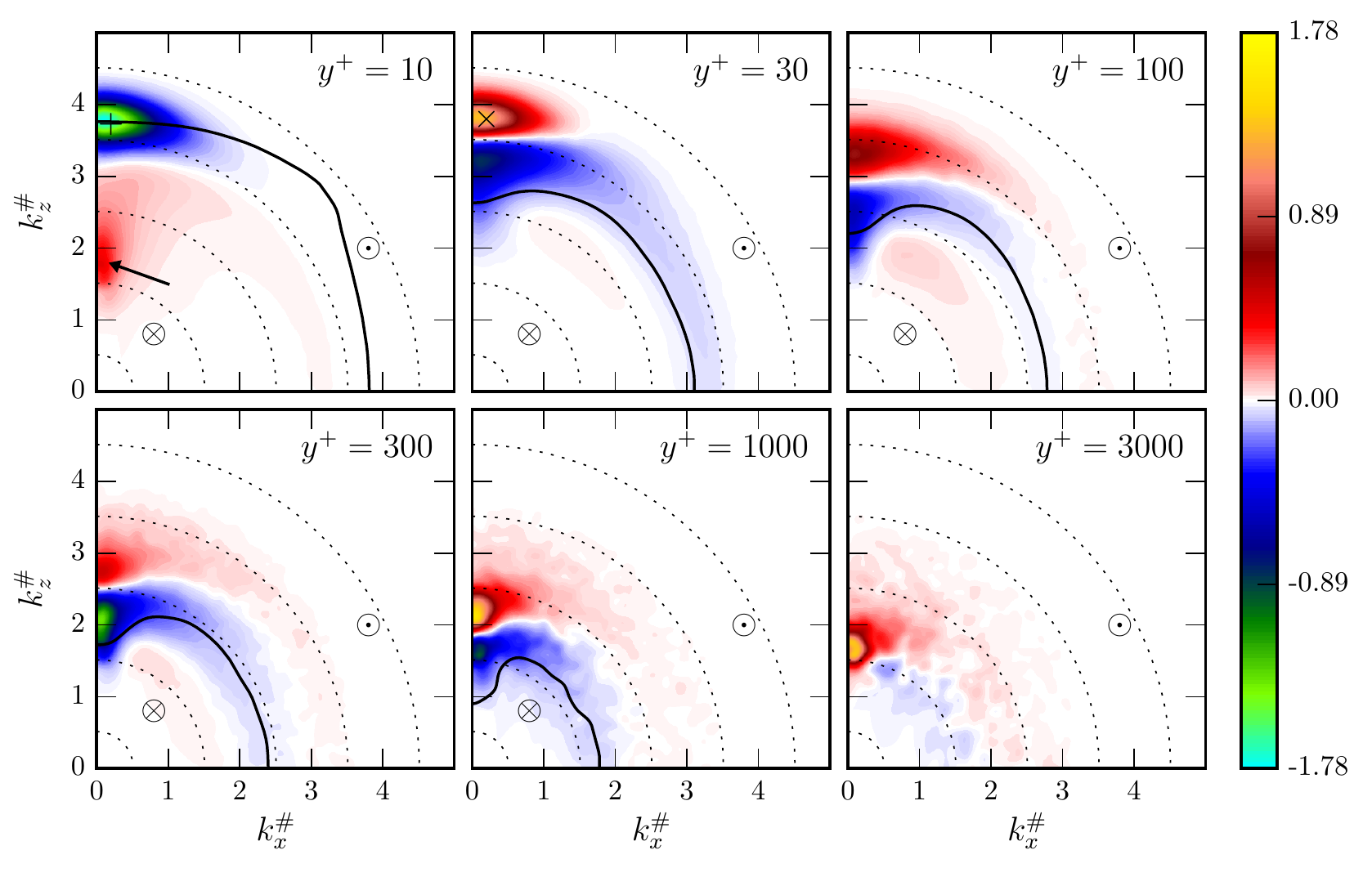}
  \caption{Two-dimensional spectral density of $\mathcal{N}_{11}$ in
    log-polar coordinates, as defined in
    figure~\ref{fig:polar_coord_explain}, from R5200. $\lambda^+=10$
    on the outer-most dotted circle and increases by a factor of 10
    for each dotted circle moving inward. The solid line is
      the $E^{\Phi_\bot}_{11}=0$ contour, dividing the regions where
      transport is toward the wall (marked with $\otimes$) and away
      from the wall (marked with $\odot$). The arrow and symbols ($+$
      and $\times$) mark features discussed in the text.}
  \label{fig:2d_nonlinear_transport_uu}
\end{figure}

\begin{figure}
  \includegraphics[width=\textwidth]{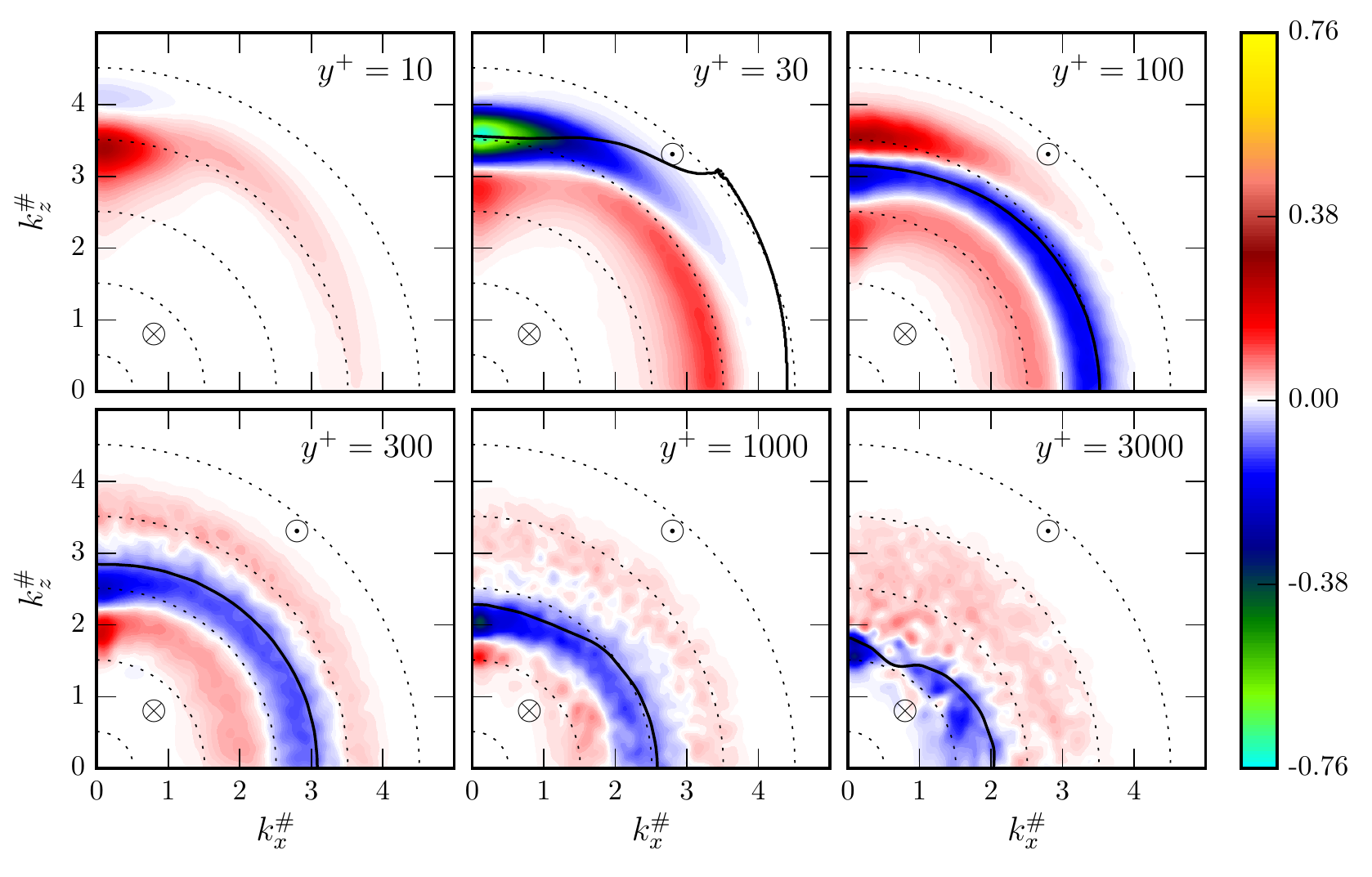}
  \caption{Two-dimensional spectral density of $\mathcal{N}_{22}$ in
    log-polar coordinates, as defined in
    figure~\ref{fig:polar_coord_explain}, from R5200. $\lambda^+=10$
    on the outer-most dotted circle and increases by a factor of 10
    for each dotted circle moving inward. The solid line is the
    $E^{\Phi_\bot}_{22}=0$ contour, dividing the regions where
    transport is toward the wall (marked with $\otimes$) and away from the
    wall (marked with $\odot$).}
  \label{fig:2d_nonlinear_transport_vv}
\end{figure}

\begin{figure}
  \includegraphics[width=\textwidth]{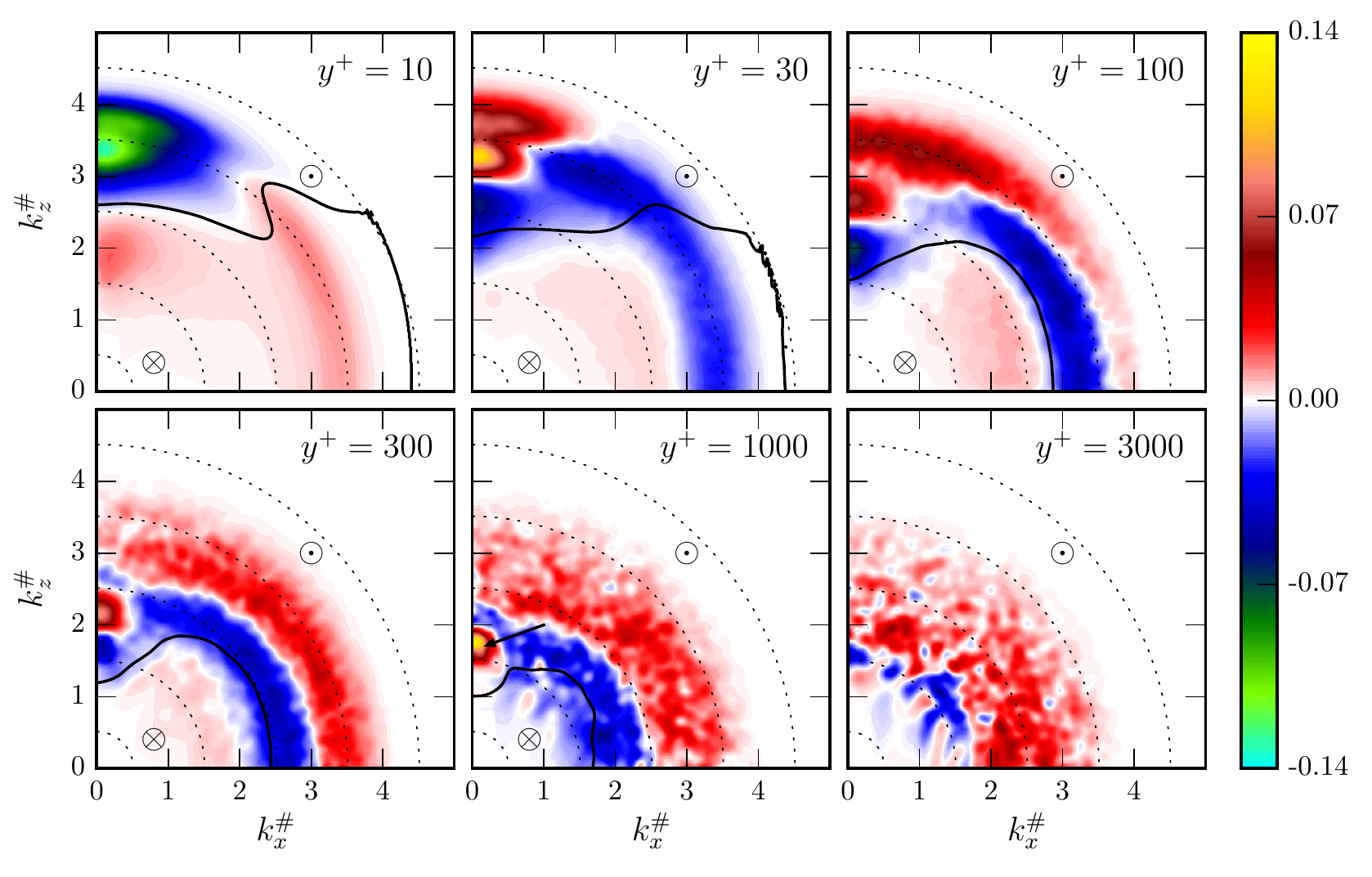}
  \caption{Two-dimensional spectral density of $\mathcal{N}_{33}$ in
    log-polar coordinates, as defined in
    figure~\ref{fig:polar_coord_explain}, from R5200. $\lambda^+=10$
    on the outer-most dotted circle and increases by a factor of 10
    for each dotted circle moving inward. The solid line is the
    $E^{\Phi_\bot}_{33}=0$ contour, dividing the regions where
    transport is toward the wall (marked with $\otimes$) and away from the
    wall (marked with $\odot$). The arrow marks a feature
      discussed in the text.}
  \label{fig:2d_nonlinear_transport_ww}
\end{figure}

\begin{figure}
  \includegraphics[width=\textwidth]{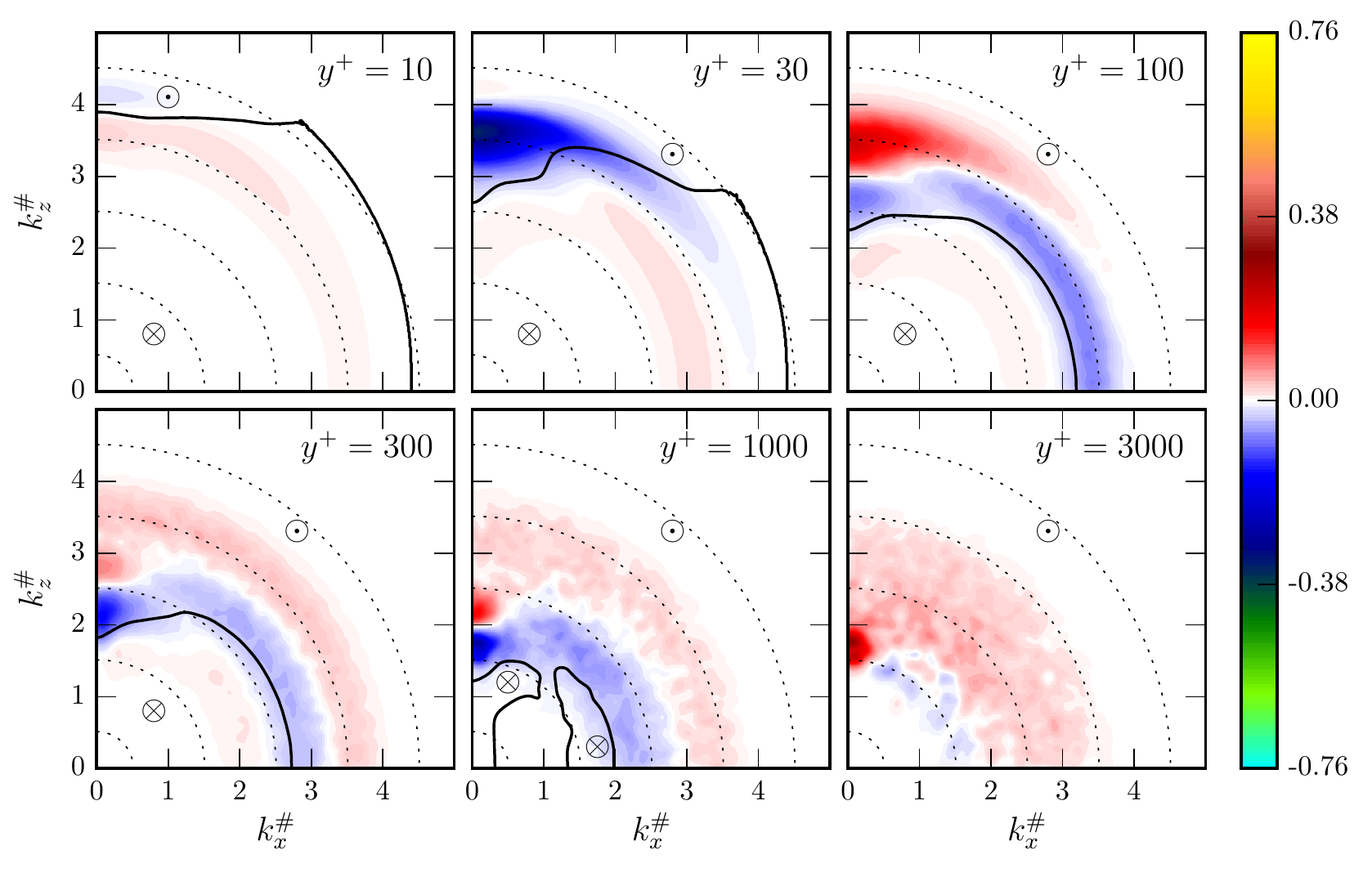}
  \caption{Two-dimensional spectral density of $T_{22}$ in
    log-polar coordinates, as defined in
    figure~\ref{fig:polar_coord_explain}, from R5200. $\lambda^+=10$
    on the outer-most dotted circle and increases by a factor of 10
    for each dotted circle moving inward. The solid line is the
    $E^{\Phi_\bot}_{33}=0$ contour, dividing the regions where
    transport is toward the wall (marked with $\otimes$) and away from the
    wall (marked with $\odot$).}
  \label{fig:2d_turbulent_transport_vv}
\end{figure}

\begin{figure}
  \includegraphics[width=\textwidth]{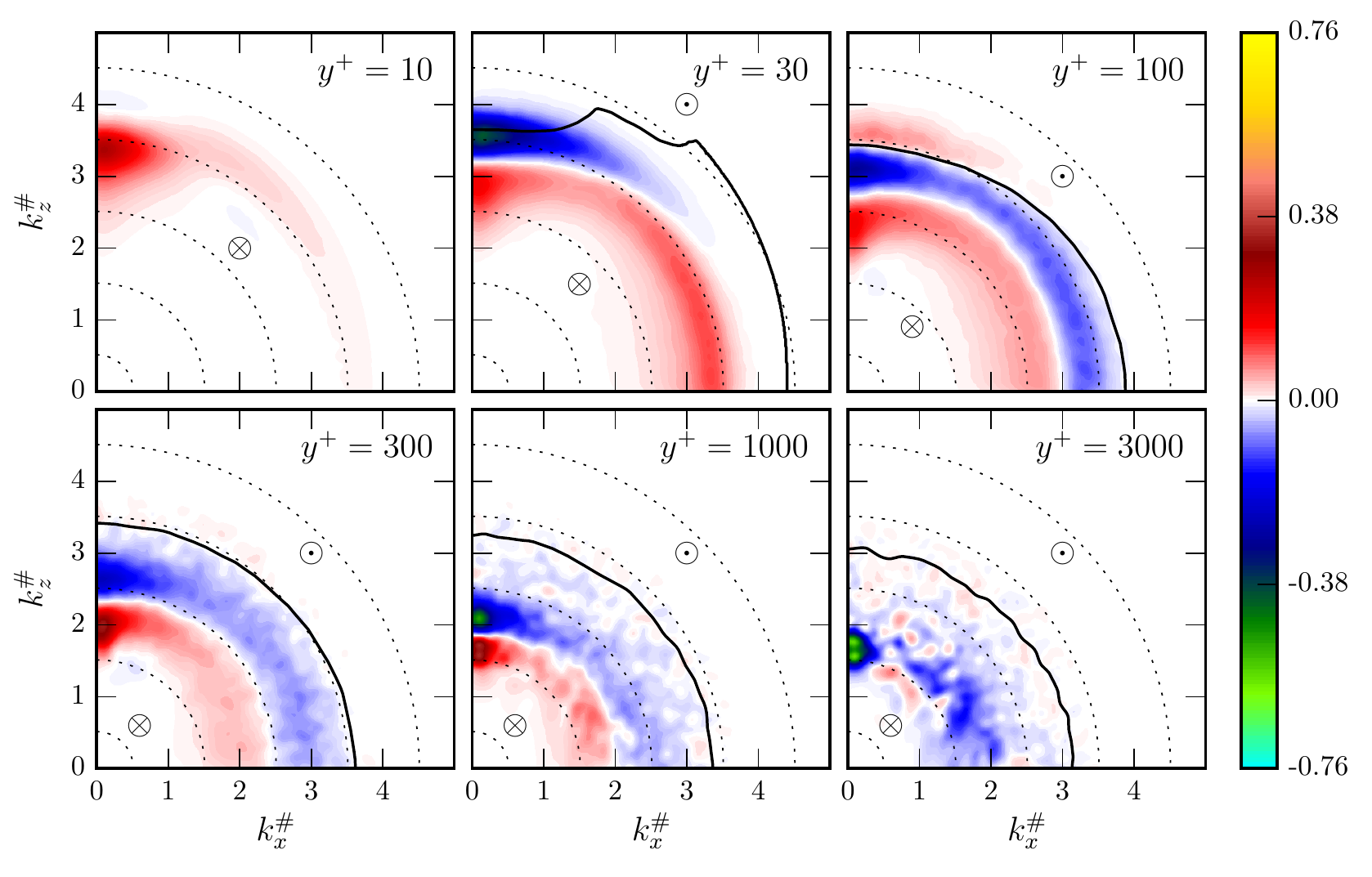}
  \caption{Two-dimensional spectral density of $\Pi^d_{22}$ in
    log-polar coordinates, as defined in
    figure~\ref{fig:polar_coord_explain}, from R5200. $\lambda^+=10$
    on the outer-most dotted circle and increases by a factor of 10
    for each dotted circle moving inward. The solid line is the
    $E^{\Phi_\bot}_{33}=0$ contour, dividing the regions where
    transport is toward the wall (marked with $\otimes$) and away from the
    wall (marked with $\odot$).}
  \label{fig:2d_pressure_transport_vv}
\end{figure}

\begin{figure}
  \centering
  \includegraphics[width=0.9\textwidth]{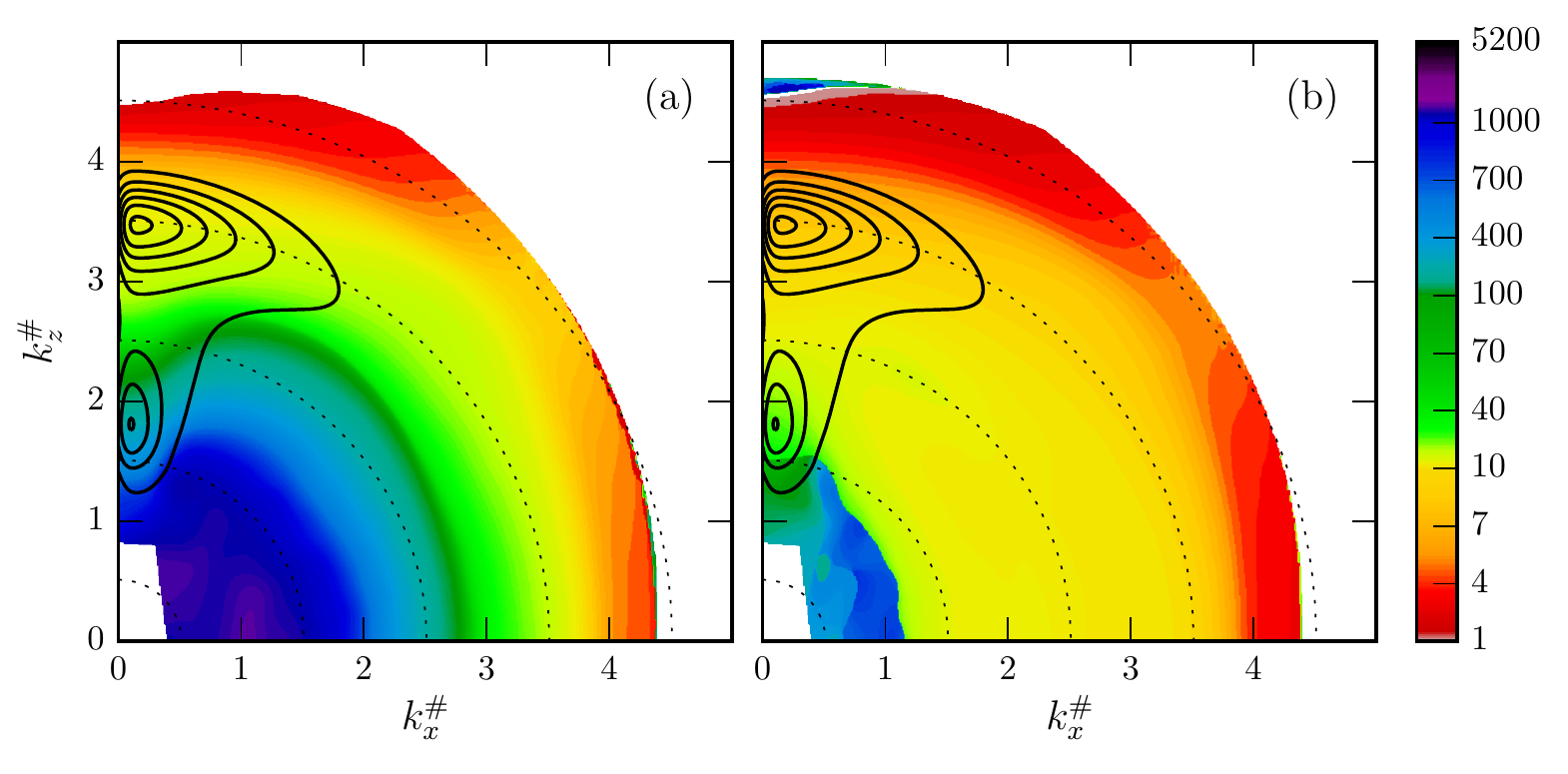} 
  \caption{Non-linear transport flux of \uu in log-polar coordinates,
    as defined in figure~\ref{fig:polar_coord_explain}, from
    R5200. Shown are: (a) $y^+$ at which $E^{\Phi_\bot}_{11}=0$, (b)
    $y^+$ at which $E^{\Phi_\bot}_{11}$ has its first local maximum
    from the wall. Solid curves are the value of
      $k^2E^{\Phi_\bot}_{11} / \left(u_\tau^3
      \log_{10}(k/k_{\textrm{ref}})\right)$ at its first local maximum from
      the wall. The contour values increment by 0.1, starting with
      0.1. $\lambda^+=10$ on the outer-most dotted circle and
    increases by a factor of 10 for each dotted circle moving inward.}
  \label{fig:2d_nonlinear_transport_flux}
\end{figure}

The other wall-normal transport mechanism is nonlinear, resulting from
the action of turbulent fluid motions. As discussed in
\S\ref{sec:method}, the non-linear transport term
$\mathcal{N}_{ij}=T^\bot_{ij}+\Pi_{ij}^d$ is the combination of the
usual turbulent transport and pressure transport terms, which we will
consider here first.  It was shown in \citet{Lee:2015er} that the
difference between production and dissipation is $Re$ dependent for
$y^+ \gtrsim 70$.  As we saw in \S~\ref{subsubsec:wall-normal_linear}
viscous transport terms do not have a significant $Re$ dependence, so
the nonlinear transport term must be $Re$-dependent in the outer
region to balance the production/dissipation miss-match.

The one-dimensional profiles of non-linear transport are shown in
figure~\ref{fig:1d_turb_transport}. As with the other terms examined
here, only modest $Re$ dependencies are observed near the wall,
particularly the slow increase of the near-wall peak of
$y^+\mathcal{N}_{11}$($=y^+T^+_{11}$) with $Re$. However, further from
the wall ($y^+\gtrsim 100$), the turbulent transport of \uu
($\mathcal{N}_{11}$) decreases with increasing $Re$ where negative
$\mathcal{N}_{\alpha\alpha}$ implies a net transport of energy from
this region to other region(s). The local minimum of
$y^+\mathcal{N}^+_{11}$ in the R5200 case occurs at
$y^+\approx300$. Beyond this local minimum, $y^+ \mathcal{N}^+_{11}$
increases logarithmically when $Re_\tau\gtrsim 2000$, as shown in
figure~\ref{fig:1d_turb_transport}, and the slope appears to be
independent of $Re$.  This logarithmic dependence ends at $y/\delta
\approx 0.6$, where $y^+\mathcal{N}^+_{11}$ increases much more
rapidly.

Unlike $y^+ \mathcal{N}^+_{11}$, the $y^+ \mathcal{N}^+_{22}$ and $y^+
\mathcal{N}^+_{33}$ profiles do not exhibit significant $Re$
dependencies. The transport of \vv by $\mathcal{N}_{22}$ is primarily
from the region where $y^+\gtrsim 15$ and $y/\delta \lesssim 0.6$ to
the near-wall region and region around the channel center line.  The
transport by $\mathcal{N}_{33}$ is weaker, with energy transported
from regions around $y^+=10$ and $y^+=50$ to the outer region
($y/\delta \gtrsim0.3$). Note that both $T^\bot_{22}$ and
$\mathcal{N}_{33}$ are nearly zero in the region from $y^+\approx 200$
to $y/\delta\approx 0.3$. Hence \vv energy is only deposited in this
region by pressure transport and \ww energy is transported through
this region without impacting the local balance of \ww.

The rather complex structures of $\mathcal{N}_{11}$ and
$\mathcal{N}_{22}$, raises the question of the direction of the net
transport of \uu and \vv. To determine this, the net flux of energy
$\Phi_{\alpha\alpha}$ due to $\mathcal{N}_{\alpha\alpha}$ was
computed, where
\begin{equation}
\Phi_{\alpha\alpha}(y) = \int_0^{y} \mathcal{N}_{\alpha\alpha} (Y)
\intd Y = - \langle u_\alpha'u_\alpha'v' \rangle (y) + \langle u_2'p' \delta_{2\alpha}\rangle (y).
\label{eq:turb_transport_flux}
\end{equation}
With this definition, positive $\Phi_{\alpha\alpha}$ signifies flux
toward the wall.  The profiles of $\Phi_{\alpha\alpha}$ are shown in
figure~\ref{fig:1d_turb_transport}. It is obvious that
$\Phi_{\alpha\alpha}$ must be zero at the wall and the center of the
channel because of the boundary conditions and statistical
symmetry. In all cases, $\Phi_{11}$ is zero at $y^+ \approx 15$, which
is where $y^+ \mathcal{N}_{11}^+$ has a local minimum. Thus, in all
cases, the net flux of \uu is toward the wall for $y^+\lesssim 15$ and
away from the wall at larger $y^+$. In all cases, $\Phi_{22}$ is zero
at $y/\delta \approx 0.2$, indicating that \vv energy is transported
toward the wall in this region. On the other hand, \ww energy is
transported away from the wall throughout the channel.

Note that there is a local maximum in $\Phi_{11}$ at $y^+\approx 100$,
which is obviously just the point where $\mathcal{N}_{11}$ changes
sign. But, the value of $\Phi_{11}$ at the local maximum is getting
larger (less negative) with increasing $Re$. The reason is the growth
with $Re$ of the near-wall peak in $\mathcal{N}_{11}$. If this trend
continues, at high enough $Re$, this local maximum will exceed zero,
introducing another region in which the net flux is toward the
wall. This complex structure of the turbulent transport is an
aggregation of transport occurring at all scales. A better
understanding requires a spectral analysis.

In \S\ref{sec:method}, we showed that the turbulent convection of the
two-point correlation can be decomposed into terms representing
transport in the wall-normal direction and scale transfer in
wall-parallel directions (see eq~(\ref{eq:turb_decomposition})).  The
two-dimensional spectra of these terms are just their Fourier
transforms, and they inherit the properties of the decomposition, such
as eqs~(\ref{eq:turb_vertical_condition},
\ref{eq:turb_horizontal_condition}). The turbulent convection spectrum
$E_{ij}^T$, which is the Fourier transform of $R_{ij}^T$, is thus
decomposed, as
\begin{equation}
E_{ij}^T (k_x,y,k_z) = E_{ij}^{T^{\|}} (k_x,y,k_z) + E_{ij}^{T^{\bot}}
(k_x,y,k_z)
\label{eq:turb_decomp_energy}
\end{equation}
where $E_{ij}^{T^{\|}}$ and $E_{ij}^{T^{\bot}}$ are the Fourier
transforms of $R_{ij}^{T^{\|}}$ and $R_{ij}^{T^{\bot}}$,
respectively. Further,
$E_{ij}^{\mathcal{N}}=E_{ij}^{T^\bot}+E_{ij}^{\Pi^d}$ is the Fourier
transform of $R_{ij}^{\mathcal{N}}$. The Fourier transforms
satisfy
\begin{subequations}
\begin{equation}
\int_0^\delta E_{ij}^{\mathcal{N}} (k_x,y,k_z) \intd y = 0, \quad \forall (k_x, k_z)
\label{eq:nonlinear_wall_normal_int_in_y}
\end{equation}
and
\begin{equation}
\iint_0^\infty E_{ij}^{T^{\|}} (k_x,y,k_z) \intd k_x \intd k_z  = 0, \quad \forall y .
\label{eq:turb_transport_int_in_xz}
\end{equation}
\label{eq:turb_transport_decomp_property}
\end{subequations}
Hence, we can interpret $E_{ij}^{\mathcal{N}}$ as non-linear transport
in the wall-normal direction at each scale defined by the wavenumbers
$(k_x,k_z)$, and interpret $E_{ij}^{T^\|}$ as the transfer in scale in
the wall-parallel directions at each wall-normal location
$y$. Further, the property (\ref{eq:turb_transport_int_in_xz}) 
implies that
\begin{equation}
\iint_0^\infty E_{ij}^{\mathcal{N}} (k_x,y,k_z) \intd k_x \intd k_z  = \mathcal{N}_{ij} (y)=T_{ij}+\Pi_{ij}^d, \quad \forall y,
\label{eq:turb_transport_homo_int_in_xz}
\end{equation}
so $E_{ij}^{\mathcal{N}}$ is essentially the spectrum of
$\mathcal{N}_{ij}$.

The one-dimensional non-linear transport spectra,
$E_{ij,x}^{\mathcal{N}}$ and $E_{ij,z}^{\mathcal{N}}$, are shown in
figure~\ref{fig:1d_E_N_y}. Note that the property
(\ref{eq:nonlinear_wall_normal_int_in_y}) is inherited by
$E_{ij,x}^{\mathcal{N}}$ and $E_{ij,z}^{\mathcal{N}}$:
\begin{equation}
\begin{split}
\int_0^\delta E_{ij,x}^{\mathcal{N}} (k_x,y) \intd y = 0, &\quad \forall k_x \\ 
\int_0^\delta E_{ij,z}^{\mathcal{N}} (y,k_z) \intd y = 0, &\quad \forall k_z
\end{split}
\label{eq:turb_transport_int_in_y_1d}
\end{equation}
These spectra provide a view of the complex interactions of the
turbulence at different distances from the wall. Focus first on the
region near the wall ($y^+\lesssim 200$). In this region,
$\mathcal{N}_{11}$ is dominant (see
figure~\ref{fig:1d_turb_transport}), and this is reflected in the
spectra as well. For both Reynolds numbers shown in
figure~\ref{fig:1d_E_N_y}, the near-wall $\mathcal{N}_{11}$ spectra
peak at wavenumbers $\lambda_x^+ \approx 500$ and $\lambda_z^+\approx
100$, which, not surprisingly, is near the peak of the two-dimensional
kinetic energy spectrum near the wall. In this wavenumber range,
$\mathcal{N}_{11}$ transports energy from around $y^+\approx 20$
toward the wall ($y^+\lesssim10$) and away from the wall
($y^+\gtrsim40$). Note that in the streamwise spectra
(figure~\ref{fig:1d_E_N_y}), these vertical locations are
approximately independent of the wavenumber, but that in the spanwise
spectra (figure~\ref{fig:1d_E_N_y}) the contours are inclined,
indicating that structures with larger spanwise scales act somewhat
further from the wall.

Another feature of $\mathcal{N}_{11}$ near the wall is that at large
scale (small wavenumber), energy is deposited near the wall from
regions far from the wall. This is particularly apparent in the
spanwise spectra in the R5200 case (figure~\ref{fig:1d_E_N_y}), where
at the lowest wavenumbers, energy is transported to the wall from as
far away as $y^+\approx 1000$. To identify the direction of the
turbulent transport at each scale, the turbulent transport flux
spectra $E_{ij}^{\Phi^{\mathcal{N}}}$ were computed. The
one-dimensional spectra are given by:
\begin{equation}
E_{ij,x}^{\Phi^{\mathcal{N}}}(k_x,y) = \int_0^y E_{ij,x}^{\mathcal{N}}
(k_x,Y) \intd Y, \quad E_{ij,z}^{\Phi^{\mathcal{N}}}(y,k_z) = \int_0^y
E_{ij,z}^{\mathcal{N}} (Y,k_z) \intd Y
\label{eq:turbulent_transport_flux_in_y}
\end{equation}
In figures~\ref{fig:1d_E_N_y}, $E_{ij,x}^{\Phi^{\mathcal{N}}} = 0$ and
$E_{ij,z}^{\Phi^{\mathcal{N}}} = 0$ contours are plotted as black
lines.  In these figures, the energy flux is toward the wall in the
region below the solid line and away from the wall in the region
above. At the largest Reynolds number (case R5200), \uu energy is
transported to the near-wall region from $y^+\lesssim 200$
($y/\delta\lesssim 0.04$) at the smallest $k_x$, and from
$y/\delta\lesssim 0.3$ at the smallest $k_z$.

The streamwise spectra of non-linear vertical transport of \vv has a
much more complex structure near the wall, with transport toward the
wall from two distinct regions, one centered at $y^+\approx 25$,
$\lambda_x^+\approx 500$ (marked with $\times$) and the other centered
around $y^+\approx 50$, $\lambda_x^+\approx 100$ (marked with
$+$). This latter corresponds in $\lambda_x$ and $y$ with the
near-wall peak in the pressure-strain source of \vv energy (see
figure~\ref{fig:1d_E_PI_s}). There is also a high wavenumber donor
region of \vv energy due to the pressure-strain sink closer to the
wall ($\lambda_x^+\approx 300$, $y^+\approx 5$, $\lambda_z^+\approx
100$, marked with $\bigcirc$) caused by the the splat effect
\citep{Perot:1995vu,Mansour:1988vz}. This is apparently fed by
wall-normal transport from above, as evidenced by the significant
donor regions in the one-dimensional spectra. The splat-effect
pressure-strain is also a significant source of \ww energy in this
region, which is presumably why the near-wall turbulent flux of \ww
energy is only away from the wall. This gives $E^{T^\bot}_{33}$ a
simpler near-wall structure than the other two components.  The
near-wall spanwise spectra are generally simpler, with transport both
toward and away from the wall across a broad range of $y$. Spectra of
\ww transport are also notable for their low magnitude relative to \uu
and \vv.

Near-wall turbulent transport at high wavenumbers, as discussed above,
presumably results from the small-scale self-sustaining near-wall
dynamics described by
\citet{Hamilton:1995vu,Jeong:1997uj,Jimenez:1999wf,Schoppa:2002dq},
which is expected to be Reynolds-number independent when scaled in
wall units. Consistent with this expectation, the high-wavenumber
near-wall structure of the one-dimensional spectra in $\lambda^+_x$
and $\lambda^+_z$ is the same in the R1000 and R5200 cases as shown in
figure~\ref{fig:1d_E_N_y}. At large-scales (low wavenumbers),
turbulent transport of energy to the near-wall region is consistent
with the modulation of the near-wall dynamics by large-scale
outer-flow structures as proposed by
\citet{Hutchins:2007kd,Marusic:2010hy}.

Farther away from the wall ($y^+\gtrsim 200$, say), the structure of
the $\mathcal{N}$ spectra appears much simpler. In all three velocity
components, and in both the $x$ and $z$ directions, there is a
self-similar structure in which energy is transported from a region
spanning approximately a half decade in $y^+$, to a region directly
above it. The $y$-location at which this occurs scales as the inverse
of the wavenumber, as can be seen by the dashed lines in
figure~\ref{fig:1d_E_N_y}. This self-similar structure extends out to
$y/\delta\approx 0.6$, and is consistent with the scaling expected in
the log region. Note that this structure is much clearer in the
spanwise ($k_z$) spectra, where it is evident for both the R5200 and
R1000 cases. In the streamwise ($k_x$) spectra, the self-similar
structure is present in the R5200 case, but it is more obscure in
R1000. This is consistent with the observation that there is a
distinct separation in scales between the near-wall and outer region
turbulence evident in the one-dimensional $z$ energy spectrum for both
the R5200 and R1000 cases, but only in the R5200 case for the
$x$-spectrum. \citep{Lee:2015er}

Another complexity in the high-Reynolds-number streamwise \uu spectrum
(figure~\ref{fig:1d_E_N_y}) is the low wavenumber (large scale) region
between $y^+\approx 100$ and $y^+\approx 1000$ ($y/\delta\approx 0.2$,
marked with $\star$) where $E^{T^{\mathcal{N}}}_{11}$ is negative,
with the energy transported primarily toward the center of the
channel. This feature is absent in the spanwise and \ww spectra, and
is very weak in the \vv spectra. This indicates that the turbulent
transport processes do not strictly adhere to the self-similar scaling
of the log region.

To better understand the complex non-linear transport spectra, and
particularly the differences between the streamwise and spanwise
spectra, consider the two-dimensional spectra shown in
figures~\ref{fig:2d_nonlinear_transport_uu}-\ref{fig:2d_nonlinear_transport_ww}.
Also plotted in these spectra is the $E^{\Phi_\bot}_{\alpha\alpha}=0$
contour, separating the spectral region where transport is toward the
wall (marked with $\otimes$) and away from the wall (marked with
$\odot$).  At all $y$ locations, the nonlinear-wall-normal \uu
transport spectrum (figure~\ref{fig:2d_nonlinear_transport_uu}) is
dominant along the $k_z^\#$ axis, as with the energy and production
spectra studied earlier. As expected, near the wall ($y^+=10$) there
is a strong donor region around $\lambda^+=100$ corresponding to the
near-wall streaks (marked with $+$), and this energy is transported
both toward the wall and away (there is a corresponding recipient
region at $y^+=30$, marked with $\times$). There is also a recipient
region along the $k_z^\#$ axis at $y^+=10$, spanning from about
$\lambda^+=1000$ to 10000 ($\lambda/\delta\approx 2$, marked with an
arrow), where flux is entirely from above. This corresponds with the
low wavenumber energetic region in the \uu energy spectrum
(figure~\ref{fig:2d_uu}), confirming wall normal transport from above
as a source of energy in this spectral region, as speculated in
\S\ref{subsec:energy_spectra}. A bit further from the wall ($y^+=30$),
the picture changes, with the short wavelength region receiving energy
(from below) and the long wavelength donating, with the flux being
away from the wall except for the very largest wavelengths
($\lambda^+\gtrsim 1000$).  The source of the long wavelength
transport into the near-wall region appears to be the range
$30\lesssim y^+\lesssim1000$. Over this range of $y$, the relevant
spectral region includes a donor with flux toward the wall.
Consistent with the one-dimensional spectra, over a range from
$y^+\approx 100$ to $y^+\approx 1000$, the transport spectra appear to
have a self-similar structure, with wavelengths approximately
proportional to $y$. In this range, there is a short wavelength
recipient region and longer wavelength donor region, with the dominant
transport being away from the wall. Also notice that the for
$30\lesssim y^+\lesssim 1000$, the region where flux is away from the
wall extends to much lower wavelengths in the streamwise elongated
modes, so that only the very widest streamwise elongated fluctuations
at any height are responsible for transport toward the wall into the
strong recipient region marked with an arrow at $y^+=10$.  Far enough
from the wall ($y^+=3000$ or $y/\delta=0.6$), the spectrum is
dominated by the recipient region, with energy transported from below
and very weak donor regions. Overall then, the dominant transport of
\uu is in streamwise elongated modes and is primarily directed away
from the wall and local in $y$. The exception is non-local in $y$
transport in the largest wavelength modes to very close to the wall.

In the spectra for $100\lesssim y^+\lesssim1000$, there is also weak
transport over a range of wavenumbers with orientations far from the
streamwise elongation of the dominant spectral features. For example,
at $y^+=100$, there is a recipient region in a band between
$\lambda^+\approx1000$ and 10000, a donor band between
$\lambda^+\approx100$ and 1000, and another recipient band with
$\lambda^+<100$. This indicates that in addition to the dominant
transport in streamwise elongated modes as discussed above, turbulence
with a more isotropic distribution of scales is responsible for local
in $y$ transport away from the wall, and to a lesser extent, toward
the wall.

The \vv and \ww nonlinear transport spectra (figures
\ref{fig:2d_nonlinear_transport_vv} and
\ref{fig:2d_nonlinear_transport_ww}, respectively) also have a
structure in which the streamwise elongated modes behave differently
than the more isotropically distributed modes, but the
streamwise-elongated transport is not dominant in these cases. Away
from the wall ($y^+\geq100$) in \vv, the isotropically distributed
modes have transport split more-or-less evenly between being directed
toward and away from the wall. These scale-isotropic modes also
deposit \vv energy near the wall ($y^+\leq15$), which does not occur
in \uu. Also far from the wall, the magnitude of the \vv transport
spectrum is stronger in the streamwise elongated modes and is directed
toward the wall, unlike the preferentially outward transport of the
streamwise elongated \uu modes. Further, the dominant wavelengths are
longer than for the other orientations, by a factor ranging from about
2 (further from the wall) to 3 (closer). The \ww transport spectra
(figure~\ref{fig:2d_nonlinear_transport_ww}) are much weaker than
either \vv or \uu, resulting in the computed spectra being much
noisier, but they have a structure that is qualitatively similar to
the \uu spectra. One apparent difference is that the isotropically
distributed part of the \ww spectrum appears more prominent, but this
is primarily due to the fact that the color scale in figure
~\ref{fig:2d_nonlinear_transport_ww} has a much smaller range (a
factor of about 13); the actual magnitudes are comparable. The scales
are different because the \ww streamwise elongated modes (along the
$k_z^\#$ axis) are much weaker than those for \uu, but their structure
is more complex.  Notice, in particular the concentrated recipient
region, representing transport from below, along the $k_z^\#$ axis at
$y^+=1000$ occurring around $\lambda^+\approx 8000$
($\lambda/\delta\approx 1.5$, marked with an arrow). This corresponds
with the longest wavelength of the dominant large scale mode in the
\uu energy spectrum discussed in \S\ref{subsec:energy_spectra}. It is
curious, however, that there is no region of high \ww energy at this
point, perhaps because on an absolute scale, this source region is not
particularly strong. This recipient region appears to coincide with a
weak donor region in the two-dimension scale transfer spectrum at
$y^+=1000$ (see \S\ref{subsec:inter-scale_transfer}).

The \vv nonlinear transport spectrum includes contributions from the
turbulent transport $T_{22}$ and the pressure transport
$\Pi^d_{22}$. These are plotted separately in
figures~\ref{fig:2d_turbulent_transport_vv} and
\ref{fig:2d_pressure_transport_vv} respectively. The near-wall
deposition of \vv ($y^+=10$) from above is dominated by the pressure
term. Indeed, throughout the channel, the pressure transport is
dominantly toward the wall while the turbulent transport is primarily
away. Thus the observation of a more-or-less equal split between
transport toward and away from the wall in $\mathcal{N}_{22}$ is a
consequence of the pressure transporting toward and the turbulence
transporting away from the wall. In addition, because the near-wall
transport near the wall is dominated by the pressure, the complex
structures of the near-wall $\mathcal{N}_{22}$ one-dimensional spectra
(figure~\ref{fig:1d_E_N_y}) are a consequence of the pressure.

The discussion above indicates that significant energy (particularly
of \uu) is deposited near the walls in streamwise elongated modes with
spanwise scale of order $\delta$.  To see the two-dimensional spectral
structure of this wall-ward transport, consider the three quantities
depicted in figure~\ref{fig:2d_nonlinear_transport_flux}.  These are:
1) the $y$-location of the boundary of the region of wall-ward
transport, which is the analog of the black lines in
figures~\ref{fig:1d_E_N_y}-\ref{fig:2d_pressure_transport_vv}
(location at which $E^{\Phi^\mathcal{N}}_{11}=0$) as color contours in
figure~\ref{fig:2d_nonlinear_transport_flux}a; 2) $y$-boundary of the
near-wall region of net energy deposition by transport (location of
the first maximum of $E^{\Phi^\mathcal{N}}_{11}$) as color contours in
figure~\ref{fig:2d_nonlinear_transport_flux}b; and 3) net rate of
energy deposition in this region (value of $E^{\Phi^\mathcal{N}}_{11}$
at the first local maximum) as line contours in both panels of
figure~\ref{fig:2d_nonlinear_transport_flux}. Notice that the contours
of deposition rate correspond closely to the structure of the
two-dimensional spectrum of \uu at $y^+=15$
(figure~\ref{fig:2d_E_D_uu}). Further, the dominant region of large
deposition rate, which extends out from the $k_z^\#$ axis at around
$\lambda^+\approx100$, occurs where the flux to the wall is in most
cases from $y^+\lesssim10$ with no case more than $y^+\lesssim 100$,
and this energy is deposited below $y^+\approx 10$. This is all
representative of the near-wall dynamics.

Along the $k_z^\#$ axis, there is a region of significant energy
deposition with wavelengths ranging from $\lambda^+\approx1000$ to
10000. In this low-wavenumber region, energy is transported from
farther from the wall, ranging from $y^+\approx100$ to 700, with
larger scale modes receiving energy from larger $y^+$. Further, in
this spectral region, the energy is deposited near the wall in a layer
characterised by $y^+<10$ at smaller wavelengths to $y^+<70$, at
larger wavelengths.  These results confirm that the near-wall energy
in these large scales is driven by transport from much farther from
the wall, consistent with suggestions by
\citet{Hutchins:2007kd,Marusic:2010hy}.

\subsection{Inter-scale energy transfer}
\label{subsec:inter-scale_transfer}
\begin{figure}
  \begin{center}
    \includegraphics[width=\textwidth]{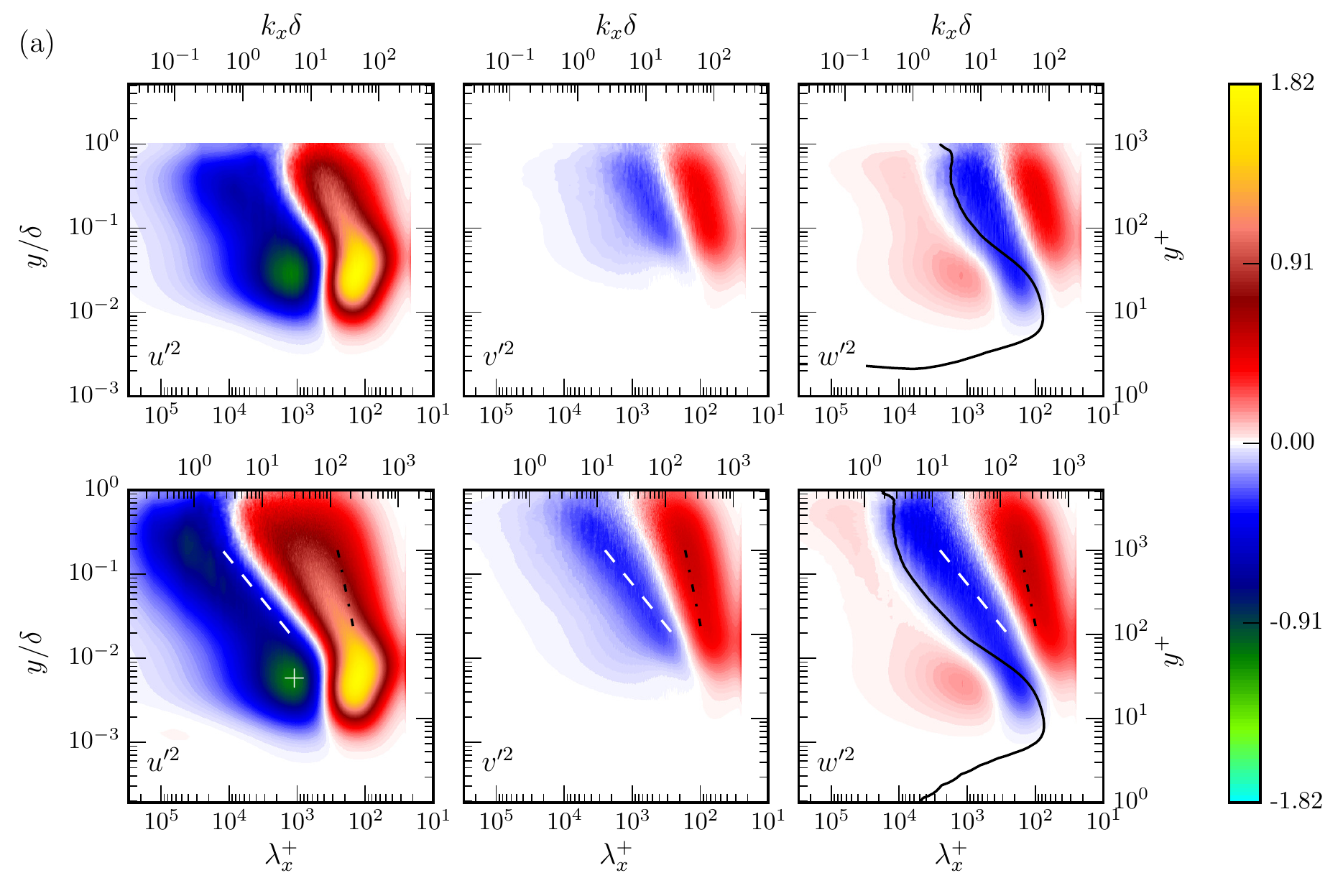}\\
    \includegraphics[width=\textwidth]{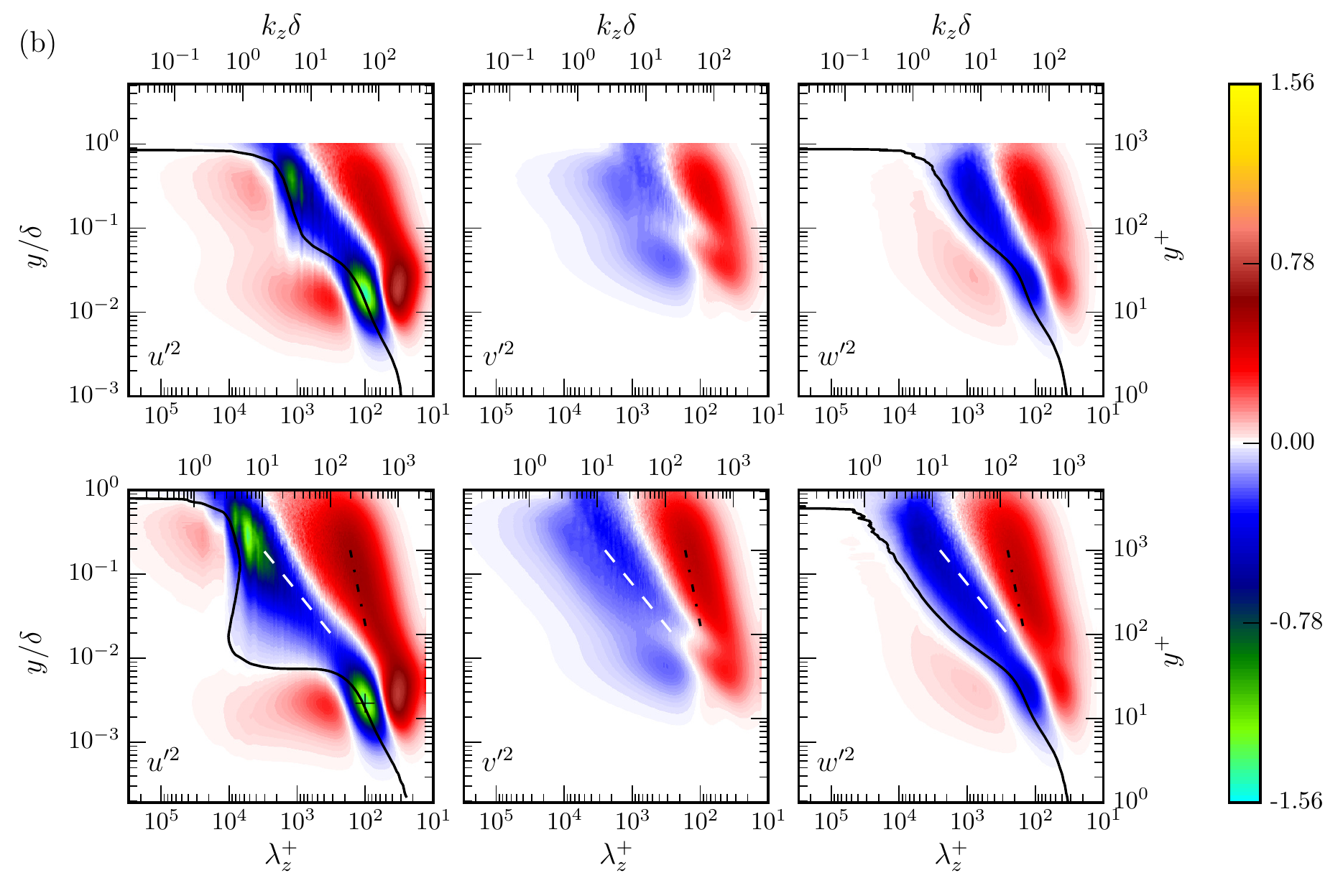}
  \end{center}
  \caption{One-dimensional spectral density of
    $T^{\|}_{\alpha\alpha}$. Solid curves are where
    $E_{\alpha\alpha,x}^{\Phi^{\|}}=0$ or
    $E_{\alpha\alpha,z}^{\Phi^{\|}}=0$. Dashed white lines are at
    $k_xy=0.5$ and $k_zy=2$ for \uu, and $ky=2.5$ for \vv and \ww,
    which correspond to the lines in figure~\ref{fig:1d_E_P_uu}
    (production of \uu) and figure~\ref{fig:1d_E_PI_s} (pressure
    strain for \vv and \ww).
    Dash-dot black lines are at
    $k_xy^{1/4}\delta^{3/4}=85$, 130 and 130 (streamwise spectra) and
    $k_zy^{1/4}\delta^{3/4}=130$, 130 and 85 (spanwise spectra) for
    \uu, \vv and \ww, respectively, which correspond to the lines in
    figure~\ref{fig:1d_E_E} (dissipation).
  }
  \label{fig:1d_E_T_xz}
\end{figure}

\begin{figure}
  \begin{center} 
    \includegraphics[width=\textwidth]{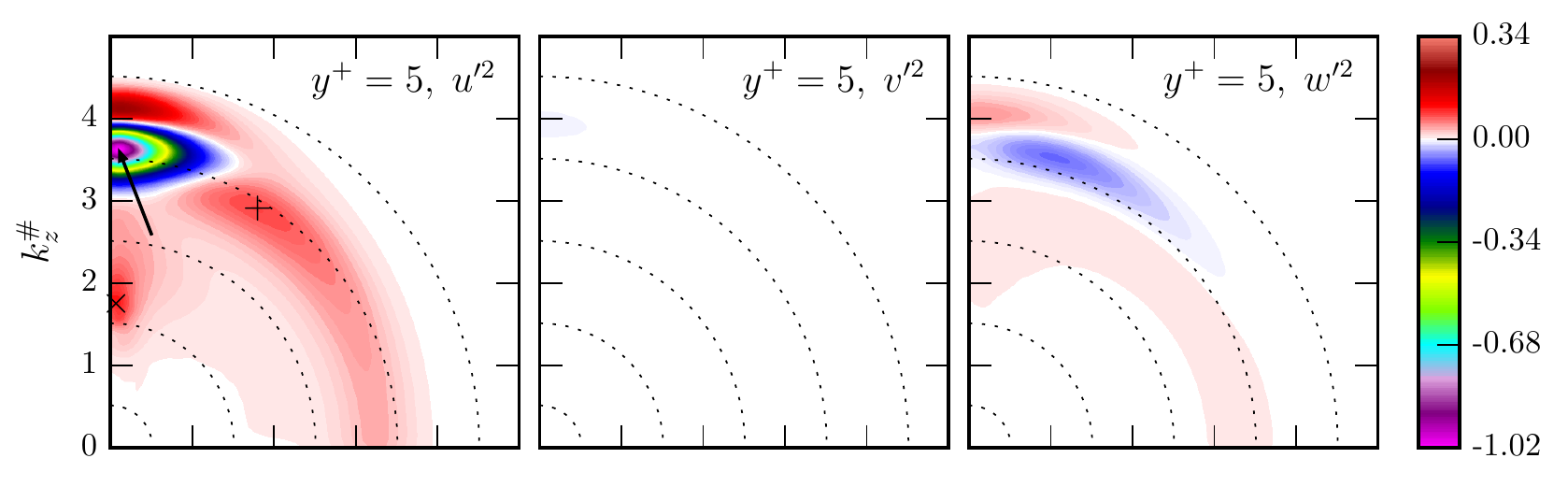} \\[-0.9em]
    \includegraphics[width=\textwidth]{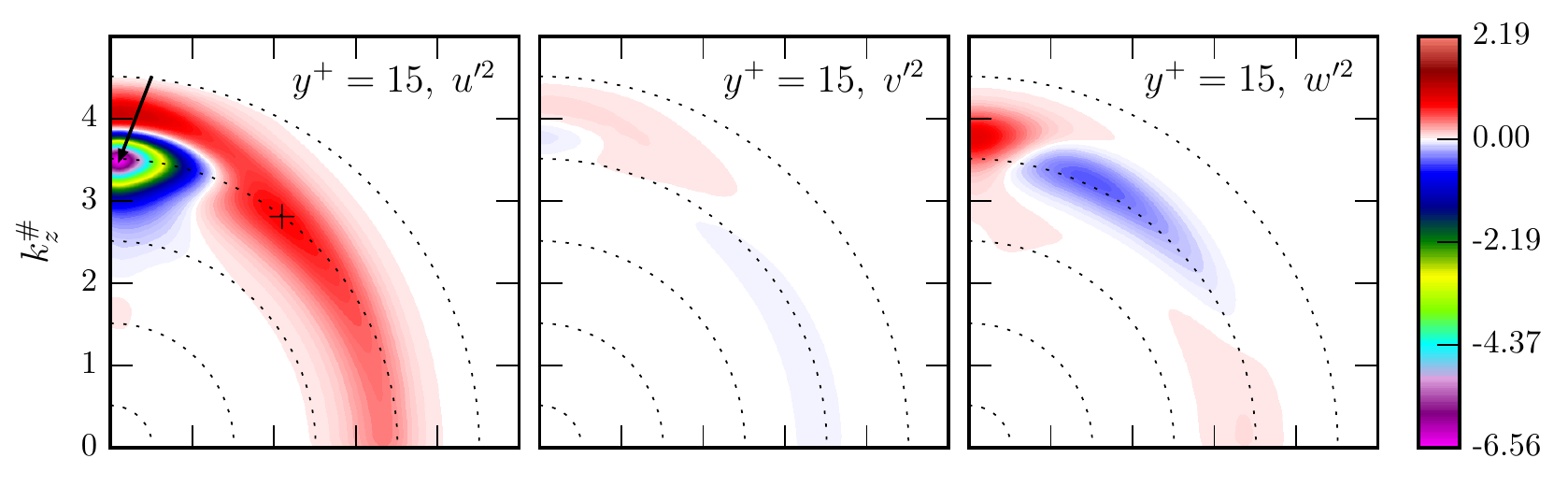} \\[-0.9em]
    \includegraphics[width=\textwidth]{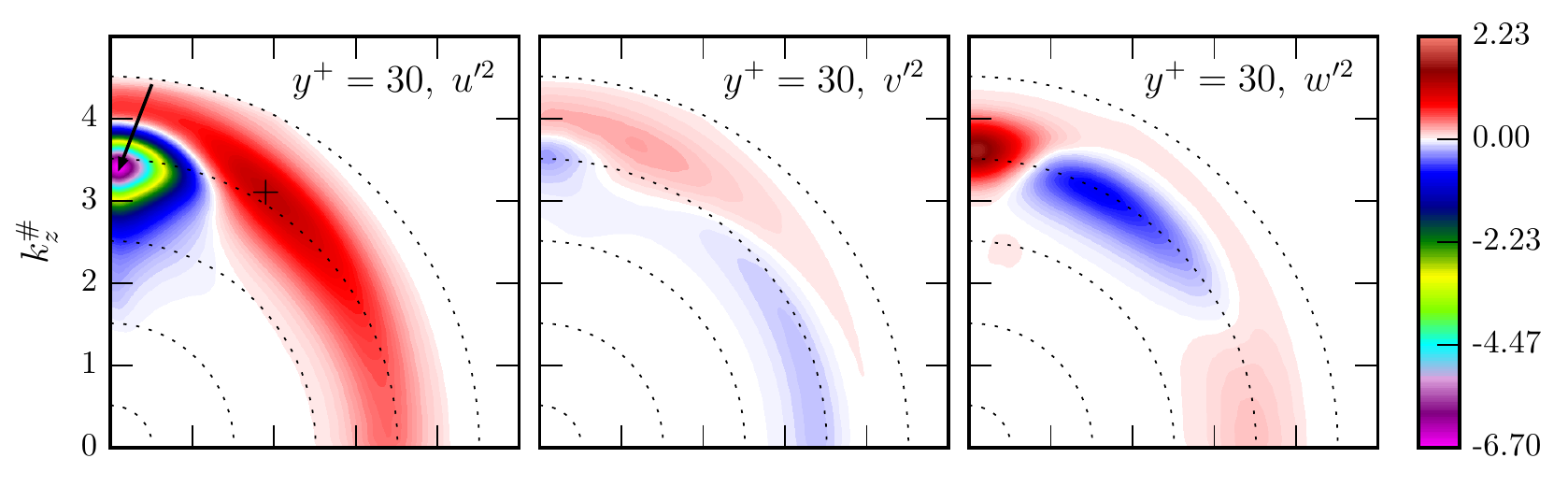} \\[-0.9em]
    \includegraphics[width=\textwidth]{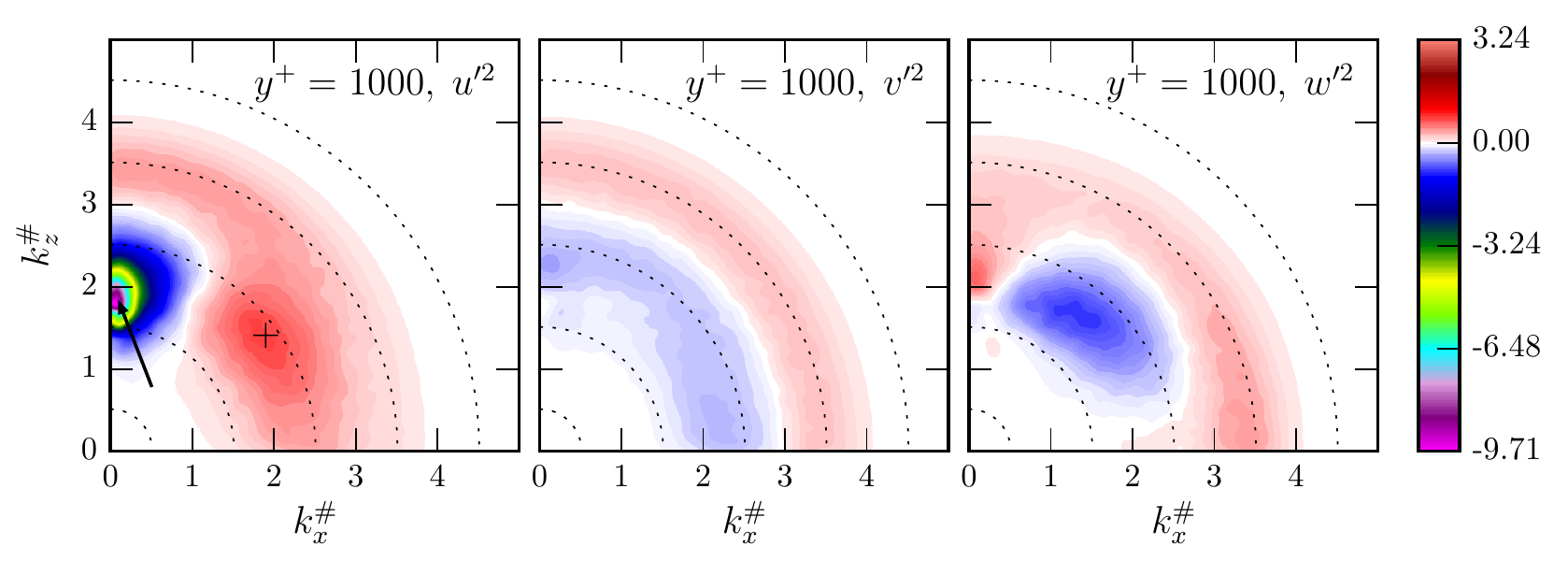}
  \end{center}
  \caption{Two-dimensional spectral density of $T^{\|}_{\alpha\alpha}$
    in log-polar coordinates, as defined in
    figure~\ref{fig:polar_coord_explain}, from R5200. $\lambda^+=10$
    on the outer-most dotted circle and increases by a factor of 10
    for each dotted circle moving inward. The arrows mark features
    discussed in the text.}
  \label{fig:2d_E_T_xz}
\end{figure}

The other part of the turbulent transport decomposition,
(\ref{eq:turb_decomp_energy}), is $E_{ij}^{T^{\|}}$. Among the terms
in (\ref{eq:E_evol_eq_final}), $E_{ij}^{T^{\|}}$ is the only one that
describes energy transfer across scales. One-dimensional $E^{T^\|}$
spectra are shown in figure~\ref{fig:1d_E_T_xz}. Note that these
one-dimensional scale transfer spectra inherit the property of the
two-dimensional spectra (\ref{eq:turb_transport_decomp_property});
that is,
\begin{equation}
\int_0^\infty E_{ij,x}^{T^{\|}} (k_x,y) \intd k_x = \int_0^{\infty}
E_{ij,z}^{T^{\|}} (y,k_z) \intd k_z = 0, \quad \forall y
\label{eq:turb_transport_inhomo_int_in_xz}
\end{equation}
One obvious feature of the scale transfer spectra $E^{T^\|}$ is that
they are significantly larger than the wall-normal transport spectra
$E^{T_\bot}$, especially for the streamwise ($k_x$) spectra. For this
reason, spectra of the total turbulent convection term $E^{T}$ are
dominated by the scale-transfer component.

For all the components, at both Reynolds numbers, and for both the
$k_x$ and $k_z$ spectra, $E_{ij}^{T^\|}$ is dominated by two
features. There is a negative region that is aligned approximately
with the source for that component; production for \uu
(figure~\ref{fig:1d_E_P_uu}), and pressure-strain for \vv and \ww
(figure~\ref{fig:1d_E_PI_s}). In the overlap region, as expected, the
wavelengths of the dominant donor region scale with $y$, which is
indicated by the dashed white lines in figure~\ref{fig:1d_E_T_xz},
which correspond to the solid lines in figure~\ref{fig:1d_E_P_uu} for
\uu, and figure~\ref{fig:1d_E_PI_s} for \vv and \ww).  There is also a
dominant positive region at higher wavenumbers that aligns
approximately with the dissipation of each component
(figure~\ref{fig:1d_E_E}). Also as expected, in the overlap region,
wavelengths in this recipient region scale approximately with
$y^{1/4}$, indicated by the black chain-dashed lines, which correspond
to the solid lines in figure~\ref{fig:1d_E_E}. Note that the alignment
of the donor and recipient regions with the source and dissipation
peaks, as indicated by the dashed lines, is imperfect for the
streamwise spectra of \uu and to a lesser extent \ww. This arises
because, to a large extent, the scale transfer acts to redistribute
energy in orientation from the streamwise elongated modes, in addition
to the Kolmogorov-like transfer to smaller scales (see discussion of
the two-dimensional spectra below).

Near the wall, the scale transfer spectra should largely reflect the
autonomous near-wall dynamics, and should thus be largely independent
of Reynolds number in wall units. Indeed this appears to be the case
in figure~\ref{fig:1d_E_T_xz}. Close to the wall, there is also a
dominant transfer of energy in the \uu spectra from small to large
wavenumbers (large to small scales) that peaks at around $y^+ =
15-30$, with donor peak occurring at about $\lambda_x^+\approx1100$
and $\lambda_z^+\approx100$ (marked with $+$). This corresponds to the
near-wall peaks in the one dimensional energy spectra in
\citet{Lee:2015er}. The recipient peak of the \uu scale-transfer
spectra occur at $\lambda_x^+\approx 120$ and $\lambda_z^+\approx
30$. The dominant feature of the streamwise velocity fluctuations at
$y^+ = 15-30$ are the near-wall streaks, which are consistent with
these scales, and so this scale transfer presumably reflects the
break-up of the streaks.

The near-wall \ww scale-transfer spectra do not have as distinct a
near-wall peak for transfer from large to small scales, but they do
occur and at approximately the same locations in $y$ and
wavelength. However, in the spanwise ($k_z$) spectra, the near-wall
high-wavenumber recipient peak is absent; more about this below.  In
the \vv spanwise spectra, the scale-transfer donor and recipient peaks
occur at larger $y$ ($y^+\approx 40$) and smaller scale for donor
(factor of 3) and about the same scale for recipient. There are no
distinct near-wall donor and recipient peaks in the streamwise ($k_x$)
\vv transfer spectra.

A striking feature of the one-dimensional scale-transfer spectra is
that there are regions in which the energy is transferred to larger
scales. This occurs in the streamwise transfer spectra of \ww, and the
spanwise spectra of \uu and \ww. This ``inverse'' energy transfer is
most noticeable near the wall, centered around $y^+\approx 15$ (for
\uu) to $y^+\approx 25$ (for \ww). Indeed, in the streamwise \ww
transfer spectrum, there is no distinct high-wavenumber recipient
peak, and the energy removed from the donor peak is primarily
deposited at lower wavenumbers. This inverse energy transfer has been
noted previously in channel flow DNS
\citep{Cimarelli:2013ke,Cimarelli:2016bt,Aulery:2016cg}.  It is
interesting because it suggests that there are mechanisms by which
fluctuations with scales much larger than the dominant scales of the
near-wall dynamics are generated. However, it is not clear whether
this inverse transfer is autonomous to the near-wall flow, or whether
it must be mediated by large-scale fluctuations imposed from the outer
flow. The structure of the terms in $R^{T^\|}$
(\ref{eq:turb_horizontal}) suggest that it may be the latter. A
definitive determination could be made using a similar spectral
analysis of a near-wall DNS, with the outer turbulence removed as in
\citet{Jimenez:1999wf}. The inverse transfer in $k_z$ of \uu in the
outer flow is similarly interesting, because of the underlying
mechanisms that could create very wide (in $z$) structures.

As with $E_{ij}^{T_\bot}$, it is useful to evaluate the scale transfer
flux $E^{\Phi^\|}$ to identify the direction of net energy flow in
scale. To this end, we compute
\begin{equation}
E_{ij,x}^{\Phi^{\|}}(k_x,y) = \int_0^{k_x} E_{ij,x}^{T^{\|}} (K_x,y)
\intd K_x, \quad E_{ij,z}^{\Phi^{\|}}(y,k_z) = \int_0^{k_z}
E_{ij,z}^{T^{\|}} (y,K_z) \intd K_z
\label{eq:turb_transport_flux_in_xz}
\end{equation}
and plot as black lines in figure~\ref{fig:1d_E_T_xz}, the location at
each $y$ at which $E_{ij,x}^{\Phi^{\|}} = 0$ or $E_{ij,z}^{\Phi^{\|}}=
0$. To the left of these lines, the net scale-transfer energy flux is
toward large scales. From this it is clear that there is \uu and \ww
energy transfer to the lowest $k_z$ wavenumbers across almost the
entire channel, and \ww transfer to the lowest $k_x$ wavenumbers
across the channel.

Interpretation of the inverse transfer in the one-dimensional spectra
is complicated by the fact these spectra represent an integral over
the wavenumber in the other horizontal direction. To obtain a more
complete picture of the phenomenon, the two-dimensional scale-transfer
spectra at several $y$-locations near to and away from the wall (from
$y^+=5$ to $y/\delta=0.2$) are shown for the R5200 case in
figure~\ref{fig:2d_E_T_xz} in log-polar plots.  While these
two-dimensional spectra provide details of the scales that are donors
and recipients of energy, the energy flux that leads to this is not
uniquely defined. Thus we cannot uniquely determine the direction in
two-dimensions of the energy transfer.

The most striking feature of these two-dimensional scale transfer
spectra is the exceptionally strongly peaked donor region in the \uu
spectra located along the $k_z^\#$ axis, with dominant spanwise
wavenumber that decreases with $y$ ($k_zy\approx 1.5$ for $y^+>30$,
marked with an arrow).  This donor region transfers energy to a
more-or-less circular band of wavenumbers, that is, wavevectors of all
orientations in a band of wavenumber magnitudes $k$. However, there is
also a broad recipient peak (marked with $+$) that is weakly elongated
in the spanwise direction away from the wall ($k_z\lesssim k_x$ at
$y^+=1000$) and weakly elongated in the streamwise direction near the
wall ($k_z\gtrsim k_x$ at $y^+=15$ and 30). In all cases, the
recipient peak is at smaller $k_z$ than the donor peak (by about a
factor of 3). These scale transfer patterns are clearly consistent
with the apparent transfer to large scale in the spanwise
one-dimensional $E^{T^\|}_{11}$ spectrum.  The structure of the \uu
transfer spectra at all $y$ appear to be consistent with the break-up
of streamwise-elongated low and high speed streaks, which appear to be
dominant feature of the \uu spectra, due to instabilities. Near the
wall, such instabilities were identified by \citet{Jimenez:1999wf} as
part of the near-wall self-sustaining mechanism. The transfer to lower
spanwise wavenumber could arise if there was a tendency for streak
instabilities to become phase coherent in the span.

In addition, at $y^+=5$, \uu energy is transferred to a region along
the $k_z^\#$ axis with $\lambda^+>1000$ (marked with $\times$). This
inverse energy transfer does not appear to be intrinsic to the
near-wall self-sustaining mechanism. However, this region coincides
with the region in which there is wall-normal energy transport from
the outer flow (figure~\ref{fig:2d_nonlinear_transport_flux} and
\ref{fig:2d_nonlinear_transport_uu}). This suggests that the inverse
energy transfer is the result of the nonlinear interaction of
large-scales driven by the outer layer and inner-layer small-scales.
Nonetheless, this inverse energy transfer in \uu only occurs for $y^+
< 15$.

The two-dimensional scale transfer spectra of \vv are rather
unremarkable. Away from the wall ($y^+=1000$), they represent a nearly
isotropic transfer of energy from larger scales to smaller scales,
with little or no dependence on orientation. Closer to the wall
($y^+=15$ and 30), there is an anisotropic structure, but the transfer
is much weaker than for the other components. Consistent with the the
one-dimensional $E^{T^\|}_{22}$ spectra in figure~\ref{fig:1d_E_T_xz},
there is no significant contribution of \vv to scale transfer for
$y^+<30$.

In the \ww transfer spectrum near the wall ($y^+ < 15$) there is weak
inverse scale transfer near the wall ($y^+=5$). As with \uu, the
recipient region in figure~\ref{fig:2d_E_T_xz} coincides with the
recipient region in wall-normal transport
(Figure~\ref{fig:2d_nonlinear_transport_ww}). Hence, the non-linear
interaction between small-scales in the inner layer and relatively
larger-scales farther away ($y^+ =30-100$) could explain the near-wall
inverse transfer. However, the magnitude of inverse transfer is
minimal and occurs only for $y^+<30$. The primary energy transfer is
thus in orientation, from the somewhat oblique orientation with
$k_z\gtrsim k_x$ (i.e. somewhat elongated in the $x$-direction)
primarily to very strongly streamwise aligned orientations, with small
$k_x$. Presumably, near the wall the spanwise velocity fluctuations
are primarily associated with streamwise vortices. So, this streamwise
elongation captured in the transfer spectrum could be a consequence of
streamwise vortex stretching, which was identified as part of the
autonomous dynamics of the near-wall turbulence
\citep{Jimenez:1999wf}.

Near the wall, energy is deposited in the donor wavenumber region by
pressure strain inter-component transfer from \uu (see
figure~\ref{fig:2d_E_PI_s}). Further, inter-component energy transfer
from \ww to \vv occurs in the recipient region along $k_x=0$. There is
thus a chain of transfers from \uu to \ww in oblique Fourier modes,
then from those oblique modes to streamwise elongated modes, and
finally, from \ww to \vv. Farther from the wall ($y^+=1000$), the same
transfers in components and orientation occur, and presumably the same
process of streamwise vortex stretching is responsible for the
transfer to streamwise aligned orientations. However, far from the
wall, this orientation transfer to streamwise elongation is not as
dominant. There is also a general more isotropic transfer to smaller
scales (larger $k$). In addition, there is a preferential transfer to
orientations that are elongated in $z$ (i.e. the $k_z=0$ region). This
is necessary if the \ww dissipation is going to be as scale isotropic
as was observed in figure~\ref{fig:2d_E_uu}, since there are no
inter-component transfer sources of \ww with $k_z$ near zero, as
discussed in section~\ref{subsec:pressure_strain}. A similar weak
transfer of \ww to spanwise elongated modes occurs near the wall as
well. It is clear that this \ww transfer in orientation toward
streamwise and spanwise elongated modes is responsible for the
apparent inverse transfer for $y^+ \gtrsim 15$ in the one-dimensional
\ww transfer spectra (figure~\ref{fig:1d_E_T_xz}).

\subsection{Universality of small-scale motions}
\label{subsec:small_scale_universality}

\begin{figure}
  \begin{center}
    \includegraphics[width=\textwidth]{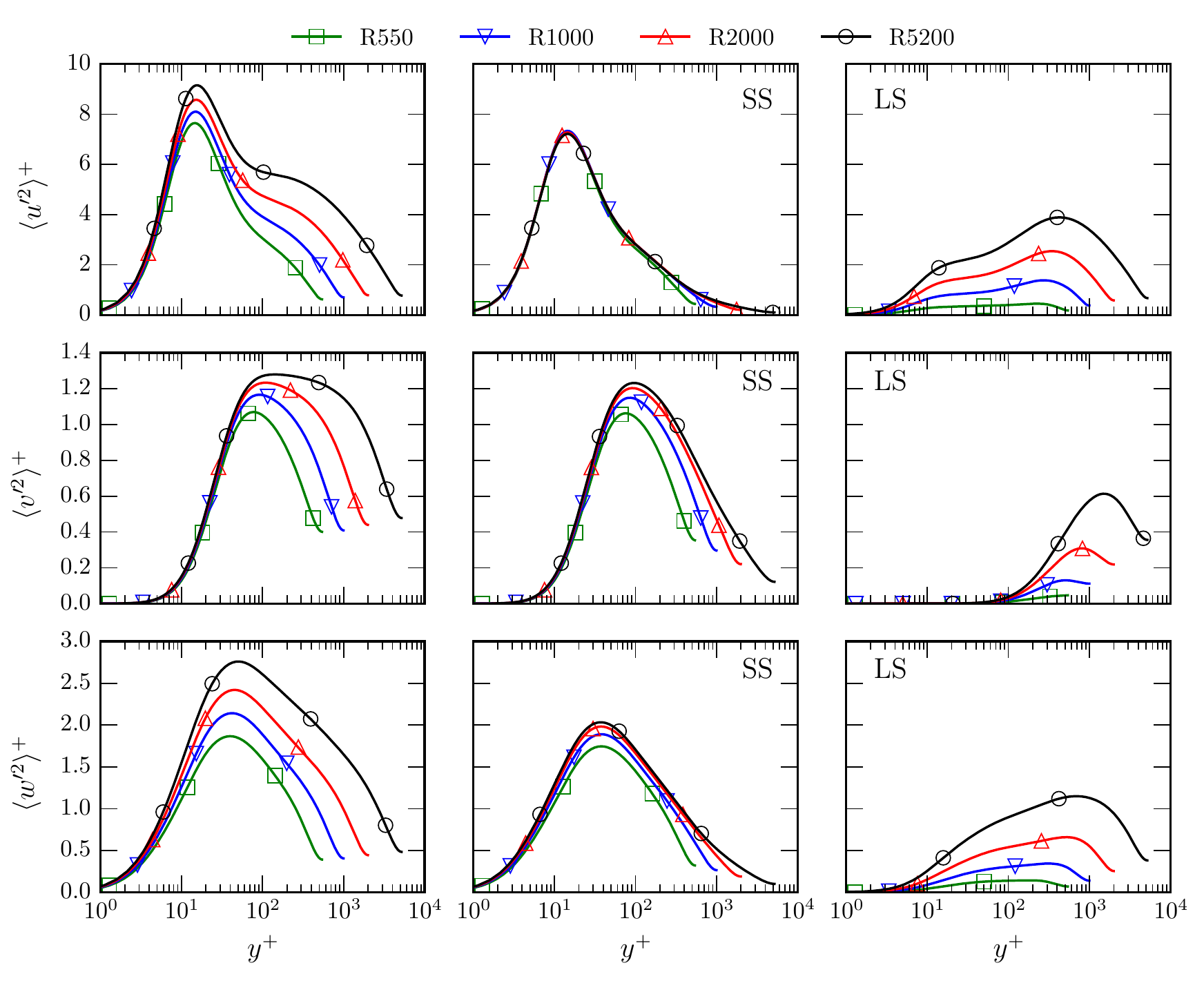}
  \end{center}
  \caption{Unfiltered, and high-pass (SS) and low-pass (LS) filtered
    velocity variances with $k_\textrm{cut-off}^+ = 0.00628$
    $\left(\lambda^+_\textrm{cut-off} = 1000\right)$.}
  \label{fig:filtering_uu}
\end{figure}

\begin{figure}
  \begin{center}
    \includegraphics[width=0.44\textwidth]{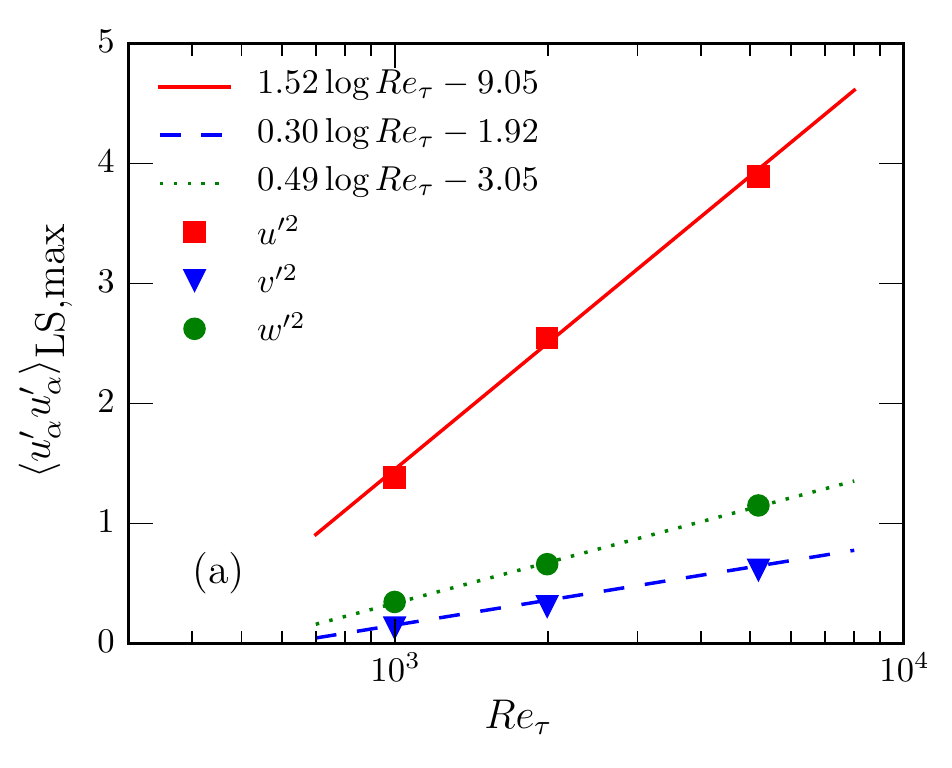}
    \includegraphics[width=0.44\textwidth]{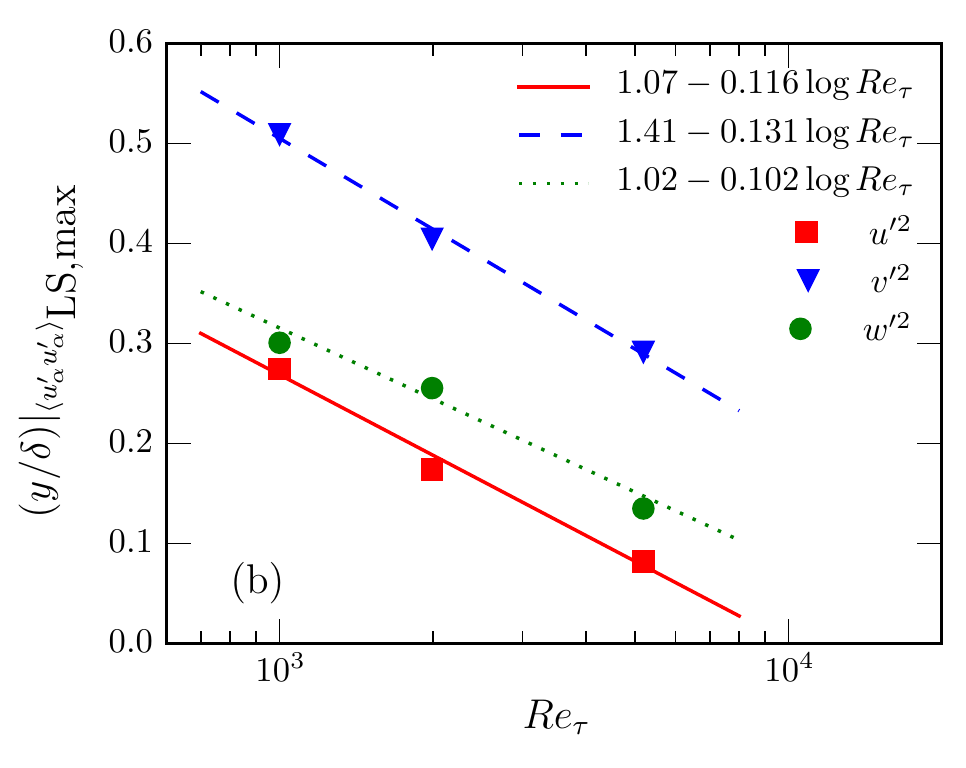}
  \end{center}
  \caption{Reynolds number dependence of $\langle u'_\alpha u'_\alpha
    \rangle_\textrm{LS}$; (a) maximum value of $\langle u'_\alpha
    u'_\alpha \rangle_\textrm{LS}$; (b) $y/\delta$ at which $\langle
    u'_\alpha u'_\alpha \rangle_\textrm{LS}$ is maximum.}
  \label{fig:uu_ls_peak}
\end{figure}

\begin{figure}
  \begin{center}
    \includegraphics[width=\textwidth]{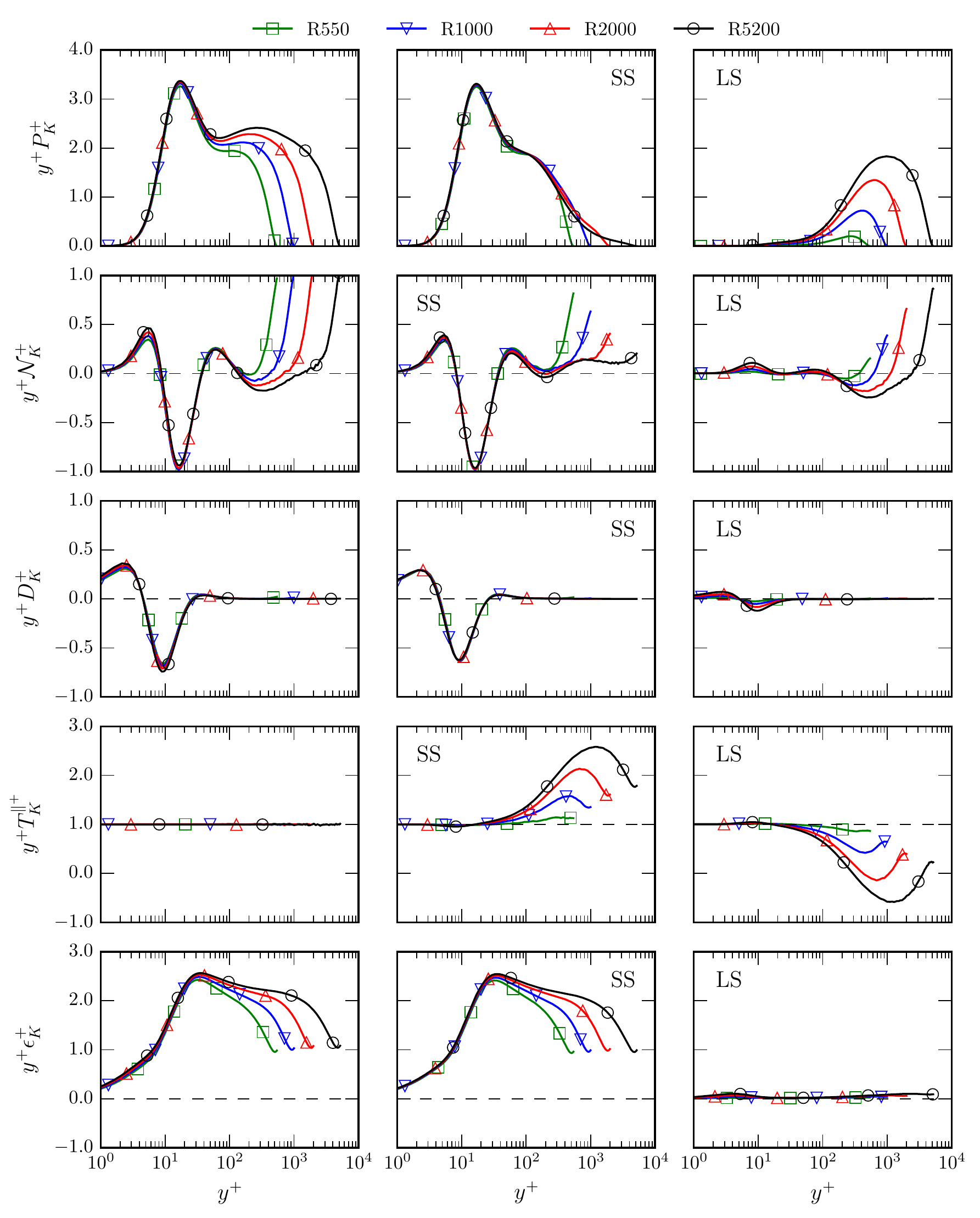}
  \end{center}
  \caption{Unfiltered, high-pass (SS) and low-pass (LS) filtered terms
    in the TKE balance equation, as log-densities;
    $k_\textrm{cut-off}^+ = 0.00628$ $\left(\lambda^+_\textrm{cut-off}
    = 1000\right)$.}
  \label{fig:filtering}
\end{figure}

\begin{figure}
  \begin{center}
    \includegraphics[width=0.4\textwidth]{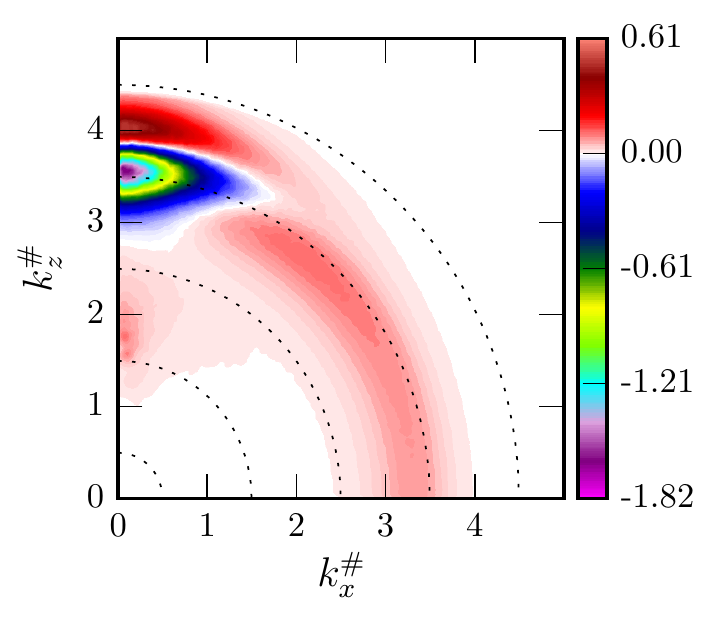}
  \end{center}
  \caption{Two-dimensional spectral density of $T^{\|}_{K}$ in
    log-polar coordinates, as defined in
    figure~\ref{fig:polar_coord_explain}, from R5200 at $y^+=8$. $\lambda^+=10$
    on the outer-most dotted circle and increases by a factor of 10
    for each dotted circle moving inward.}
  \label{fig:TKE_T_xz_0008}
\end{figure}

\def\LS#1{#1_{\textrm{LS}}}
\def\SS#1{#1_{\textrm{SS}}}
\def\bk{\mathbf{k}}
\def\uupss{ $\langle u'^2 \rangle^+_{\textrm{SS}}$ }
\def\vvpss{ $\langle v'^2 \rangle^+_{\textrm{SS}}$ }
\def\wwpss{ $\langle w'^2 \rangle^+_{\textrm{SS}}$ }
\def\uupls{ $\langle u'^2 \rangle^+_{\textrm{LS}}$ }
\def\vvpls{ $\langle v'^2 \rangle^+_{\textrm{LS}}$ }
\def\wwpls{ $\langle w'^2 \rangle^+_{\textrm{LS}}$ }

The results in previous sections indicate that the Reynolds number
dependence of the velocity variances and the terms in Reynolds
stress transport equations are a consequence of the large-scale
motions. For example, only the low wavenumber regions in the
one-dimensional spectra depend on $Re$. To quantify this, the
large and small-scale contributions to the turbulent kinetic energy
and the kinetic energy transport terms were determined. In particular,
if $\Psi(y)$ is a one-dimensional statistical profile, such as the TKE
or a transport equation term, and $E_{\Psi}(\bk,y)$ is the
two-dimensional spectral density of $\Psi$ ($\bk$~is the
two-dimensional wave vector in $x$ and $z$), the large-
and small-scale contributions
($\LS\Psi$ and $\SS\Psi$, respectively) are given by  
\begin{subequations}
\begin{equation}
\Psi = \LS\Psi + \SS\Psi
\end{equation}
where
\begin{equation}
\LS\Psi(y) = \int_{|\bk|<k_{\textrm{cut-off}}} E_\Psi (\bk,y)
\intd \bk
\label{eq:lpf}
\end{equation}
\begin{equation}
\SS\Psi(y) = \int_{|\bk|>k_{\textrm{cut-off}}} E_\Psi
(\bk,y) \intd \bk .
\label{eq:hpf}
\end{equation}
\label{eq:filtered_stat}
\end{subequations}
\def\LS#1{#1_\textrm{L}}
\def\SS#1{#1_\textrm{S}}

Selecting a cut-off wavenumber is somewhat arbitrary and controversial
in the context of defining the large-scales
\citep{Monty:2009cd,Hutchins:2009ew,Marusic:2010bn,Ganapathisubramani:2012dh,WuEtal2012,ChinEtal2014,AhnEtal2017}. However,
as discussed in \S\ref{subsec:energy_spectra} the two-dimensional
spectra of \uu at $y^+=15$ (near the peak of the \uu profile) has two
distinct features (figure~\ref{fig:2d_uu_y_plus_15}). One is the
dominant small-scale feature ($\lambda^+\lesssim 1000$) that is
associated with the autonomous near-wall dynamics, and as is clear in
the figure, is approximately Reynolds number independent. The other is
the large scale feature that arises from interaction with the outer
flow ($\lambda^+\gtrsim 1000$) which is very much Reynolds number
dependent, and all but absent at $Re=550$. Therefore, it is expected
that defining the large and small scales using the cutoff wavenumber
$k^+_\textrm{cut-off}=0.00628$ ($\lambda^+_\textrm{cut-off} = 1000$)
will result in a Reynolds number independent peak in the small-scale
\uu and turbulent kinetic energy (TKE) profiles. This is the cut off
used here. Further, notice in figure~\ref{fig:2d_uu_y_plus_15} that a
one-dimensional cut off (filtering only in the $x$ direction), as done
in many of the references above, would include some of the
Reynolds-number-independent parts of the two-dimensional spectrum with
the large scales. It appears that, at least for the purpose of
separating Reynolds number independent and dependent scales in the
near-wall layer, a filter that is isotropic in the $x$--$z$ plane
(surprisingly), with $\lambda^+_\textrm{cut-off}\approx 1000$, is most
appropriate.

The contribution of small- and large-scale motions to the velocity
variances are shown in figure~\ref{fig:filtering_uu}.  As expected,
the peak of \uupss (at $y^+=15$) is independent of Reynolds number for
all cases, indeed the \uupss profile is Reynolds number independent
for $y/\delta\lesssim 0.2$.  The \vvpss and \wwpss profiles have weak
$Re$ dependencies, which become weaker as the $Re$
increases. Particularly, \wwpss from R2000 and R5200 are minimally
different for $y/\delta \lesssim 0.2$.  Interestingly, \uupss and
\wwpss decrease logarithmically with $y$ in the interval $y^+ > 50$
and $y/\delta < 0.1$.  The attached eddy hypothesis predicts
logarithmic dependence of \uu and \ww in this region at high Reynolds
number
\citep{Townsend:1976uj,Perry:1986vs,Woodcock:2015ji,deSilva:2015he},
and indeed this was observed in \ww in these channel simulations
\citep{Lee:2015er}, but not in \uu. In the spectra shown in
figure~\ref{fig:2d_uu}, the primary qualitative differences between
\uu and \ww in this $y$-region are the highly energetic
long-wavelength streamwise-elongated streaky modes. But these modes
are precisely those eliminated in \uupss and \wwpss. It appears then
that the small scale modes with $\lambda^+<1000$ are consistent with
the attached eddy hypothesis. The large-scale streaky modes that
dominate the \uu spectra, on the other hand, are apparently different
in character from the random distribution of wall-attached
scale-similar eddies assumed by the hypothesis. That is, they are not
``attached eddies.''

Several experimental studies have found that \uup at very high
$Re_\tau$ (say $Re_\tau > 20000$) decreases logarithmically with $y$
for $y^+>3Re_\tau^{1/2}$ and $y/\delta<0.15$
\citep{Marusic:2013hf,Hultmark:2013dz}. Even for the R5200 case, this
would imply a very short range of logarithmic decrease
($220<y^+<780$), and indeed no convincing logarithmic decrease in \uup
was observed \citep{Lee:2015er}.  Presumably this is because \uupls is
not varying logarithmically in this region. Perhaps at higher Reynolds
number, the \uupls profile will develop a logarithmically increasing
region as occurs in \wwpls, which in combination with the
logarithmically decreasing \uupss would yield a logarithmic \uup. Even
if so, the difference in character between the small- and large-scale
contributions to this logarithmic dependence of \uu and \ww remains.

As expected, $\langle u'_\alpha u'_\alpha\rangle^+_\textrm{LS}$ are
strongly Reynolds number dependent. The maximum values of $\langle
u'_\alpha u'_\alpha \rangle^+_\textrm{LS}$ increase logarithmically
with $Re_\tau$ as shown in figure~\ref{fig:uu_ls_peak}a. Recall that
the maximum values of \uup at $y^+ \approx 15$ also increases
logarithmically with $Re_\tau$ \citep{Lee:2015er}, though \uupss does
not.  The peaks of $\langle u'_\alpha u'_\alpha \rangle^+_\textrm{SS}$
occur at approximately constant $y^+$ because the near-wall
small-scale turbulence scales in wall units. Similarly, one might
expect that the peaks of $\langle u'_\alpha u'_\alpha
\rangle^+_\textrm{LS}$ would occur at approximately constant
$y/\delta$ because the large scales are dominated by outer-layer
turbulence, scaling with $\delta$. However, the $y/\delta$ at which
$\langle u'_\alpha u'_\alpha \rangle^+_\textrm{LS}$ is maximum
decreases logarithmically with $Re_\tau$, as shown in
figure~\ref{fig:uu_ls_peak}b. This suggests that a mixed scaling may
be needed for the outer turbulence. Indeed, mixed scaling of both the
$y$-location of the outer streamwise spectral peak, and outer region
streamwise spectral peak wavenumbers have been observed in boundary
layers and pipes at up to $Re_\tau\approx20000$
\citep{MathisEtal2009,VallikiviEtal2015}.  Also, recall that the
choice of the cut-off wavenumber defining the large and small scales
is somewhat arbitrary and that the coefficients in the logarithmic
relationships in figure~\ref{fig:uu_ls_peak} will vary with the
cut-off wavenumber.

The small- and large-scale contributions to the terms in TKE transport
equation are shown in figure~\ref{fig:filtering} (recall that $\Pi_K$
is identically zero). Near the wall ($y/\delta\lesssim 0.2$), the weak
Reynolds number dependence of the transport terms is largely absent
from the small-scale contribution to those terms. The exception is the
non-linear transport term $\mathcal{N}$, in which the Reynolds number
dependence is reduced, but not eliminated. The small-scale
contribution to $\mathcal{N}$ includes net transport from the
near-wall region to $y^+>200$. Because the near-wall profiles are
nearly Reynolds number independent, so is the rate of transport to
$y^+>200$. With increasing Reynolds number, the range of $y^+$ over
which this energy is deposited increases, so that actual values of
$y\SS{\mathcal{N}}$ decrease. This is the reason for the reduction and
near-elimination of the centerline peak in $\SS{\mathcal{N}}$. It
appears that with even higher Reynolds number, this centerline peak
would be eliminated entirely, and the value of $y\mathcal{N}$ would go
to zero for large enough $y^+$. If so, it would be consistent with the
notion that the direct effect of the autonomous near-wall dynamics on
the outer region turbulence can only extend to some finite distance
from the wall, perhaps of order $10^4$ wall units. Such a large (in
wall units) zone of influence would suggest that low Reynolds number
effects will persist until the the Reynolds number is high enough for
the small-scale transport to be confined to a thin region near the
wall; perhaps confined to the log region and below, which would
require $Re_\tau\gtrsim5\times10^4$ or more.

As expected from the spectra examined in \S\ref{subsec:prod}, the
large-scale production of TKE occurs primarily in the outer region,
and increases with $Re$. Also as expected from
\S\ref{subsubsec:wall-normal_nonlinear}, $\mathcal{N}$ transports some
of this large-scale energy to a near-wall region centered at
$y^+\approx 10$, and the rate of this transport, while small, is
increasing with Reynolds number. This is consistent with large-scale
modulation of the near-wall turbulence
\citep{Hutchins:2007kd,Marusic:2010hy}, which increases with Reynolds
number. Another expected result is that the small-scale contribution
dominates the viscous terms (the linear transport, $D$, and the
dissipation, $\epsilon$). The small contribution of the large scales
to the viscous terms near the wall arises because near the wall, even
low wavenumber modes will have large gradients in the wall-normal
direction. The fact that these large-scale viscous effects increase
with Reynolds number is due to the increasing transport of large-scale
energy to the wall, as described above, consistent with the
observations of \citet{Hoyas:2008jl}.

The small- and large-scale contributions to the scale transfer term
$T^\|_K$ are necessarily equal and opposite, since their sum must be
zero. These are simply the net transfer of energy between large and
small scales defined by $\lambda^+_{\textrm{cut-off}}=1000$. This
transfer occurs primarily away from the wall ($y^+>20$), is primarily
from large to small scales, and is growing with Reynolds number in
wall units, as expected. However, in a region centered around
$y^+\approx 8$ there is a mild net transfer from small to large scales
at $Re_\tau=5200$.  As is clear in the two-dimensional (log-polar)
spectrum of $T_K^\|$ at $y^+=8$ (figure~\ref{fig:TKE_T_xz_0008}), this
energy transfer to large scales is depositing energy in the streamwise
elongated streaky modes that dominate the large scale \uu spectrum
(figure~\ref{fig:2d_uu}).  This is apparently due to a nonlinear
response of the autonomous near-wall dynamics to the large-scale
modulation imposed by the streamwise elongated modes in \uu in the
outer flow.

\section{Discussion and Conclusion}
\label{sec:conclusion}
Spectral analysis of the terms in the Reynolds stress transport
equations was conducted to investigate the flow of turbulent energy in
space, scale and components for turbulent channel flow up to
$Re_\tau\approx 5200$. It has long been understood that in a
wall-bounded parallel shear flow like this there is only production of
\uu, which gets redistributed to \vv and \ww by pressure
inter-component transfer; that production is primarily at large scales
and there is transfer to small scales where dissipation occurs; and
that there is interaction between the near-wall and outer turbulence
through wall-normal transport. However, the detailed spectral analysis
reported here provides a much more detailed picture of these energy
flows. This analysis also yielded some remarkable insight into the
interaction between the near-wall and out-layer turbulence, and the
dominant features of the outer-layer turbulence.

One of the more striking outcomes of the current spectral analysis is
evident in the log-polar two-dimensional spectra shown in
figures~\ref{fig:2d_uu} and \ref{fig:2d_uu_y_plus_15}. Here the
spectra of \uu away from the wall ($y^+>300$) are dominated by
streamwise elongated modes, with spanwise wavelength that increases
with distance from the wall.  These elongated modes away from the wall
are driven by production that is also dominant in these elongated
modes (figure~\ref{fig:2d_P_uu_new}). Further the production appears
to have sharp spectral peaks that occur at a set of discrete spanwise
wavelengths at different wall distances. These peaks are separated by
a factor of 1.5 in wavelength, with long wavelengths dominant at
larger distances from the wall. The wall-normal turbulent transport of
\uu away from the wall is also dominated by these streamwise elongated
modes, with transport acting to project this large-scale streamwise
elongated structure on to the near-wall layer. Thus, the modulation of
the near-wall autonomous dynamics by the outer layer described by
\citet{Hutchins:2007kd,Marusic:2010hy} is primarily through these
large scale streamwise elongated modes. The result is the distinctive
\uu spectral structure near the wall, shown at $y^+=15$ as a function
of Reynolds number in figure~\ref{fig:2d_uu_y_plus_15}. Here there is
a clear distinction between the nearly Reynolds number independent
spectral structure produced by the autonomous near-wall dynamics, in
modes with wavelength $\lambda^+<1000$, and the Reynolds number
dependent streamwise elongated spectral structure imposed by transport
from the outer layer, with $\lambda^+>1000$.

While the dominance of the streamwise elongated modes in the log and
outer layers is remarkable, it appears to be consistent with a number
of previous observations. For example, very long streamwise
wavelengths of order $8R$ to $16R$ were observed in pipe flow by
\citet{Guala:2006dd}. These observations of very-large-scale motions
(VLSM) are consistent with the elongated modes observed here, which
have wavelengths of order $12\delta$ to $25\delta$. The VLSM of
\citet{Guala:2006dd} also made a significant contribution to the
Reynolds stress, consistent with the current streamwise elongated
modes. Furthermore, the current streamwise elongated modes are
inclined from streamwise by $6^\circ$ or less, consistent with VLSMs
observed in pipe flow simulations by \citet{BaltzerEtal2013} and
\citet{AhnSung2017}.

The imprint of these streamwise elongated modes on the near-wall layer
appears to correspond to the near-wall ``inactive motions'' that do
not carry Reynolds stress, as hypothesised by Townsend
\citep{Townsend:1976uj}, and analysed by many others
\citep{Perry:1986vs,Perry:1995jx,Kunkel:2006ju,Hutchins:2007ty,Hwang:2015iz}. This
is indicated by the two-dimensional production spectrum
(figure~\ref{fig:2d_P_uu_new}) where the $\lambda^+>1000$ streamwise
elongated modes are absent near the wall (in any constant $y$ plane,
production is proportional to Reynolds stress).

Many of the features of the energy flow in space, scale and components
arise because of the streamwise-elongated structure of \uu described
above. Consider that in the relatively large Reynolds number R5200
case, there is sufficient scale separation between the inner and outer
regions for there to be a significant ``scaling region'' from
$y^+\approx300$ to $y/\delta\approx0.6$ in which the structure of the
energy flows is approximately scale-similar. Here, the length scales
at which the production occurs scale with $y$, and the dissipative
length scales approximately like $(y\nu^3/u_\tau^3)^{1/4}$, which is
consistent with expectations from Kolmogorov. The dissipation is
increasingly isotropic with respect to components and scale as $y$
increases, but the production is decidedly not, since it occurs only
in \uu and primarily in streamwise-elongated modes. Energy must be
transferred from \uu to the other components by pressure-strain, but
streamwise-elongated \uu modes cannot contribute significantly to
pressure-strain. Instead, there is a strong transfer in orientation of
\uu energy from the streamwise-elongated modes.  Energy is then
transferred from \uu to the other components over about a decade-wide
band of wavenumber magnitudes, and over a range of orientations, but
is strongest for transfer to \ww in oblique orientations with
$k_x\approx k_z$. Direct transfer to \vv is weaker and more evenly
distributed in orientation. Orientation transfer in \ww deposits
energy in streamwise elongated modes where it is transferred to \vv
through pressure strain. Scale transfer to smaller scales can then
occur in all components and across a broad distribution of
orientations. The population of streamwise elongated modes and the
inter-component transfer from \ww to \vv may be associated with
streamwise vortex stretching. Further the transfer to oblique modes
with $k_x\gtrsim k_z$ (inclination from streamwise greater than about
$45^\circ$) is consistent with break-down of low and high speed
streaks. Completing the picture in the scaling region is wall-normal
transport, which in all components is dominated by a scale-similar
transport away from the wall, though at large scales there is also
transport from the scaling region to the near-wall region.

In the near-wall region, many of the features of the inter-component
and scale transfer spectra are similar to those described above. The
major differences are that in the near-wall region, the scale transfer
to smaller scales over a broad range of orientations only occurs in
\uu, and that there is a more complex wall-normal transport. The
former is consistent with the intrinsically low Reynolds number of the
near-wall turbulence and the latter is presumably due to the presence
of the wall. In the near-wall region, many of the features of the
transfer spectra can be connected to the processes that have been
identified as part of a self-sustaining near-wall turbulence mechanism
\citep{Jimenez:1999wf}, specifically the instability and break-down of
low- and high-speed streaks and streamwise vortex stretching. The
similarity of the features of the transfer spectra near the wall and
in the scaling region suggest that the same processes are active in
the scaling region. This would be consistent with observations of
streaks and streamwise vortices in homogeneous shear flows
\citep{Lee:1990tl}, and suggests an intrinsic dynamic mechanism in the
scaling region, as hypothesised by
\citet{Jimenez:2012ef,Jimenez:2018hg}.

Both near the wall and in the scaling region, there is scale transfer
in orientation, i.e. transfer between modes with approximately the
same wavenumber magnitude, but different orientation. This implies
that one of the wavenumber components is smaller in the recipient
region than the donor region, and is responsible for the apparent
inverse energy transfer in one-dimensional spectra observed here and
previously by others
\citep{Cimarelli:2013ke,Cimarelli:2016bt,Aulery:2016cg}. Thus, this
does not really represent energy transfer to larger scales. However,
very near the wall ($y^+<15$) there is a true inverse scale transfer
of \uu from the dominant streaky structures with $\lambda^+\approx
100$ to streaky modes with $\lambda^+>1000$, which appears to be
driven by interaction of the streaks with large-scale out-layer
structures.

The wall-normal transport spectra provide insight into the interaction
between the near-wall region and the turbulence farther from the
wall. Transport of \uu dominates this interaction, and in a spectral
region with $\lambda^+>1000$, energy transport it toward the wall from
as far away as $y^+=700$ ($y/\delta=0.13$) and is primarily deposited
in the near-wall region where $y^+\lesssim 70$. This impact of
large-scale outer-region turbulence on the near-wall is responsible
for the Reynolds number dependence of near-wall statistics, since when
a high-pass filter is used to remove these large-scale modes, the
Reynolds number dependence is eliminated.  This is consistent with the
suggestion of large-scale modulation of the near-wall turbulence by
large-scale outer turbulence by
\citet{Hutchins:2007kd,Marusic:2010hy}. It also implies that in a
large eddy simulation (LES) representing turbulence at large enough
horizontal scale, the interaction between the near-wall and out
turbulence will be primarily one-way from outer to near-wall. This
suggests that a near-wall model for such an LES based on universal
small-scale near-wall turbulence statistics should be possible.

Finally, the spectral analysis of the Reynolds stress transport
processes pursued here provide a rather intricate picture of the
workings of wall-bounded turbulence. However, there are features of
this picture that are not well understood in terms of underlying
turbulent flow mechanisms. As such, the results presented here
represent a challenge to conceptual models of wall-bounded turbulence
(e.g. attached eddy models \citep{Townsend:1976uj}) to be able to
represent the spectral phenomena observed here.

\section{Acknowledgment}
The work presented here was supported by the National Science
Foundation under Award Number [OCI-0749223]. An award of computer time
was provided by the Innovative and Novel Computational Impact on
Theory and Experiment (INCITE) program. This research used resources
of the Argonne Leadership Computing Facility, which is a DOE Office of
Science User Facility supported under Contract DE-AC02-06CH11357.

\bibliographystyle{jfm}

\end{document}